\documentclass[aps,prx,twocolumn,unsortedaddress,floatfix,nofootinbib,superscriptaddress]{revtex4-2}
\usepackage{amsthm}
\usepackage{amsfonts}
\usepackage{siunitx}
\usepackage{amsmath}
\usepackage{amssymb}
\usepackage{graphicx}
\usepackage{verbatim}
\usepackage[colorlinks]{hyperref}
\usepackage{tikz}
\usepackage{braket}
\usepackage{xcolor}
\usepackage{lineno}
\setcitestyle{super,open={},close={}}

\makeatletter%
\def\@ssect@ltx#1#2#3#4#5#6[#7]#8{%
  \def\H@svsec{\phantomsection}%
  \@tempskipa #5\relax
  \@ifdim{\@tempskipa>\z@}{%
    \begingroup
      \interlinepenalty \@M
      #6{%
       \@ifundefined{@hangfroms@#1}{\@hang@froms}{\csname @hangfroms@#1\endcsname}%
       {\hskip#3\relax\H@svsec}{#8}%
      }%
      \@@par
    \endgroup
    \@ifundefined{#1smark}{\@gobble}{\csname #1smark\endcsname}{#7}%
  }{%
    \def\@svsechd{%
      #6{%
       \@ifundefined{@runin@tos@#1}{\@runin@tos}{\csname @runin@tos@#1\endcsname}%
       {\hskip#3\relax\H@svsec}{#8}%
      }%
      \@ifundefined{#1smark}{\@gobble}{\csname #1smark\endcsname}{#7}%
      \addcontentsline{toc}{#1}{\protect\numberline{}#8}%
    }%
  }%
  \@xsect{#5}%
}%
\makeatother

\definecolor{linkcolor}{RGB}{0,83,166}
\hypersetup{
  colorlinks = true,
  allcolors = {linkcolor}
}

\begin{document}
\newcommand{\mytitle}{Quantum critical dynamics in a 5000-qubit programmable spin glass}
\title{\mytitle}

\newcommand{\affildw}{D-Wave Systems, Burnaby, British Columbia, Canada}
\newcommand{\affilsfu}{Department of Physics, Simon Fraser University, Burnaby, British Columbia, Canada}
\author{Andrew D.~King}
\email[]{aking@dwavesys.com}
\affiliation{\affildw}

\author{Jack Raymond}
\affiliation{\affildw}
\author{Trevor Lanting}
\affiliation{\affildw}
\author{Richard Harris}
\affiliation{\affildw}
\author{Alex Zucca}
\affiliation{\affildw}
\author{Fabio Altomare}
\affiliation{\affildw}
\author{Andrew J.~Berkley}
\affiliation{\affildw}
\author{Kelly Boothby}
\affiliation{\affildw}
\author{Sara Ejtemaee}
\affiliation{\affildw}
\author{Colin Enderud}
\affiliation{\affildw}
\author{Emile Hoskinson}
\affiliation{\affildw}
\author{Shuiyuan Huang}
\affiliation{\affildw}
\author{Eric Ladizinsky}
\affiliation{\affildw}
\author{Allison J.R.~MacDonald}
\affiliation{\affildw}
\author{Gaelen Marsden}
\affiliation{\affildw}
\author{Reza Molavi}
\affiliation{\affildw}
\author{Travis Oh}
\affiliation{\affildw}
\author{Gabriel Poulin-Lamarre}
\affiliation{\affildw}
\author{Mauricio Reis}
\affiliation{\affildw}
\author{Chris Rich}
\affiliation{\affildw}
\author{Yuki Sato}
\affiliation{\affildw}
\author{Nicholas Tsai}
\affiliation{\affildw}
\author{Mark Volkmann}
\affiliation{\affildw}
\author{Jed D.~Whittaker}
\affiliation{\affildw}
\author{Jason Yao}
\affiliation{\affildw}
\author{Anders W.~Sandvik}
\email[]{sandvik@buphy.bu.edu}
\affiliation{Department of Physics, Boston University, 590 Commonwealth Avenue, Boston, Massachusetts 02215, USA}

\author{Mohammad H.~Amin}
\email[]{mhsamin@dwavesys.com}
\affiliation{\affildw}
\affiliation{\affilsfu}

\date{\today}
\begin{abstract}
  Experiments on disordered alloys~\cite{Brooke1999,Aeppli2005,Das2008} suggest that spin glasses can be brought into low-energy states faster by annealing quantum fluctuations than by conventional thermal annealing.  Due to the importance of spin glasses as a paradigmatic computational testbed, reproducing this phenomenon in a programmable system has remained a central challenge in quantum optimization~\cite{Kadowaki1998,Santoro2002,Harris2010,Roennow2014,Katzgraber2014,Hen2015,Heim2015,Boixo2016,Denchev2016,Albash2018}.  Here we achieve this goal by realizing quantum critical spin-glass dynamics on thousands of qubits with a superconducting quantum annealer.
  We first demonstrate quantitative agreement between quantum annealing and time-evolution of the Schr\"odinger equation in small spin glasses.  We then measure dynamics in 3D spin glasses on thousands of qubits, where simulation of many-body quantum dynamics is intractable.  We extract critical exponents that clearly distinguish quantum annealing from the slower stochastic dynamics of analogous Monte Carlo algorithms, providing both theoretical and experimental support for a scaling advantage in reducing energy as a function of annealing time.
\end{abstract}

\maketitle

\def\title#1{\gdef\@title{#1}\gdef\THETITLE{#1}}

\begin{figure*}
  \includegraphics[scale=1]{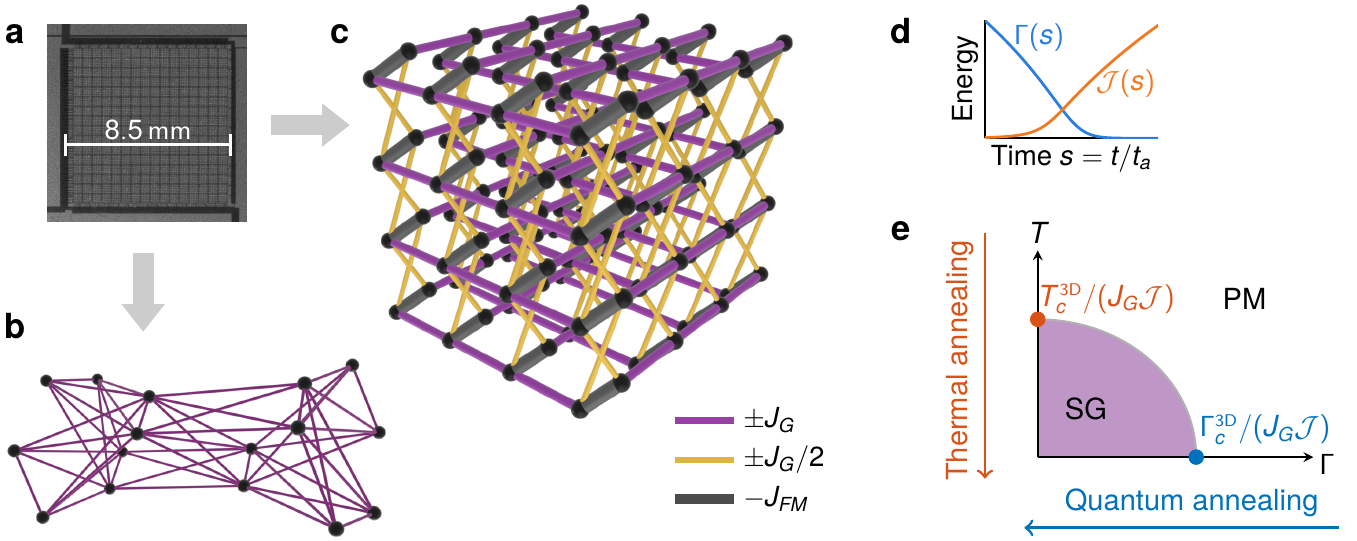}
  \caption{{\bf Programmable quantum spin glasses.} {\bf a}, QA processor realizing a transverse-field Ising model in pairwise-coupled superconducting flux qubits, into which various lattice geometries can be programmed.  {\bf b}, 16-spin graph used for small-scale studies of Schr\"odinger evolution.  Each line represents a coupling, whose energy is set to $J_G$ or $-J_G$ at random.  {\bf c}, 3D structure, where gray bonds represent ferromagnetically coupled dimers and any two dimers have total coupling $J_G$, $-J_G$, or $0$ between them.  {\bf d}, The QA schedule guides the system from a quantum paramagnet toward a classical state by annealing $\Gamma(s)/\mathcal J(s)$ over time $t_a$.  {\bf e}, 3D quantum spin-glass phase diagram.  Paramagnet and spin-glass phases are separated by a thermal phase transition at temperature $T=T_c^{\text{3D}}>0$ when $\Gamma=0$, and a quantum phase transition at $\Gamma_c^{\text{3D}}$ when $T=0$.}\label{fig:1}
\end{figure*}

The study of spin glasses initiated an enormously productive exchange between physics and computer science~\cite{Mezard2009,Stein2013}.  One key byproduct of this exchange was the invention of simulated annealing (SA) \cite{Kirkpatrick1983}, a method of optimization that simulates a gradually cooling system as it settles into a low-energy state.  Recent decades have seen annealing brought to bear against countless multivariate optimization applications, seeking low-energy states that translate to low-cost solutions~\cite{Tan2008}.

Passing through a thermal phase transition---as in SA---is not the only way to evolve a spin system from an ``easy'' disordered phase into a ``hard'' glassy phase.  One can also pass through a quantum phase transition (QPT), where the ground state undergoes a macroscopic shift in response to changing quantum fluctuations.  Both experiments~\cite{Brooke1999} and simulations~\cite{Kadowaki1998,Santoro2002} suggested that quantum annealing (QA) can guide a spin glass toward equilibrium faster than thermal annealing.  Thus originated QA as a means of both studying quantum critical phenomena and optimizing quadratic objective functions~\cite{Albash2016,Das2005}.  Simulating the Schr\"odinger dynamics of QA with a classical computer is an unpromising optimization method, since memory requirements grow exponentially with system size.  But, as Feynman famously asked~\cite{Feynman1982}, ``Can you do it with a new kind of computer---a quantum computer?''

This question motivated not only the development of programmable QA processors \cite{Harris2010,Johnson2011,Lechner2015,Weber2017,Novikov2018,Hauke2020}, but a more general effort to probe the capabilities of near-term quantum devices via quantum simulation, including in trapped ions~\cite{Blatt2012,Monroe2021}, ultracold atoms~\cite{Gross2017}, and Rydberg arrays~\cite{Scholl2020} (the latter was recently explored as an annealing optimizer~\cite{Ebadi2022}).  While D-Wave QA processors have already been used to simulate quantum spin glasses in a decohering thermal bath~\cite{Harris2018}, it was only recently shown that they can simulate QPTs with negligible interaction with the thermal environment~\cite{King2022}: coherent dynamics of quantum-critical phenomena despite the finite temperature of the apparatus itself.  Here we use a QA processor to study the critical dynamics of a spin-glass QPT.  The exceedingly slow dynamics of the spin-glass state make this phase transition vitally important in the study of quantum optimization.  We compare these dynamics against SA and simulated quantum annealing (SQA), an algorithm based on path-integral Monte Carlo \cite{Suzuki1976} that reproduces thermal equilibrium statistics of QA~\cite{Roennow2014, Denchev2016, Isakov2016}.

\section*{Quantum annealing and Schr\"odinger dynamics}

We use a D-Wave Advantage QA processor (Fig.~\ref{fig:1}a) whose pairwise-coupled superconducting flux qubits can be programmed to realize a transverse-field Ising model described by the Hamiltonian
\begin{eqnarray}
  \mathcal H(s) &=& \Gamma(s)\mathcal H_D + \mathcal J(s) \mathcal H_I\label{eq:ham}\\
  \mathcal H_D &=& -\sum_i \sigma_i^x \label{eq:hamd}\\
  \mathcal H_I &=&  \sum_{i,j}J_{ij} \sigma_i^z\sigma_j^z\label{eq:hamp}
\end{eqnarray}
Here $\sigma_i^x$, $\sigma_i^z$ are Pauli operators on qubit $i$, $s$ is a unitless normalized time, the transverse field $\Gamma(s)$ imparts quantum fluctuations through the driver Hamiltonian $\mathcal H_D$, and $\mathcal J(s)$ is the energy scale of the classical Ising Hamiltonian $\mathcal H_I$.  Over annealing time $t_a$, $s=t/t_a$ increases from $0$ to $1$, annealing the system from a quantum paramagnet dominated by $\mathcal H_D$, to a classical Ising model dominated by $\mathcal H_I$, following an annealing schedule as in Fig.~\ref{fig:1}d. The coupling coefficients $J_{ij}$ can be programmed into a variety of 2D and 3D geometries, among others~\cite{Harris2018,King2018,Denchev2016,Nishimura2020,Weinberg2020,Zhou2020} (the QA processor also provides programmable biases, which we set to zero in this study).

Although our main focus is on large spin glasses, we first seek evidence of coherent quantum dynamics in an ensemble of small spin glasses.  Taking the 16-spin graph in Fig.~\ref{fig:1}b, we generate spin-glass realizations with each coupling set to $J_{ij}=+1$ or $-1$ uniformly at random.  We select 100 spectrally unique realizations in which $\mathcal H_I$ has two ground states and many first excited states.  At this scale we can numerically evolve the time-dependent Schr\"odinger equation
\begin{equation}
i\hbar\frac d{dt}\ket{\psi(t)} = \mathcal H(t/t_a)\ket{\psi(t)},
\end{equation}
for the wavefunction $\ket{\psi(t)}$, where $\mathcal H(s)$ is given by Eq.~(\ref{eq:ham}).  Let $\ket{\phi_i(s)}$ denote instantaneous eigenstates of $\mathcal H(s)$ with eigenvalues $E_i(s)$.  Sufficiently slow evolution results in adiabatic quantum optimization (AQO)~\cite{Farhi2001} into the twofold-degenerate final ground states $\ket{\phi_0(1)}$ and $\ket{\phi_1(1)}$. The relevant timescale is proportional to $\delta_{\text{min}}^{-2}$, where
\begin{equation}
  \delta_{\text{min}} = \min_s\left|E_2(s)-E_0(s)\right|
\end{equation}
is the minimum parity-preserving eigengap. Faster anneals have a higher probability of excitation.

\begin{figure*}
  \includegraphics[scale=1]{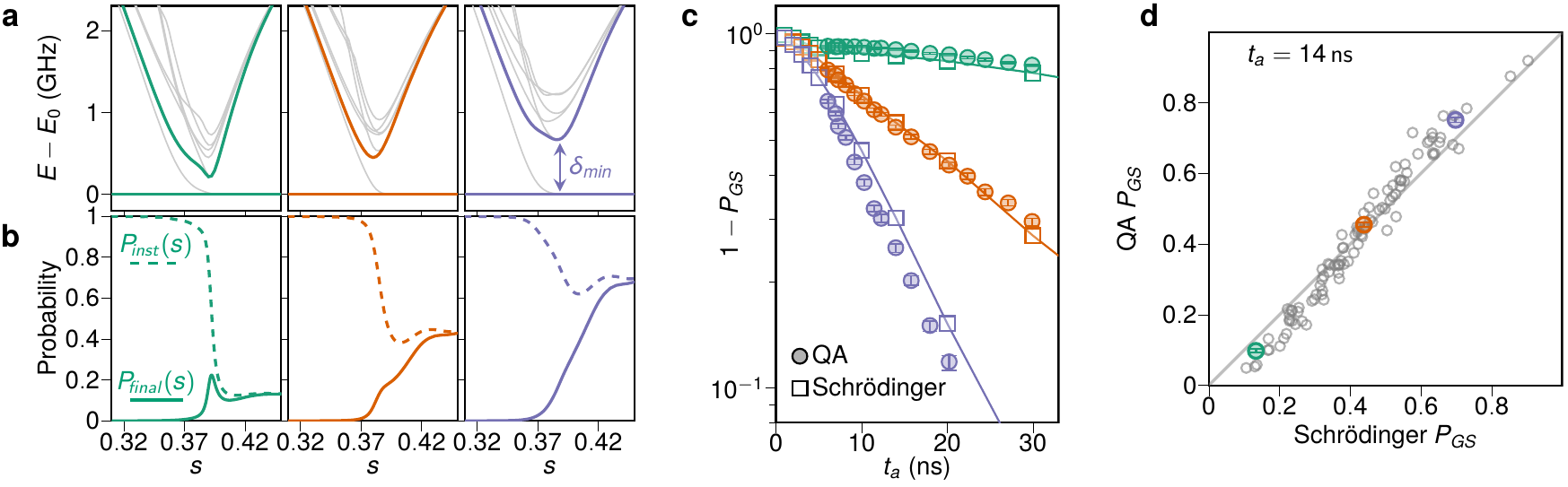}
  \caption{{\bf Coherent Schr\"odinger dynamics.} {\bf a}, For three exemplary 16-qubit spin glasses, we show the eight lowest eigengaps of the time-dependent QA Hamiltonian.  {\bf b}, We evolve the Schr\"odinger equation for QA with $t_a=\SI{14}{ns}$, tracking the wavefunction's probability of collapse onto the final (classical) ground-state manifold ($P_{\it{final}}(s)$) and the two lowest-energy instantaneous eigenstates ($P_{\it{inst}}(s)$).  $P_{\it{final}}(1)$ is the final ground-state probability $P_{\text{GS}}$ of the Schr\"odinger evolution.  {\bf c}, $1-P_{\text{GS}}$ for the same three spin glasses, in experimental QA and Schr\"odinger evolution for a range of annealing times.  {\bf d}, Comparison of $P_{\text{GS}}$ for QA and Schr\"odinger evolution for an ensemble of 100 16-spin glasses, for $t_a=\SI{14}{ns}$.  Error bars indicate 95\% confidence intervals for the average over parallel QA experiments using 192 different sets of qubits.}\label{fig:2}
\end{figure*}

In Fig.~\ref{fig:2}a, we show spectral gaps $E_i(s)-E_0(s)$ for three 16-qubit examples: one with a small gap, one with a moderate gap, and one with a large gap.  For fixed $t_a$ we define
\begin{eqnarray}
  P_{\it{inst}}(s) = \sum_{n\in\{0,1\}}\left|\braket{\phi_n(s)|\psi(s)}\right|^2, \\
  P_{\it{final}}(s) = \sum_{n\in\{0,1\}}\left|\braket{\phi_n(1)|\psi(s)}\right|^2,
\end{eqnarray}
which measure instantaneous probabilities of being in the ground or first excited state of $\mathcal H(s)$ and $\mathcal H(1)$, respectively.  Since the classical ground states are twofold degenerate, $P_{\it{final}}(1)$ gives the success probability of QA under Schr\"odinger dynamics.  Fig.~\ref{fig:2}b tracks  $P_{\it{inst}}(s)$ and $P_{\it{final}}(s)$ through anneals with $t_a=\SI{14}{ns}$.  The wavefunction begins concentrated on the easily-prepared ground state at $s=0$, and this probability decreases via Landau-Zener excitation in the vicinity of a small gap.

We run QA experiments on each of these 100 instances using 192 disjoint sets of qubits in parallel.  The Schr\"odinger excitation probability $1-P_{\text{GS}} = 1-P_{\it{final}}(1)$ is compared against experimental QA excitation probability in Fig.~\ref{fig:2}c; the probabilities are in close agreement with no fitting parameters used, showing an approximately exponential form.  Fig.~\ref{fig:2}d compares $P_{\text{GS}}$ for $t_a = \SI{14}{ns}$ across the entire 100-instance ensemble.  In Extended Data Fig.~\ref{fig:ed_fes} we compare probability distributions among ground and first excited states for QA, SA, SQA, and Schr\"odinger dynamics, and find that experimental QA data are better explained by Schr\"odinger dynamics than by SA and SQA.  The quantitative agreement between QA experiment and Schr\"odinger evolution up to $t_a=\SI{30}{ns}$ provides strong evidence for coherent quantum dynamics at small scale.  We now consider critical dynamics in large 3D spin glasses.

\section*{Critical spin-glass dynamics}

When a system is brought slowly (annealed) through a continuous phase transition, its dynamics slow down due to diverging correlations, and its macroscopic properties follow universal behavior described by critical exponents.  Here we use extensions of the Kibble-Zurek (KZ) mechanism~\cite{Kibble1976,Zurek1985,Polkovnikov2005,Dziarmaga2005,Zurek2005,Deng2008,Degrandi2011,Chandran2012,Liu2015}, which describes the generation of excitations as an annealed system falls out of equilibrium.  We use a dynamic finite-size scaling (DFSS) ansatz \cite{Degrandi2011} (SM \ref{sec:fss}) to relate time and the growth of correlations as functions of two critical exponents: $\nu$, which describes the divergence of correlation length at a phase transition, and $z$, which describes divergence of the characteristic timescale of domain fluctuations.  We investigate three annealing dynamics---QA, SQA, and SA---and their corresponding phase transitions.

Spin-glass order is quantified via the overlap between two replicas (independently annealed $N$-spin states $S$ and $S'$) of a given realization (set of couplings $J_{ij}$):
\begin{equation}
  q=\frac 1N\sum_{i=1}^NS_iS'_i~~~~~S_i,S'_i\in \{-1,1\}.
\end{equation}
The mean-squared Edwards-Anderson order parameter is given by $\langle q^2\rangle$, with $\langle\cdot\rangle$ denoting an average over both independent replicas and realizations.

Due to restrictions on the available coupling geometry, we program spin glasses in the 3D layout shown in Fig.~\ref{fig:1}c, which differs from the simple cubic lattice: it has two qubits at every $(x,y,z)$ coordinate, coupled as a dimer with strong ferromagnetic coupling $J_{ij}= -J_{\text{FM}}=-2$.  Between neighboring cubic coordinates, we program a total coupling of $\pm J_G$ ($0<J_G\leq 1$)---this coupling uses one coupler in the $x$ and $y$ directions, and in the $z$ direction the coupling is equally divided between two couplers.  We use open $x$- and $y$-boundaries and periodic $z$-boundaries.

In this model, contracting the strongly-correlated dimers into individual spins yields the $\pm J_G$ Ising spin glass on a simple cubic lattice, so we expect experimental results to reflect criticality in the same universality class.  Fig.~\ref{fig:1}e shows the model's phase diagram in the $\Gamma$, $T$ plane, where a spin-glass (SG) phase is separated from a disordered paramagnetic (PM) phase.  At $T=0$ there is a quantum critical point $\Gamma=\Gamma_c>0$.  At $\Gamma=0$ there is a finite-temperature classical transition at $T=T_c>0$.  Tuning $J_G$ while keeping $J_{\text{FM}}=2$ varies the details of the phase diagram, but not the qualitative picture.

\begin{figure*}
  \includegraphics[scale=1]{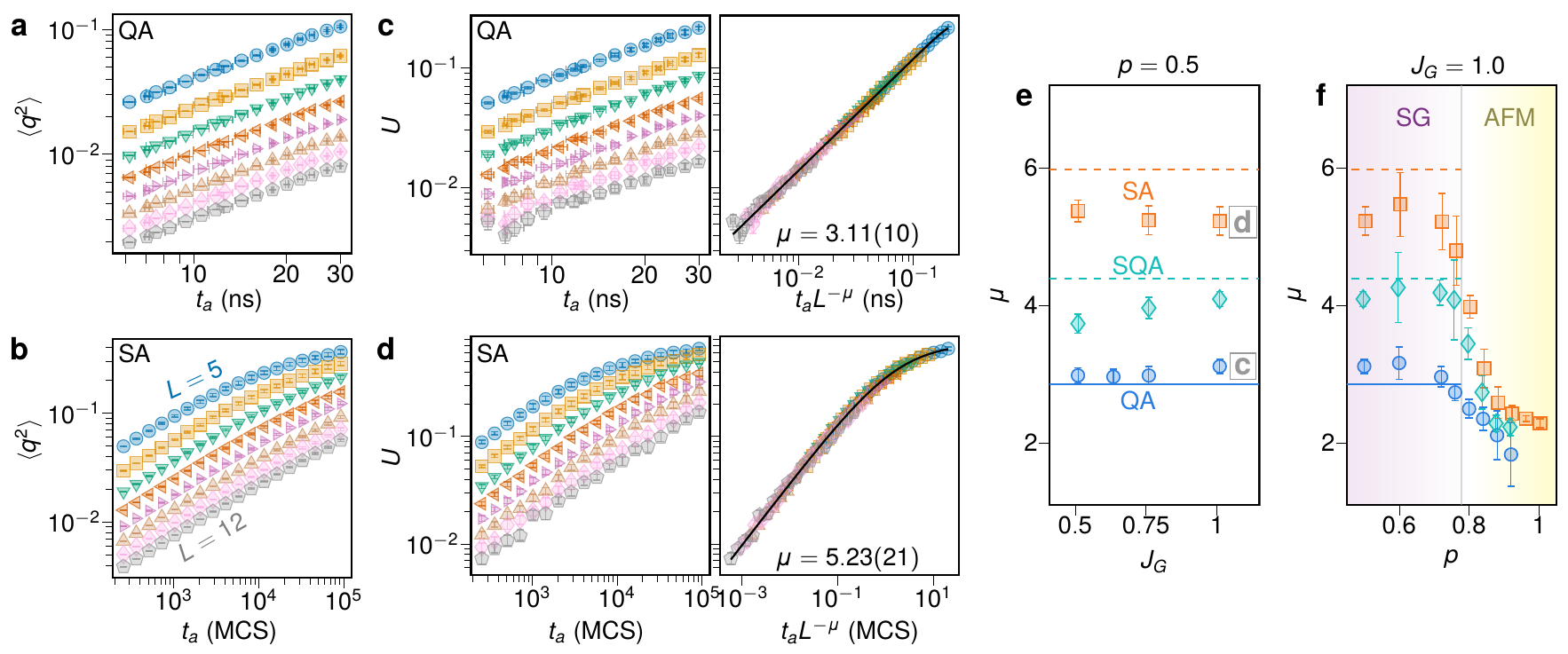}
  \caption{{\bf Dynamic finite-size scaling in 3D spin glasses.} {\bf a--b}, Squared overlap $\langle q^2\rangle$ varies as a function of $L$ and $t_a$ in QA ({\bf a}) and SA ({\bf b}) for 3D spin glasses.  {\bf c--d}, Binder cumulant $U$ scales similarly; rescaling $t_a$ by $L^\mu$ collapses data onto a single curve for a fit parameter $\mu$, providing estimates of the KZ exponent $\mu= z+1/\nu$.  {\bf e}, Estimates of $\mu$ for varying spin-glass coupling strengths $J_G$, where dimer coupling $J_{\text{FM}}=2$, in QA, SA, and SQA.  Solid and dashed horizontal lines indicate literature values \cite{Guo1994} and estimates from simple cubic lattices (Extended Data \ref{fig:ed_collapse_logical}) respectively.  {\bf f}, Tuning the doping probability $p$ reveals a finite-size crossover between the SG phase and the AFM phase, separated by a critical doping $p_c\approx 0.778$ at $T=0$, $\Gamma=0$ (vertical line).  Observed dynamics in these phases are characteristic of critical dynamics and coarsening dynamics, respectively.
    Vertical error bars are 95\% statistical confidence intervals and horizontal error bars indicate measurement uncertainty in $t_a$.
  }\label{fig:3}
\end{figure*}

We measure $\langle q^2\rangle$ for a range of $t_a$, with linear system size $L$ ranging from $5$ to $12$, in QA (Fig.~\ref{fig:3}a), SA (Fig.~\ref{fig:3}b), and SQA (Extended Data Fig.~\ref{fig:ed_collapse_full}), for $J_G=1$.  For Monte Carlo (MC) methods, $t_a$ is measured in MC sweeps (MCS).  Although in all cases we anneal through the critical point rather than stopping at the critical point (see Methods), the system experiences a critical slowing down at the QPT. Due to slower dynamics in the glass phase, measurements in the final state approximately reflect the relevant critical dynamics. The Binder cumulant
\begin{equation}
U = \frac 12\left( 3-\frac{\langle q^4\rangle}{\langle q^2\rangle^2}\right)
\end{equation}
provides a statistical signature of phase transitions, and like $\langle q^2\rangle$, also grows with $t_a$ and $1/L$, as seen in \mbox{Fig.~\ref{fig:3}c--d}.  For $U$, any post-critical effective scaling dimensions in $\langle q^2\rangle^2$ and $\langle q^4\rangle$ cancel, but the functional dependence on $t_a$ remains.  Thus, under the DFSS ansatz, $U(L,t_a)$ is expected to collapse onto a common curve for all system sizes when $t_a$ is rescaled by $L^{-z-1/\nu}$ (SM \ref{scaling:kz}), reflecting the fact that the annealing time required for the system to remain adiabatic up to a correlation length of $L$ scales as
\begin{equation}
t_a(L)\sim L^{\mu},\qquad \mu = z+1/\nu.
\end{equation}
Thus we estimate the KZ exponent $\mu$ via best-fit collapse of $U$ horizontally along the time axis.  Fig.~\ref{fig:3}c--d show collapses of $U$, from which we extract $\mu$ as a fitting parameter.

In Fig.~\ref{fig:3}e we show these estimates for QA, SA and SQA for a range of coupling energies $J_G$. In all cases, $U(L,t_a)$ approximately follows power-law scaling when far from equilibrium \cite{Liu2015}.  The extracted KZ exponents are clearly distinct, with smaller values indicating faster dynamics.  We find $\mu_{\text{QA}}$ between $2.9$ and $3.1$ depending on $J_G$.  Adding our MC estimate $1/\nu_{\text{QA}}\approx 1.55$~(SM~\ref{sec:qmc}) to $z_{\text{QA}}\approx 1.3$ \cite{Guo1994} gives $\mu_{\text{QA}}\approx 2.85$, in close agreement with experimental results. 

For SA we find $\mu_{\text{SA}} \approx 5.3$, much larger than QA but smaller than estimates from simple cubic lattices; for SA we found a value $6.0$ (dashed line) compared to previously reported\cite{Liu2015} value $6.3$.  These deviations are largely explained by finite size effects, and dimers delaying the onset of asymptotic behavior (see Figs.~\ref{fig:ed_collapse_full}---\ref{fig:ed_collapse_logical} and SM Section \ref{sec:freezing}).  No previous estimate of $\mu_{\text{SQA}}$ has been reported; the dashed line in Fig.~\ref{fig:3}e indicates the extracted value $\mu_{\text{SQA}}=4.39$ for simple cubic lattices (Fig.~\ref{fig:ed_collapse_logical}).  Since QA and SQA share an equilibrium exponent $\nu$, this corresponds to $z_{\text{SQA}}\approx 2.8$, which reflects the MC dynamics.

To better understand the role of frustration in the glassy dynamics, we increase the ``doping'' probability $p$ of a random inter-dimer coupling being antiferromagnetic ($+J_G$). 
The SG phase has been shown to persist until $p_c=0.778$ (vertical line) for $T=0$ and $\Gamma=0$, beyond which the system is a disordered antiferromagnet \cite{Hartmann1999,Hasenbusch2007,Hasenbusch2007a}---we might expect that $\Gamma>0$ slightly reduces $p_c$ \cite{Nishimori1992}.  Fig.~\ref{fig:3}f shows the $p$-dependence of $\mu$ extracted by collapsing for even values of $L$; all three dynamics are insensitive to changes in $p$ until it approaches $p_c$.  For $p$ close to $1$, QA, SA, and SQA all have $\mu\approx 2$, consistent with coarsening dynamics in the AFM phase \cite{Chandran2012}.  In this scenario, the dynamics that occurs after the critical point eliminates small domains, replacing the KZ exponent $\mu$ with a universal exponent $2$.  The latter corresponds to correlation length scaling as $\xi \propto t_a^{-1/2}$, expected for diffusive dynamics present in the AFM phase. Due to the rough potential landscape in the SG phase, the post-critical dynamics have negligible effect on $\mu$, although they affect energy decay as we discuss next.

\begin{figure*}
  \includegraphics[scale=1]{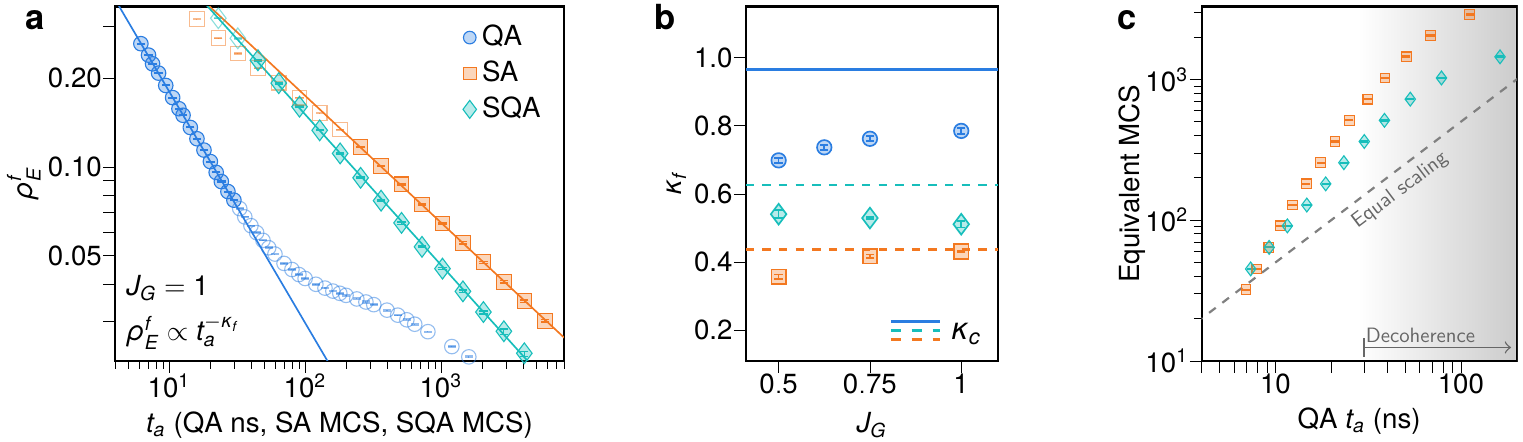}
  \caption{{\bf Critical scaling of final residual energy.}
    {\bf a}, Scaling of final residual energy density $\rho_E^f$ versus $t_a$, with lines showing power-law fits to the form $\rho_E^f \propto t_a^{-\kappa_f}$.  Empty markers are excluded from power-law fit (see text).
    {\bf b}, Markers show extracted exponents $\kappa_f$ for QA, SA, and SQA for varying $J_G$ in 3D spin glasses; horizontal lines show estimates of $\kappa_c$ (Eq.~(\ref{eq:kappa})) using estimates of $\mu$ shown as horizontal lines in Fig.~\ref{fig:3}e.
    {\bf c}, Number of MCS required for SA and SQA to match the residual energy achieved by QA in a given QA annealing time.  Dashed line is a guide to the eye showing equal scaling.  Shading represents the onset of thermal effects outside the coherent QA regime.  Error bars represent 95\% bootstrap C.I.~over spin-glass realizations.}\label{fig:4}
\end{figure*}

\section*{Energy decay}

The smaller values of critical exponents $z$ and $\mu$, obtained via data collapse, indicate faster critical dynamics in QA compared to SA and SQA.  We now ask whether this leads to a speedup in approximating the ground-state energy in classical Ising models. 
We first answer this question theoretically by considering a hypothetical annealing protocol that is measured at the critical point. Although this is not the real schedule of QA, it makes the energy decay dependent only on the dynamics approaching the critical point, and therefore allows connecting the relevant critical exponents. We define dimensionless residual Ising energy density at the critical point as
\begin{equation}\label{eq:rhoEc}
\rho_E^c =  \big\langle \mathcal H_I - E_c \big\rangle/(NJ_G).
\end{equation}
Here $N$ is the number of spins, and $E_c$ is the equilibrium expectation of $\mathcal H_I$ at $T=T_c$ and $\Gamma = 0$ for SA, and at $T=0$ and $\Gamma = \Gamma_c$ for QA and SQA. Notice that $\rho_E^c \to 0$ as $t_a \to \infty$. It is shown in SM~\ref{sec:fss} that $\rho_E^c$ follows a power-law relation:
\begin{equation}\label{eq:kappa}
 \rho_E^c\propto t_a^{-\kappa_c},\qquad \kappa_c = (d_s -1/\nu)/\mu,
\end{equation}
where $d_s=d$ for SA and $d_s=d+z_{\text{QA}}$ for QA and SQA. As expected, $\kappa_c$ is inversely related to the KZ exponent $\mu$, therefore faster critical dynamics (smaller $\mu$) leads to faster decay of energy. Moreover, $d_s$ is larger for quantum than classical, making a larger contribution to the numerator.

However, $\rho_E^c$ is not very relevant to optimization because $E_c$ is far from the ground-state energy $E_0$ of $\mathcal H_I$.  We therefore consider the corresponding final quantities $\rho_E^f$ and $\kappa_f$ obtained by annealing to the low-temperature classical point $T\ll T_c$, $\Gamma=0$:
\begin{equation}
\rho_E^f = \big\langle \mathcal H_I - E_0 \big\rangle /(NJ_G),
\end{equation}
and fitting $\rho_E^f\propto t_a^{-\kappa_f}$.  Again the average is over realizations and samples, with annealing according to the full QA schedule. For very long (adiabatic) anneals, we expect $\rho_E^f \to 0$, thus optimal solutions are asymptotically reached.

  Fig.~\ref{fig:4}a shows $\rho_E^f$ as a function of annealing time for QA, SQA, and SA, for 3D spin glasses on $N=5374$ spins ($15\times 15\times 12$ dimers, with some vacancies). The ground-state energy $E_0$  is estimated by exchange MC (see SM \ref{sec:gse}). Each dynamics follows a power-law scaling within a window of $t_a$ (SM~\ref{sec:windows}) but deviates outside the window, most notably for QA due to the onset of thermal effects.  For experimental QA, deviation from coherent behavior is expected for longer anneals due to the onset of thermal excitations \cite{King2022}.  For SA, the decay of $\rho_E^f$ settles onto a consistent exponent only after several hundred MCS.

  We estimate $\kappa_f$ for the three dynamics with varying $J_G$ from power-law fits. Figure \ref{fig:4}b shows, as a function of $J_G$, the fit values of $\kappa_f$ (symbols) as well as the critical values $\kappa_c$ (horizontal lines) obtained using independent MC estimates of $z$ and $1/\nu$, corresponding to the lines in Fig.~\ref{fig:3}e.  Deviations $\kappa_f < \kappa_c$ are expected beyond the critical point (SM~\ref{sm:kappa}); we find a modest correction $\kappa_c-\kappa_f \approx 0.1$ for both SA and SQA (SM~\ref{sm:kappac}), which one might also see in QA if it could be measured at the critical point.

  QA shows the fastest energy decay $\kappa_f$, followed by SQA and SA. For sufficiently large $J_G$, $\rho_E^f$ decays roughly quadratically faster in QA than in SA, with SQA in between the two.  This experimentally-observed scaling advantage is consistent with the theoretical speedup in critical ordering dynamics (smaller $\mu$ in Fig.~\ref{fig:3}e), and faster critical energy decay (larger $\kappa_c$ in Fig.~\ref{fig:4}b).  In Fig.~\ref{fig:4}c we show the annealing time (in MCS) required by SA and SQA to match the energy achieved by a given $t_a$ in QA; within the coherent QA regime, the approximation speedup of QA over SA and SQA increases as a function of annealing time.

\section*{Outlook}

We have experimentally demonstrated quantum critical dynamics in programmable spin glasses on thousands of qubits, observing the expected scaling in system size and annealing time.  Simulation accuracy was confirmed via comparison to numerical simulation of Schr\"odinger dynamics at the 16-qubit scale.  For large 3D spin glasses, the simulated many-body quantum dynamics are far beyond the reach of current exhaustive or tensor-based techniques; the former is limited to tens of qubits and the latter can be applied to moderately-sized 2D models~\cite{Ebadi2022,Schmitt2022}.  We therefore appeal to critical exponents via finite-size scaling analysis, finding good agreement with independent MC estimates. Thus we have presented both microscopic and macroscopic evidence for a coherently annealed programmable quantum spin glass.  
These exponents indicate, in theory and experiment, that quantum annealing has a dynamical advantage over simulated annealing and simulated quantum annealing in penetrating the spin-glass phase.  The predicted speedup was experimentally demonstrated through a scaling advantage in approach to the ground-state energy. These results point to the utility of programmable quantum annealers both as quantum simulators and optimization tools.

For sufficiently large spin systems, the extent of ideal quantum critical scaling is limited in time by qubit decoherence, disorder and noise, and the results shown here indicate that improvements in these areas would pay great dividends.  Extending the region of critical scaling would not only facilitate the further study of these dynamics, but also extend their utility in real-world applications, helping QA reach lower-energy solutions.  These efforts must be balanced with improvements in qubit connectivity, which allow more flexible problem embeddings, and high coupling energy, which can protect against control error and thermal excitation outside the coherent limit.

Spin glasses represent a paradigmatic hard optimization problem, and provide a robust theoretical framework for understanding and demonstrating  quantum critical dynamics.  They were instrumental in motivating, via magnetic experiments, the field of quantum annealing itself---here we have answered in the affirmative the foundational question raised over 20 years ago: Is it possible to engineer a programmable quantum system, in which quantum annealing imparts a dynamical advantage over simulated annealing?  Extending this characterization of quantum dynamics to industry-relevant optimization problems, which generally do not allow for analysis via universal critical exponents or finite-size scaling, would mark an important next step in practical quantum computing.

\section*{Acknowledgments}

The authors are grateful to Peter Young, Hidetoshi Nishimori, Sei Suzuki, and Jonas Charfreitag for helpful discussions. A.W.S.~was supported by the the Simons Foundation under Simons Investigator Award No.~511064. Some of the numerical simulations were carried out using the Shared Computing Cluster managed by Boston University's Research Computing Services.

\section*{Author Contributions}

A.D.K., J.R., T.L., R.H., A.Z., A.W.S., and M.H.S. conceived and designed the experiments and analyzed the data.  A.D.K., J.R., T.L., and A.W.S.\ performed the experiments and simulations.  T.L., R.H., F.A., A.J.B., K.B., S.E., C.E., E.H., S.H., E.L., A.J.R.M., G.M., R.M., T.O., G.P.-L., M.R., C.R., Y.S., N.T., M.V., J.D.W., J.Y., and M.H.A.\ contributed to the design, fabrication, deployment, and calibration of the quantum annealing system.  A.D.K., J.R., R.H., A.W.S., and M.H.A.\ wrote the manuscript.

\section*{Competing Interests}

A.W.S.~declares no competing interests.  All other authors have received stock options in D-Wave as current or former employees, and declare a competing financial interest on that basis.

\let\oldaddcontentsline\addcontentsline
\renewcommand{\addcontentsline}[3]{}
\bibliography{paper_3dsg}
\let\addcontentsline\oldaddcontentsline

\clearpage

\section*{Methods}

\subsection*{Spin-glass instances}

The 3D lattices have open $x$ and $y$ dimensions and periodic $z$ dimension.  Instances up to size $9\times 9 \times 9 \times 2$ ($L=9$) are fully yielded, with no site vacancies.  Larger instances have over $99.5\%$ site yield.  The inputs are heuristically embedded into the qubit connectivity graph of the QA processor, with a structure shown in Extended Data~\ref{fig:sm_3d_embeddings}.  Input construction is discussed in detail in the supplementary material.

\subsection*{Quantum annealing system and methods}

All QA data, except specifically indicated temperature variation experiments discussed in the supplementary material, were taken using one D-Wave Advantage system operating at $\SI{12}{mK}$ (QPU1).  The variable-temperature experiments were performed using a second D-Wave Advantage system of the same design, between $\SI{12}{mK}$ and $\SI{21}{mK}$ (QPU2).

Calibration refinement methods were used to balance qubits at degeneracy and to synchronize annealing lines.  We follow the same method as described in the supplementary materials for Ref.~\cite{King2022}, but with no tuning of coupling strengths.  We describe these methods in detail in SM~\ref{sec:shim}.

Each call to the QA system resulted in 200 anneals.  QA data on 3D spin glasses are generated from 900 calls, cycling through 300 disorder realizations. For MC methods, between 100 and 300 realizations were used. Thus the data points for $\langle q^2\rangle$, $U$, and $\rho_E$ represent the average over between 100 and 300 spin-glass realizations, with error bars capturing variation between instances.  QA Binder cumulants were computed by comparing overlap between annealing samples generated in different QA calls but the same seed and embedding, thus suppressing the effect of correlated biases.  The experiments were sufficiently extensive that the Binder cumulants and overlaps computed from samples within the same QA programming give similar results.

\subsubsection*{Measuring annealing time}

The anneal of the Hamiltonian from $s=0$ to $s=1$ over annealing time $t_a$ is achieved through a rapid change in qubit control current.  Of the factors limiting the minimum value of $t_a$, two are most important: First is the ability to reliably quench $s$ at a known rate and with tolerable nonlinearity and distortion from filtering; second is the ability to synchronize the qubits to within a reasonable deviation in terms of $s$.

Annealing times slower than $\SI{20}{ns}$ are reliably realized.  Annealing times faster than $\SI{20}{ns}$ deviate significantly from their requested values due to lowpass filter bandwidth and resolution of digital control electronics, and must be measured independently.  We do so in two ways.  The recent demonstration of the KZ mechanism in a 1D Ising chain \cite{King2022} showed very good agreement with theory, in particular a scaling of residual energy $\rho_E^f \propto t_a^{-1/2}$.  Therefore extrapolating 1D chain data from $t_a=\SI{20}{ns}$ to faster anneals provides a reliable measurement of effective annealing time.  The second measurement we perform is a direct measurement in which one qubit is quenched and measured by a witness qubit, allowing us to estimate the effective annealing time.  Data for these two measurements are compared in Extended Data~\ref{fig:sm_annealtime}.  We take the average of the two measurements as our value of $t_a$.  Error bars are obtained by adding, in quadrature, the difference between the two measurements and the deviation in annealing slope before and after software filtering ($<0.05$ relative error).

\subsubsection*{Quantum annealing schedule}

Simulating the time-dependent Schr\"odinger equation requires an accurate annealing schedule $(\Gamma(s),\mathcal J(s))$.  To achieve this, we diagonalize a time-dependent many-body flux-qubit Hamiltonian for a small representative system of coupled qubits, using a flux-qubit model \cite{Harris2010a} whose parameters are given in Extended Data~\ref{tab:parameters}.  Then, for each $s$ in our range of interest, we compute $\Gamma(s)$ and $\mathcal J(s)$ as best-fit parameters that give an Ising model whose eigengaps closely match those of the diagonalized superconducting quantum interference device (SQUID) Hamiltonian.

Diagonalizing the many-body flux-qubit Hamiltonian to high accuracy is computationally challenging even at small scales because the flux qubits have multiple energy levels, unlike the model two-level Ising spins.  Thus as a representative system we take eight qubits, and divide them up into four dimers such that the two qubits in each dimer have similar interactions with the other qubits.  We then treat each dimer as a six-level object, and diagonalize the system of four dimers.

Reflecting the frustration in spin glasses, our eight-spin system, shown in Extended Data~\ref{fig:sm_schedule}, has frustration in its twofold-degenerate classical ground state.  Performing a two-parameter fit with $\Gamma(s)$ and $\mathcal J(s)$ becomes numerically unstable for $s>0.40$ due to the small first gap.  For $s>0.40$ we fit $\Gamma(s)$ to the expected exponential decay form and extract only $\mathcal J(s)$ as a fitting parameter.  Extended Data~\ref{fig:sm_schedule} shows the nominal and extracted schedules.

\subsection*{Classical Monte Carlo dynamics}

\subsubsection*{Simulated annealing}

SA uses a geometric annealing schedule from $\beta= 0.001J_G$ to $\beta=10J_G$.  Each MCS processes $N$ spins, sampled randomly with replacement.  Spin updates are accepted or rejected in a Metropolis-Hastings algorithm.

\subsubsection*{Simulated quantum annealing}

SQA uses an annealing schedule based on the QA schedule; to accelerate our experiments we begin when $\Gamma/\mathcal J \approx 6$---a fast-equilibrating regime where we assume the system is quasistatic with respect to the changing Hamiltonian---and perform 10 MCS before proceeding through the QA schedule until $\Gamma/\mathcal J \approx 1/25$, far into the ordered phase for the models studied.  Inverse temperature $\beta$, which is 64 except where specified, is relative to $\Gamma(s)=\mathcal J(s)$ at the crossing point of $\Gamma$ and $\mathcal J$.  Swendsen-Wang cluster updates are used in imaginary time.

\subsection*{Time-dependent Schr\"odinger evolution}

For the data in Fig.~\ref{fig:2} we used an iterative method that follows the annealing schedule from $s=0.1$, where the QA wavefunction $\psi$ is concentrated on the instantaneous ground state, and $s=0.7$, where dynamics are negligible.  We track populations for only the lowest $80$ of $2^{16}$ eigenstates; this exceeds the number of ground and first excited states for the classical models studied.  Step sizes in $s$ are determined adaptively based on the minimum eigengap, and range between $0.00008$ and $0.01$.

\subsection*{Statistical methods and error analysis}

Error bars on the order parameter $\langle q^2\rangle$, Binder cumulant $U$, and residual energy $\rho_E^f$ were generated by treating each random seed as an independent experiment and performing a resampling bootstrap on the set of statistics.  This bootstrapping method gives a population for each statistic estimate.  We use this population in two ways.  First, we take the middle 95\% of the population as our confidence interval for the statistic.  Second, from the population we compute a variance on the logarithm, which we use to determine error weights for our data collapses.

\subsection*{Data collapse}

To collapse measurements of the Binder cumulant $U$ for varying system sizes onto a common target curve, we need to find a best-fit value for $\mu$.  This fit minimizes a weighted sum of squared distances (in the logarithm) from the target curve.  Weights in the sum are inversely proportional to the variance of the logarithm of the estimator.  The form of the target curve must capture a crossover between the power-law form of the KZ regime and the equilibrium ($t_a\rightarrow \infty$) limit.  To achieve this, for each putative value of $\mu$ we find a best-fit target curve (nested within the $\mu$-optimizing method) whose power-law slope varies as a logistic function.  Our target curve has the form
\begin{equation}
  f(x) = a_0+a_1\log(1+\exp(a_2(x-a_3))), 
\end{equation}
which fits $\log(U)$ as a function of $\log(t_a)$.

To generate error estimates for $\mu$ we perform a jackknife on the measurements in $L$ and $t_a$ and add the resulting standard errors in quadrature.  To approximate the 95\% C.I.~in the data points we use error bars that span $\pm 2\sigma$.

Collapse in $\langle q^2\rangle$ is achieved with the same approach.  Selection of ranges of $t_a$ over which we collapse data is described in the supplementary material.

\section*{Data Availability}

The data supporting the findings are being prepared for deposition in an online repository, and are currently available upon request.

\section*{Code Availability}

An open-source version of the SQA code used in this work is available at  \url{https://github.com/dwavesystems/dwave-pimc}.

\clearpage

\widetext
\setcounter{figure}{0}
\renewcommand{\figurename}{Extended Data}

\renewcommand{\thefigure}{E\arabic{figure}}
\renewcommand{\theHfigure}{E\arabic{figure}}
\makeatother

\begin{figure*}\includegraphics[scale=1]{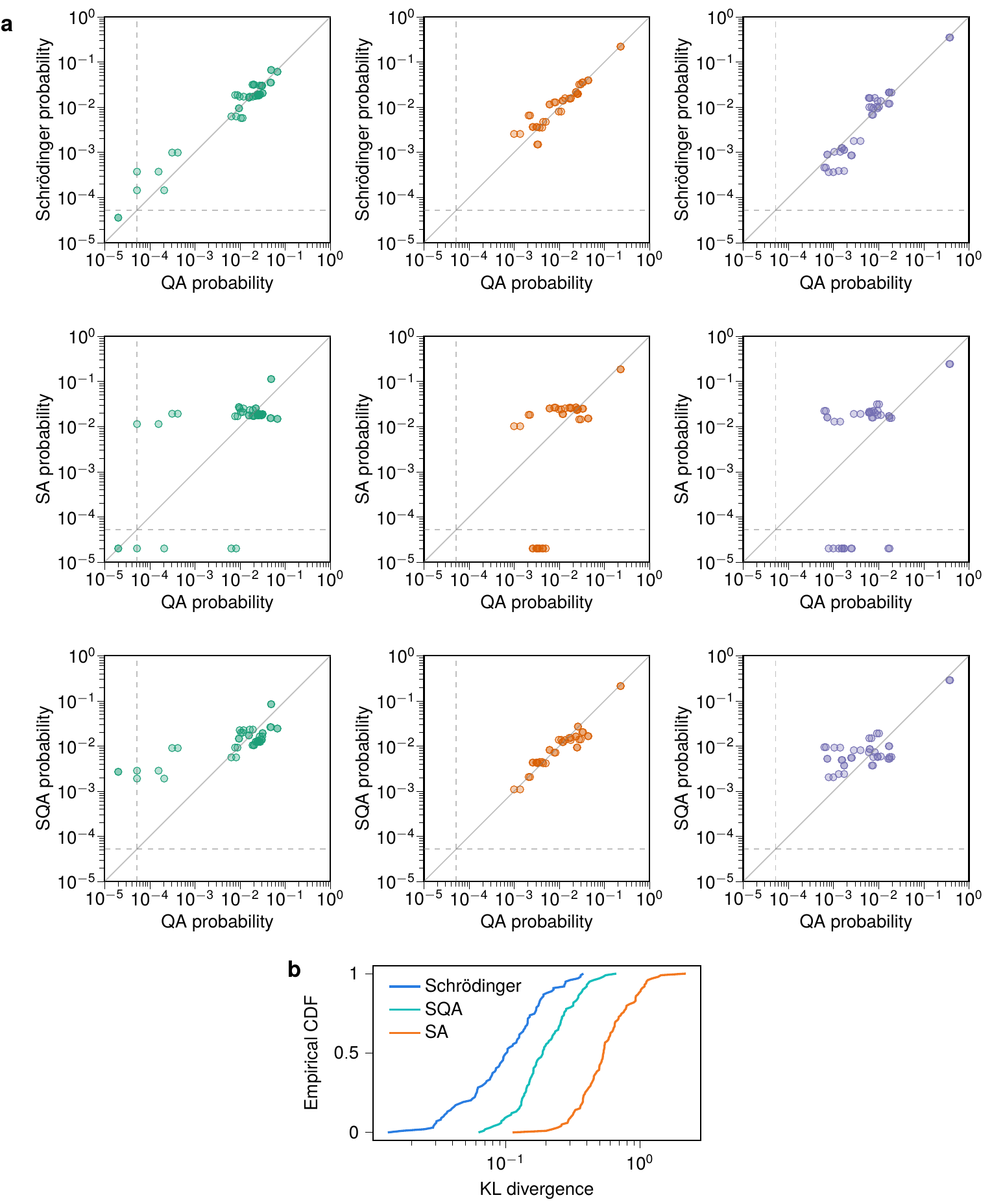}\caption{{\bf State probabilities for 16-spin glasses.} {\bf a}, We show observed probabilities for ground states and first excited states in Schr\"odinger evolution ($t_a=\SI{14}{ns}$), SA ($t_a=200$ MCS), and SQA ($t_a=100$ MCS) compared with experimental measurements from QA ($t_a=\SI{14}{ns}$).  The three columns contain data for the three exemplary instances, with colors corresponding to those in Fig.~\ref{fig:2}.   Annealing times for SA and SQA were chosen to have good agreement with Schr\"odinger evolution in average ground state probability.  Each dynamics was run 19,200 times; dashed lines indicate the statistical floor, i.e., the probability if a state is seen exactly once.  Unobserved states are represented by points below the statistical floor.  {\bf b}, Kullback-Leibler (KL) divergence in the probability distribution among first excited states, with QA used as a reference distribution, measures deviation between two dynamics for a given realization.  Empirical CDF (proportion of 100 realizations below a given KL divergence) is shown.  Data indicate that coherent quantum (Schr\"odinger) dynamics agrees more closely with experimental data better than does SA or SQA.
}\label{fig:ed_fes}
\end{figure*}

\begin{figure}
\includegraphics{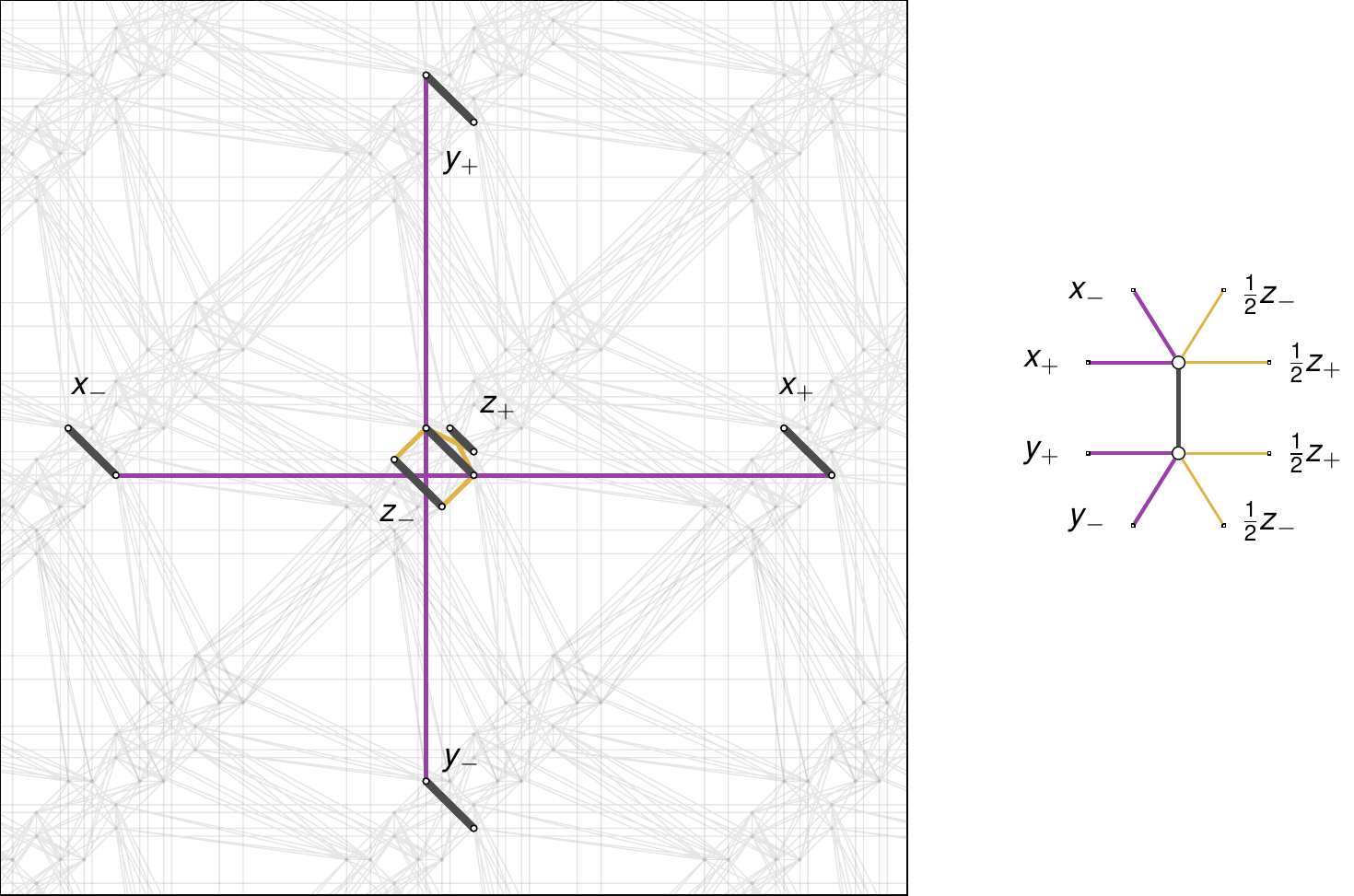}
  \caption{{\bf 3D lattice structure in qubit connectivity graph.} $L\times L\times (\max (L,12))\times 2$ lattices are found by heuristic search given a basic structure in which horizontal and vertical (long) couplings form two dimensions, and the interior of unit cells forms the third dimension.  One dimer (thick gray line) and its six neighboring dimers are shown as thick gray lines.  As in Fig.~\ref{fig:1}, purple and yellow lines represent glass couplings of $\pm J_G$ and $\pm J_G/2$ respectively.}\label{fig:sm_3d_embeddings}
\end{figure}

\begin{figure}
\includegraphics[scale=0.8]{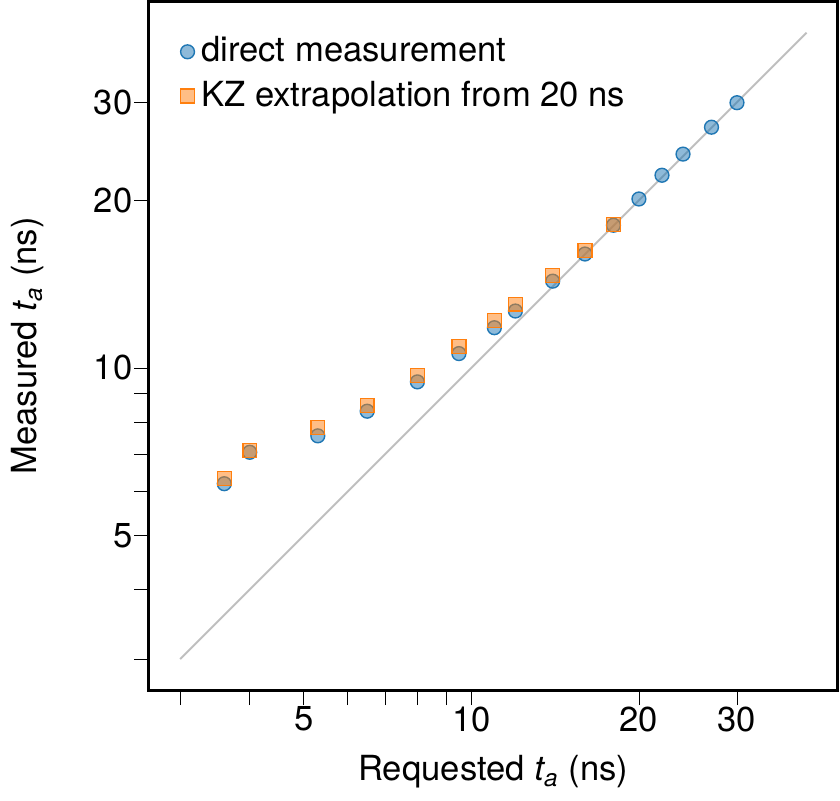}
  \caption{{\bf Measurement of effective QA annealing time.} Two independent measurement methods are used to estimate $t_a$ for fast anneals ($<\SI{20}{ns}$).  First is a direct measurement using a witness qubit.  Second is an extrapolative measurement that assumes a quantum KZ scaling in a 1D chain and assumes a kink density $n \propto t_a^{-1/2}$ for $t_a < \SI{20}{ns}$.  For the fastest anneals, $t_a$ deviates significantly from the values requested from the control electronics.  However, the two independent measurement approaches give consistent results.}\label{fig:sm_annealtime}
\end{figure}

\begin{figure}
  \begin{tabular}{|l|r|l|}
    \hline
   Parameter & Symbol & Value\\
    \hline
  Mutual inductance for $J=1$ &  $M_{\text{AFM}}$ & \SI{1.65}{pH}\\
  Initial external applied flux &  $\Phi^i_{\text{CCJJ}}$ & \SI{-0.621}{\Phi_0}\\
  Final external applied flux &  $\Phi^f_{\text{CCJJ}}$ & \SI{-0.717}{\Phi_0}\\
  Qubit inductance &  $L_q$ & \SI{371}{pH}\\
Qubit capacitance&    $C_q$ & \SI{118}{fF}\\
CJJ loop capacitance&    $C_l$ & \SI{5}{fF}\\
CJJ junction capacitance&    $C_{\text{CJJ}}$ & \SI{25}{fF}\\
Qubit critical current&    $I_c$ & \SI{2.10}{\micro A}\\
\hline
    \end{tabular}
  \caption{Physical properties for the QA processor.}\label{tab:parameters}
\end{figure}

\begin{figure}
\includegraphics{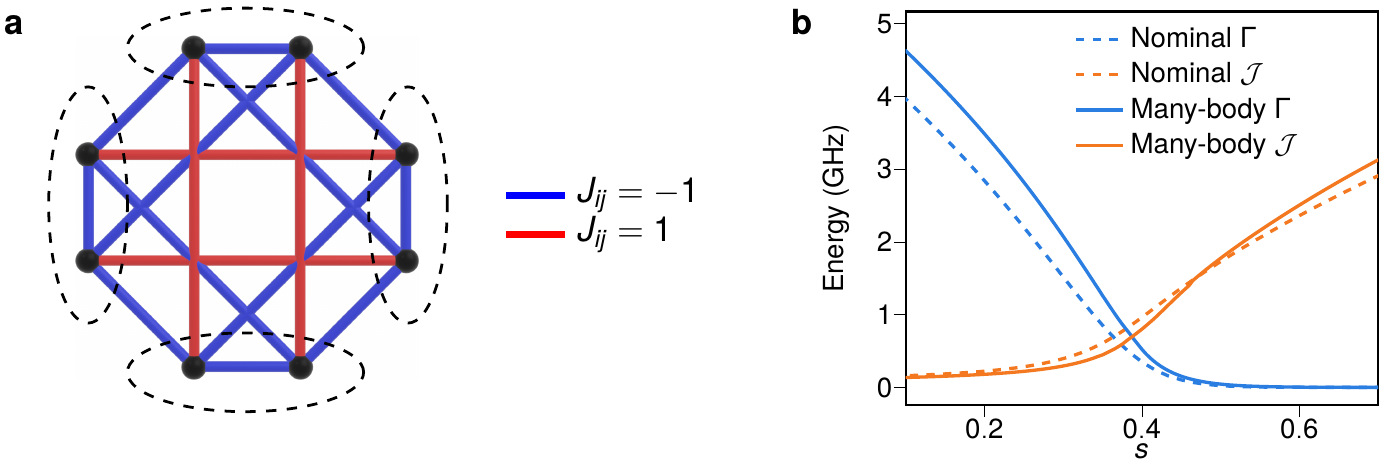}
  \caption{{\bf Extracting Ising model from flux-qubit model.} {\bf a}, Eight-qubit gadget used to extract an effective annealing schedule in the transverse-field Ising model based on a many-body flux-qubit Hamiltonian.  Dimers indicated by dashed ellipses are treated as six-level objects and combined to diagonalize a many-body Hamiltonian.  {\bf b}, General-purpose (nominal) annealing schedule based on single-qubit measurements, and extracted many-body effective schedule, used for Schr\"odinger evolution.}\label{fig:sm_schedule}
\end{figure}

\begin{figure}
\includegraphics[width=\linewidth]{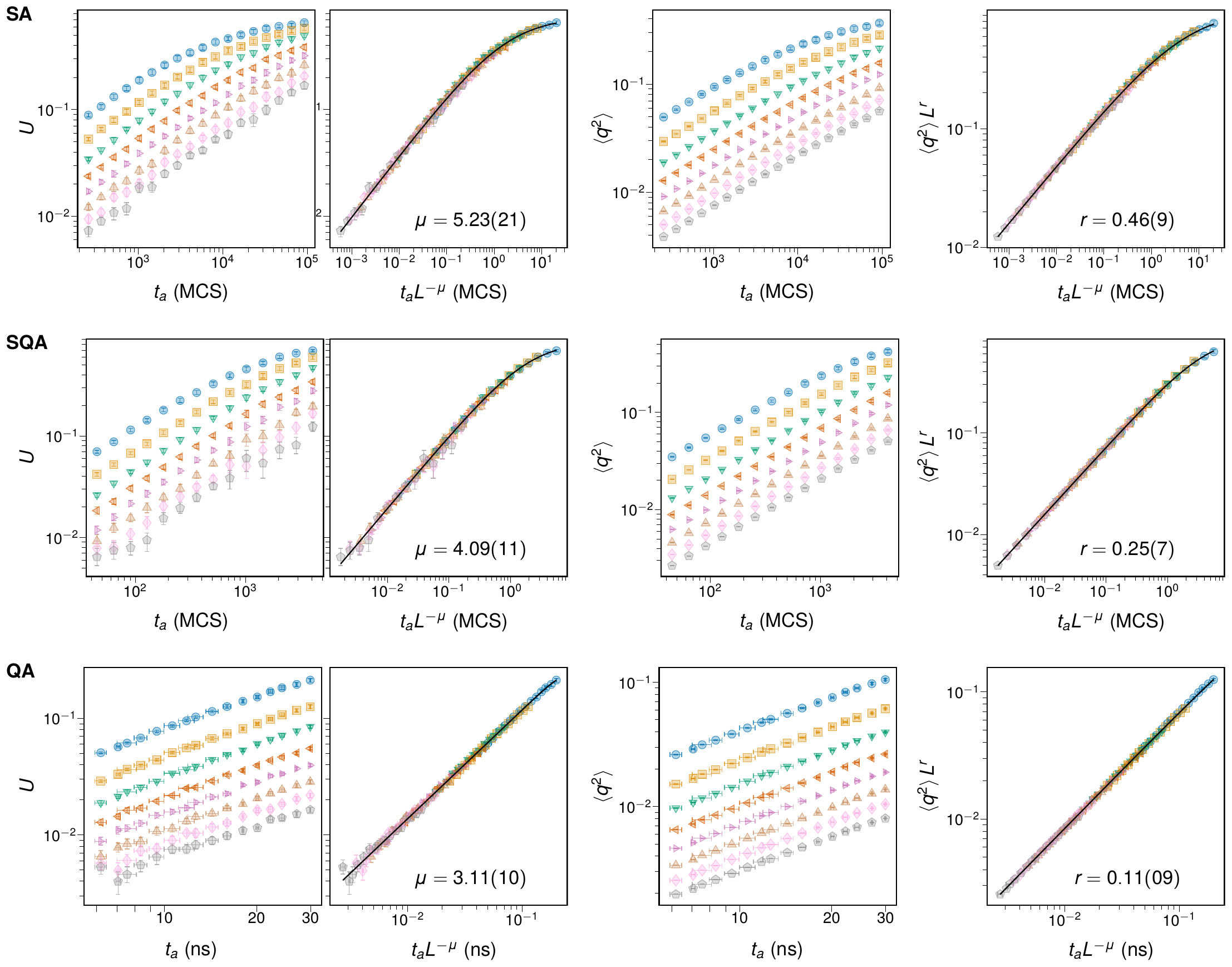}
\caption{{\bf Data collapse for 3D spin glasses with $J_G=1$.}  Best-fit exponent $\mu$ collapses $U$ by rescaling data horizontally based on $L$, and $r$ collapses $\langle q^2\rangle$ by scaling vertically based on $L$, given a horizontal rescaling by $L^{-\mu}$.}\label{fig:ed_collapse_full}
\end{figure}

\begin{figure}
\includegraphics[width=\linewidth]{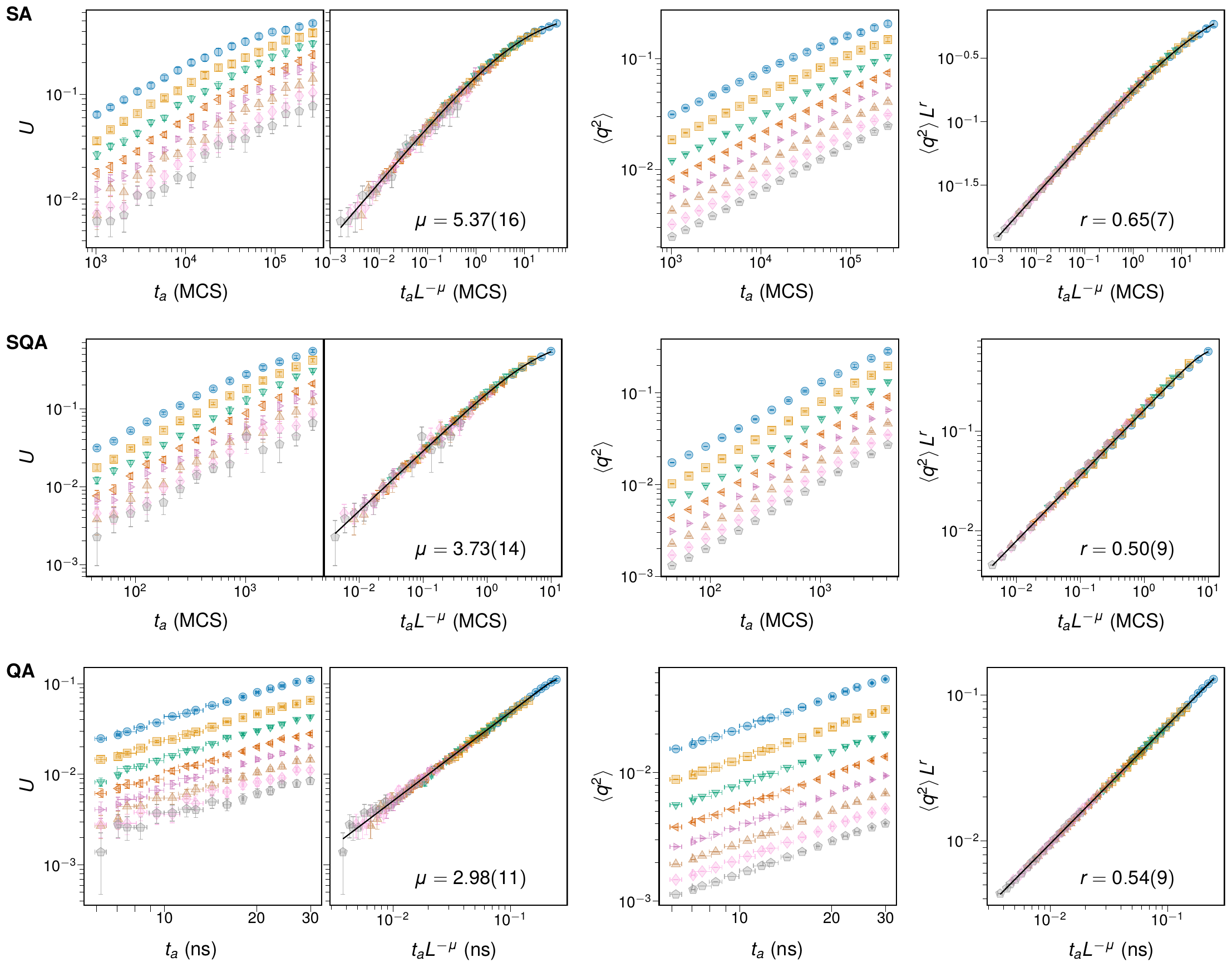}
  \caption{{\bf Data collapse for 3D spin glasses with $J_G=1/2$.}  Best-fit exponent $\mu$ collapses $U$ by rescaling data horizontally based on $L$, and $r$ collapses $\langle q^2\rangle$ by scaling vertically based on $L$, given a horizontal rescaling by $L^{-\mu}$.}\label{fig:ed_collapse_half}
\end{figure}

\begin{figure}
\includegraphics[width=\linewidth]{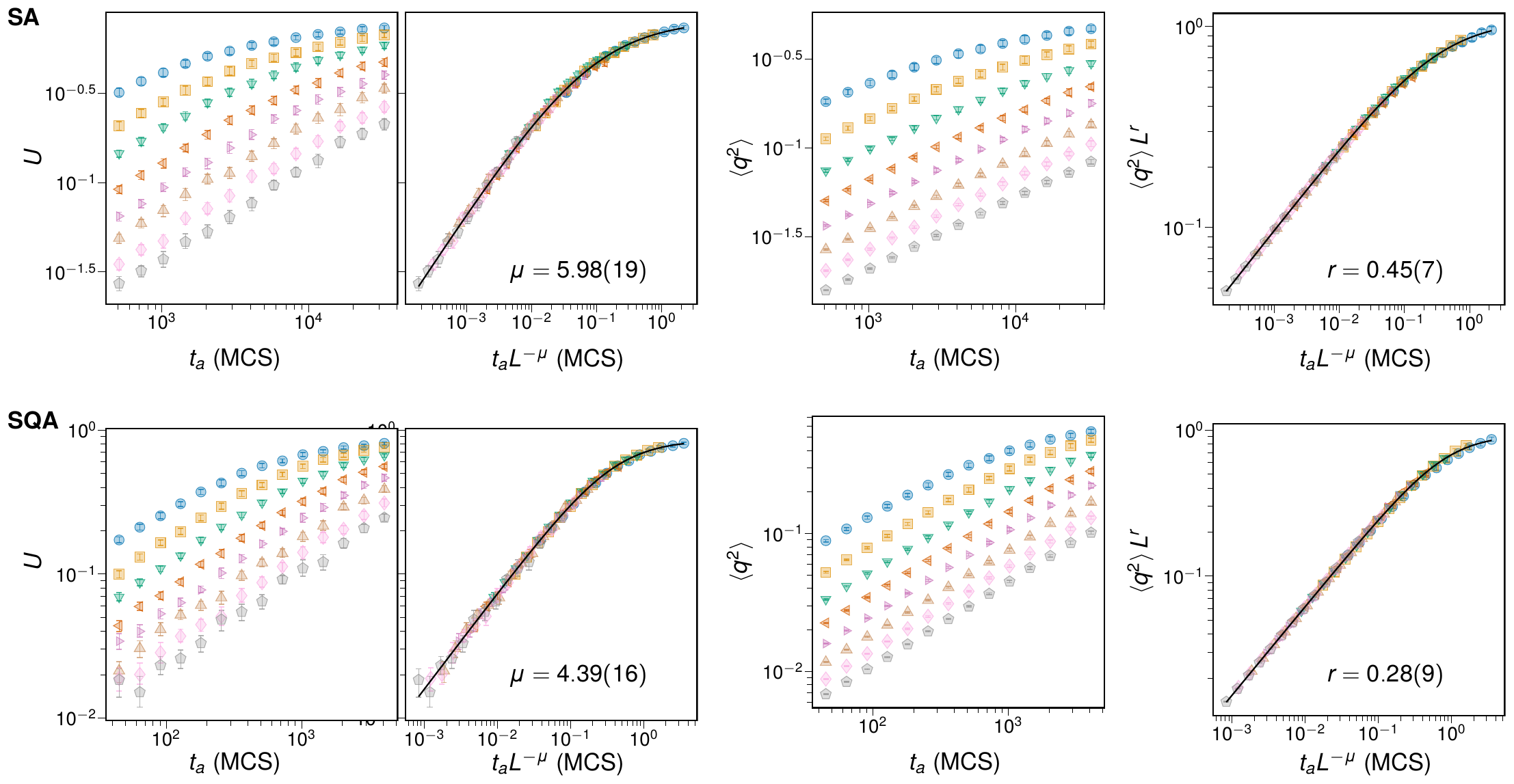}
  \caption{{\bf Data collapse for 3D spin glasses on simple cubic lattices.}  Best-fit exponent $\mu$ collapses $U$ by rescaling data horizontally based on $L$, and $r$ collapses $\langle q^2\rangle$ by scaling vertically based on $L$, given a horizontal rescaling by $L^{-\mu}$.}\label{fig:ed_collapse_logical}
\end{figure}

\widetext
\clearpage

\begin{center}
\textbf{\large Supplementary Materials:\\ \mytitle}
\end{center}

\tableofcontents
\setcounter{equation}{0}
\setcounter{figure}{0}
\renewcommand{\figurename}{FIG.}
\renewcommand{\thefigure}{S\arabic{figure}}
\renewcommand{\theequation}{S\arabic{equation}}
\renewcommand{\theHfigure}{S\arabic{figure}}
\makeatother

\clearpage

\section{Finite-size scaling}\label{sec:fss}

Here we provide the required background on the dynamic finite-size scaling (FSS) ansatz used in this work. The main goal is to justify the extraction
of the KZ exponent $\mu$ by collapsing, as in Fig.~\ref{fig:2}, the Binder cumulant $U$ for multiple annealing times and system sizes, as well as
the exponent $\kappa_f$ of the power law quantifying the reduction in excess energy with increasing annealing time in Fig.~\ref{fig:4}. Since the annealing
process stops inside the glassy phase not only for QA, but also in this work for SA and SQA, our analysis here must go beyond the simplest case of
dynamic scaling at a critical point. We will show that many aspects of KZ scaling remain valid under the circumstances prevailing in the annealing
experiments conducted here, including the power-law forms governed by the exponent $\mu$, even though the order parameter evolves beyond its
critical form.

There are many excellent reviews of conventional critical phenomena and finite-size scaling, e.g., Refs.~\cite{Barber1983,Privman1990}, but for
ease of reference in the later sections we begin in Sec.~\ref{scaling:equil} by a concise summary. In particular, we explain the relationships
between the conventional critical exponents and the scaling dimensions appearing within the renormalization group (RG) framework. The classical
case of transitions driven by thermal fluctuations at critical temperature $T_c>0$ is considered first, followed by the formally simple generalization
to quantum phase transitions, where the ground state ($T_c=0$) of a system changes as a function of a model parameter regulating the quantum
fluctuations. In Sec.~\ref{scaling:kz} we review the extensions of the finite-size scaling formalism to the case of a system brought through a
classical or quantum phase transition by an annealing process, where the annealing rate regulates the correlation length by the KZ mechanism
and there is a generalized FSS ansatz involving both the system size and the annealing rate. There are also extensive reviews of this topic,
e.g., the very recent Ref.~\cite{Rossini2021}; here we outline the formalism underlying our analysis of simulations and experimental data.
In Sec.~\ref{scaling:qa}, we focus on the particular conditions pertaining to the quantum annealing device and how some aspects of critical
scaling can persists also when driving the system past the quantum phase transition into an ordered (here glassy) state.

\subsection{Equilibrium finite-size scaling}
\label{scaling:equil}

\subsubsection{Classical phase transitions}

Consider a system described by some Hamiltonian $H$ at a short distance $\delta = T-T_c$ away from a classical continuous phase transition
with critical temperature $T_c$. There is a characteristic length scale in the infinite system, the correlation length $\xi$ governed by
the exponent $\nu>0$,
\begin{equation}
\xi \propto \delta^{-\nu},
\label{xiscaling}
\end{equation}
where the divergence takes place for $\delta \to 0$ both from above and below (and for $\delta < 0$ the absolute value $|\delta|$ is implied here and
henceforth in similar power laws). Also consider some other quantity $n$ (an expectation value of, typically, a volume-normalized density) that in
the thermodynamic limit is governed by another critical exponent $\sigma$,
\begin{equation}
n \propto \delta^\sigma,
\label{ainfty}
\end{equation}
where $\sigma$ depends on the quantity $n$ and the universality class of the transition. For example, if $n$ is the order parameter,
$n=\langle m\rangle$, then in the conventional nomenclature $\sigma \equiv \beta$. The expectation value $\langle .\rangle$ involves also
averaging over realizations in a system with some type of intrinsic disorder, like the spin glasses of interest here.

On a finite lattice with $d$ spatial dimensions and volume $N=L^{d}$, the singular behavior breaks down when the
correlation length becomes of the order of $L$. According to FSS theory \cite{Fisher1972,Barber1983,Privman1990}, for
finite $L$ the critical form (\ref{ainfty}) should be replaced by
\begin{equation}
n = L^{-\sigma/\nu}g(\delta L^{1/\nu}),
\label{afinitel}
\end{equation}
where the scaling function $g(x)$ has the property $g(x) \to {\rm constant}$ when $x \to 0$. In order for the form (\ref{ainfty}) in the
thermodynamic limit to be recovered when $L \to \infty$ (at fixed small $\delta$), the large-$x$ limit of $g(x)$ must be of the form
\begin{equation}
g(x) \to x^\sigma h(x),
\label{fxsigma}
\end{equation}
where $h(x) \to {\rm constant}$ when $x \to \infty$. Then there is no longer any $L$ dependence of Eq.~(\ref{afinitel}) and the correct
exponent on $\delta$ in Eq.~(\ref{ainfty}) is obtained.

In the case $n=\langle m\rangle$, it should be noted that a finite system does not break any symmetry spontaneously; thus $\langle m\rangle=0$
for any finite $L$ at any temperature. This apparent problem can be easily circumvented by considering the squared order parameter $\langle m^2\rangle$,
which vanishes with increasing $L$ for $T \ge T_c$ and approaches the square of the infinite-size order parameter for $T < T_c$. The FSS form 
Eq.~(\ref{afinitel}) holds for $\langle m^2\rangle$ in the neighborhood of $T_c$, with $\sigma=2\beta$.

The critical exponents can be related to the RG scaling dimensions appearing in field theory \cite{Cardy1996}. Consider an
extensive operator $P$ (a function of the system's degrees of freedom) defined as a sum over the entire
system of local density operators $p({\bf r})$;
\begin{equation}
P = \sum_{\bf r}  p({\bf r}).
\label{msum}
\end{equation}
If this operator is added  with some factor $\delta_p$ to the Hamiltonian at $T=T_c$, $H \to  H + \delta_p P$, the system would typically
be driven away from the critical point and critical scaling forms such as those discussed above for $\xi$ and $n$ will apply with $\delta \to \delta_p$.
The factor $\delta_p$ is referred to as the field conjugate to $P$.

Classical statistical mechanics relies on the Boltzmann weight ${\rm e}^{-H/T}$. At $T=T_c$, if the perturbing operator $P$ is just the Hamiltonian
itself, $H \to H + \delta H$, the offset $\delta H$ is equivalent to just shifting the inverse temperature, $1/T = (1/T_c)(1 + \delta)$ with $H$ kept
unchanged. Therefore, instead of shifting $H$ we consider deviations from the critical temperature $T_c$ and refer to $\delta= T-T_c$ as the thermal
field. Strictly speaking $\delta$ is the conjugate field to the entropy $S$, which appears in the free energy $F=E-TS$ along with the internal energy
$E=\langle H\rangle$. By the reverse of the above arguments, we can also regard $1/T-1/T_c \approx -\delta/T^2_c$ as the conjugate field to $H$.
Not worrying about factors, we can say that $\delta$ is the field conjugate to both the entropy and the internal energy.

A field of strength $\delta_p$ conjugate to an arbitrary operator $P$ is associated with a scaling dimension $y_p=1/\nu_p$ governing finite-size forms
such as Eq.~(\ref{afinitel}),
\begin{equation}
n = L^{-\sigma/\nu}g(\delta_p L^{y_p}),
\label{afinitel1}
\end{equation}
when perturbing the critical system by an arbitrary interaction $\delta_pP$. In the above case of just changing the temperature, the thermal exponent
$\nu_t$ is conventionally just called $\nu$ as in Eqs.~(\ref{xiscaling}) and (\ref{afinitel}), and for simplicity we also do not attach any subscript
on $\delta=t \equiv T-T_c$. It is important to note that the symmetry of the system does not change when the temperature is changed, and $\delta$ is therefore
also referred to as a symmetric scaling field.

For generic perturbations $P$, the symmetry of $H(\delta_p \not=0$) can be lower than that of $H(0)$, e.g., in the common case of $P=M$, where
$\langle m\rangle =\langle M\rangle/N$ is the order parameter density of a system hosting a symmetry-breaking transition. For instance, in the
ferromagnetic Ising model, $M$ is the total magnetization and $\delta_m$ is the external magnetic field. Thus, the spin-inversion symmetry
is violated when $\delta_m \not =0$, and when $\delta_m=0$ the system spontaneously breaks the symmetry, $\langle m\rangle \not = 0$, for $T < T_c$
in the thermodynamic limit.

It should be pointed out that a microscopic field strength $\delta_p$ is never strictly speaking equivalent to the scaling field defined in
a continuum theory. The term $\delta_pP$ that we add (or imagine adding) to the Hamiltonian can be thought of as consisting of an infinite
sum of operators $O_j$ multiplied by their conjugate scaling fields $\lambda_j$,
\begin{equation}
\delta_p P = \sum_j \lambda_j O_j,  
\label{POsum}
\end{equation}
and these fields enter scaling functions in the form $\lambda_j L^{y_j}$ in the same way as in Eq.~(\ref{afinitel1}). More precisely, the equality
above should be interpreted as a correspondence between the lattice operator $P$ and operators $O_j$ in a continuum field theory. If $y_j>0$ the field
$\lambda_j$ (and operator $O_j$) is said to be {\it relevant}, as its effect increases with the system size ($\lambda_j L^{y_j} \to \infty$ when
$L \to \infty$), and, conversely, $y_j < 0$ for an irrelevant field ($\lambda_j L^{y_j} \to \infty$, with the marginal case $y_j=0$ is associated
with logarithmic scaling corrections, which we will not discuss here).

Naturally the fields $\lambda_j$ in Eq.~(\ref{POsum}) are proportional to $\delta_p$ when the latter is small, but they also depend on
other parameters of a given model and the full form is nonlinear \cite{Aharony1983}, which causes corrections to scaling when taken into account
in FSS forms such as (\ref{afinitel1}). Here we neglect the scaling corrections that can (depending on the kind of analysis performed) be generated
by the nonlinear scaling field, and, since constants of proportionality are not important, we can use $\delta_p$ as the scaling parameter. The
scaling dimension $y_p$ in Eq.~(\ref{afinitel1}) corresponds to the largest among the $y_j$ values of the fields $\lambda_j$ in (\ref{POsum}), and
normally we just refer to this $y_p$ as the scaling dimension of $\delta_p$ even though, strictly speaking, this is only the leading scaling dimension.

We typically consider a transition tuned by the thermal field $\delta$, but scaling theory in principle relies on the possibility of arbitrary
perturbations of the critical system. A key fact is that, in the entire space of possible infinitesimal deformations of $H$, there is only a small number
of relevant fields (just $\delta \equiv\delta_t$ and $\delta_m$
in the most common case, with multi-critical points having additional relevant fields), whose infinitesimal presence destabilize the critical
point. Adding an irrelevant field $\delta_i$ just extends the critical point at ($\delta=0,\delta_m=0$) to a line of critical points in the
parameter space, with the universality class not affected. In practice, most lattice operators would contain
all the scaling fields (conjugate to continuum field operators) allowed by symmetry, and adding them at weak strength to $H$ only changes the
location of the critical point, again without changing the universality class. A simple example is the Ising model with next-nearest-neighbor
interactions $J'$ added to the conventional model with nearest-neighbor interactions $J$, where $T_c$ depends on $J$ and $J'$ but the universality
class of the $J$-only model persists (unless $J$ and $J'$ are strongly frustrated, whence other phases and transitions are realized \cite{Jin2013}).

A related important point is that it is normally only by symmetry that a given lattice operator $P$ can be guaranteed to contain a single relevant
field. In the Ising model, a change in temperature (conjugate to $P=H$) at zero external field only corresponds to ``touching'' the thermal field.
However, at the critical point of water (which is also of Ising type), changing the temperature (or the pressure) touches both the thermal field and
the order-parameter field, because there is no microscopic symmetry like the spin-inversion symmetry of the Ising model (though such symmetry effectively
emerges close to the critical point). Spin models in general have the advantage that the thermal field and the order parameter field can be
separated thanks to symmetry.

In addition to the scaling fields $\delta_p$, the local operators $p({\bf r})$ in Eq.~(\ref{msum}) are also associated with their own scaling
dimensions $\Delta_p$, which can be concretely defined in terms of the asymptotic distance dependence of the corresponding real-space 
correlation function at the critical point,
\begin{equation}
C_p(r) = \langle p({\bf r})p(0) \rangle - \langle p\rangle^2 \propto \frac{1}{r^{2\Delta_p}},
\label{cmdef}
\end{equation}
where, depending on the operator, $\langle p\rangle$ (same as $\langle p({\bf r})\rangle$ in a uniform system or in a system with randomness
when averaged over realizations) may or may not vanish. More precisely, each operator in the sum (\ref{POsum}) has its own scaling dimension and
$\Delta_p$ is the smallest of these, i.e., (\ref{cmdef}) represents the leading form of the correlations.

In the case of the order parameter, $P=M$, $\langle m\rangle=0$ at $T_c$ and the exponent $2\Delta_m$ is related to the conventional critical
exponents by
\begin{equation}
2\Delta_m = 2\beta/\nu = d-2+\eta,  
\label{ydrel1}
\end{equation}
where $\eta$ is called the anomalous dimension. Normally this kind of relationship is not used for any other operator besides the order
parameter, and no subscript is therefore attached to $\beta$ and $\eta$.

The first equality in Eq.~(\ref{ydrel1}) follows from writing the squared order parameter $\langle m^2\rangle = \langle M^2\rangle/N^2$ as a sum
(converted to an integral) of the order-parameter correlation function $\langle m({\bf r})m(0) \rangle$ over ${\bf r}$ up to a cut-off length $L$,
which gives $\langle m^2\rangle \propto L^{-2\Delta_m}$ from the form analogous to Eq.~(\ref{cmdef}). Then, comparing with Eqs.~(\ref{ainfty}) and
(\ref{afinitel}), where by definition $\sigma=2\beta$, the equality follows. The second equality in Eq.~(\ref{ydrel1}) can be regarded as the
definition of $\eta$, but this exponent is also fundamentally related to the fractal dimensionality of critical domains in the system.

One can show, as we will below, that the scaling dimensions $y_p$ and $\Delta_p$ are related to each other by
\begin{equation}
y_p \equiv \frac{1}{\nu_p} = d - \Delta_p.
\label{ydrel2}
\end{equation}
If $\delta_p$ drives the transition (typically but not necessarily by changing the temperature, whence $\delta_p=\delta$ and $y_p=1/\nu$),
the FSS ansatz (\ref{afinitel1}) of an observable $n$ in the neighborhood of the critical point can be expressed using the scaling dimension
$y_p$ of the driving field and the scaling dimension $\Delta_n$ of the observable $n$;
\begin{equation}
n(L,\delta_p) =  L^{-\Delta_n}g(\delta_p L^{y_p}).
\label{afinitel2}
\end{equation}
Again, more precisely, $\Delta_n$ is the smallest of the scaling dimensions of all operators in field theory that may be contained
in the lattice operator $n$.

It should be noted that the correlation function (\ref{cmdef}) from which a scaling dimension can be extracted may also include a phase depending on
${\bf r}$. As a simple example, in a square-lattice antiferromagnet the sign of the spin correlation function $\langle S({\bf r})S(0)\rangle$ is
$(-1)^{r_x+r_y}$, where $r_x$ and $r_y$ are the (integer) lattice coordinates.
For simplicity we do not consider any such phases here, but they can be easily handled by, e.g., considering absolute values.
For the order parameter $M$ defined using a sum of local operators as in Eq.~(\ref{msum}), a corresponding phase should then also be included, i.e.,
the Fourier transform should be taken at the relevant ordering wave-vector, which is $(\pi,\pi)$ in the above case of the antiferromagnet with local 
operators $S({\bf r})$. These phases do not in any other way affect the scaling formalism discussed here, and we will assume that they are
taken into account if needed.

In the common case of a simple critical point with only two relevant perturbations, we can extend the FSS form Eq.~(\ref{afinitel2})
of an observable to
\begin{equation}
n = L^{-\Delta_n}g(\delta L^{1/\nu},\delta_m L^{y_m},\delta_i L^{-|y_i|},\ldots),
\label{afinitel3}
\end{equation}
where $\delta_i$ is just the first of an infinite number of irrelevant scaling fields that are normally present, $\delta_i \not =0$,
and additional relevant fields are present at multi-critical points. An irrelevant field has scaling dimension $y_i < 0$ and
only produces finite-size scaling corrections in Eq.~(\ref{afinitel2}), as seen when Taylor expanding in the argument
$\delta_i L^{y_i}=\delta_i L^{-|y_i|}$ when
$L$ is sufficiently large. Here we will henceforth neglect irrelevant fields and scaling corrections.

The most fundamental among the finite-size scaling forms in classical statistical mechanics is that of singular part of the free-energy density $f$, which
has scaling dimension equal to the system's dimensionality, $\Delta_f=d$ (for reasons that we do not further motivate here but refer to the
literature \cite{Fisher1972,Fisher1998,Cardy1996});
\begin{equation}
f = L^{-d}g(\delta L^{1/\nu},\delta_m L^{y_m},\ldots).
\label{afinitel4}
\end{equation}
Scaling forms for arbitrary physical observables can be derived from this free-energy density by taking appropriate derivatives with respect
to added perturbations. As an example, the magnetization (the order parameter of a ferromagnet for simplicity) is the derivative of $f$ with respect
to the magnetic field $h = \delta_m$,
\begin{equation}
\langle m\rangle = \frac{\partial f}{\partial  \delta_m} = L^{-d+y_m}g'(\delta L^{1/\nu},\delta_m L^{y_m}),~~~g'(x,y)=\frac{\partial g(x,y)}{\partial y},
\label{afinitel5}
\end{equation}
where we see that, as implied by Eq.~(\ref{ydrel2}), indeed $\Delta_m=d-y_m$. Setting $\delta=0$ and taking $L \to \infty$ we can argue, in analogy
with Eq.~(\ref{fxsigma}), that the function $g'$ must become a power law; $g' \to (\delta_m L^{y_m})^x$, where the exponent must take the value
$x=\Delta_m/y_m=\Delta_m/(d-\Delta_m)$ so that the size dependence is eliminated. Thus, $\langle m\rangle=\delta_m^x$. In the conventional 
nomenclature $x$ is called $1/\delta$, which conflicts with our notation for the thermal field. Therefore, continuing to use $x$ for this
exponent here, we have the well known relationship $1/x=(d\nu-\beta)/\beta$.

Another important case is when $n$ in Eq.~(\ref{afinitel3}) is the internal energy density $E/N=\langle H\rangle/N$. In this case,
the infinite-size critical value (which by definition is its regular part) also has to be subtracted,
and we define the singular part of the energy density as (now for simplicity considering the tuning only of the thermal field $\delta$)
\begin{equation}
\rho_E(L,\delta)=E(L,\delta)/L^d -\lim_{\ell\rightarrow \infty}E(\ell,0)/\ell^d.
\label{rhoedef}
\end{equation}
This residual energy density has the finite-size critical form
$\rho_E = L^{-\Delta_{e}}g(\delta L^{1/\nu})$. Since, as discussed above, the energy is conjugate to the thermal field, we must have
$\Delta_{e} = d-1/\nu$ by Eq.~(\ref{ydrel2}). Therefore, in the thermodynamic
limit $\rho_E \propto \delta^{d\nu -1}$, from which we can easily obtain the more often considered critical form of the specific heat:
$c = {d\rho_E}/{d\delta} \propto \delta^{d\nu-2}$. The exponent governing the specific heat is customarily called $\alpha$, and
we see that the well known exponent relation $\alpha=d\nu-2$ is obtained.

An energy-energy correlation function $C_e(r)$ defined with local energy density operators $e({\bf r})$ can in principle be used
to determine the correlation-length exponent $\nu$ in numerical model simulations. At the critical point
\begin{equation}
C_e(r) \sim \frac{1}{r^{2\Delta_e}} = \frac{1}{r^{2(d-1/\nu)}}. 
\label{ecorrel}
\end{equation}
Since the expectation value of the energy does not vanish, the correlation function is defined with a subtraction of $\langle E(\infty,0)/N\rangle^2$, or,
equivalently, the operators $e({\bf r})$ can be defined with $E(\infty,0)/N$ subtracted.

In practice, it is often easier to determine $\nu$ using the finite-size scaling form (\ref{afinitel2}), preferably with some dimensionless
quantity $n$ ($\Delta_n=0$), e.g., one of the Binder ratios $R_k$ defined by 
\begin{equation}
R_k=\frac{\langle m^{2k}\rangle}{\langle m^{k}\rangle^2},
\label{rkdef}
\end{equation}
for which the scaling dimensions cancel. Data for $R_k$ (normally with $k=2$) versus $x=(T-T_c)L^{1/\nu}$ for different $T$ (sufficiently close to $T_c$)
and $L$ (sufficiently large)
then collapse onto the common scaling function $g(x)$, provided that $T_c$ and $\nu$ have their correct values. These values can be determined by
optimizing the quality of the data collapse. Alternatively, $T_c$ can be obtained by investigating the flow of crossing points $T_{12}=T(L_1,L_2)$
for which $R_k(L_1,T_{12})=R_k(L_2,T_{12})$ for different system size pairs $(L_1,L_2)$, e.g., $L_1=L$, $L_2=2L$. The slope $d R_k/d\delta$ at $T_c$
scales as $L^{1/\nu}$ and can be used once $T_c$ has been determined, or the maximum slope for each $L$ can be used in the same way even without
prior knowledge of $T_c$. The location (temperature) of the maximum, or any feature in a quantity governed by a scaling form such as Eq.~(\ref{afinitel}),
shifts with $L$ as $L^{1/\nu}$. Details of the practicalities of finite-size scaling have been discussed extensively in the literature,
and in Sec.~\ref{sec:qmc} we illustrate procedures in our application to the classical 3D spin glass. 

The above formalism and scaling methodology also apply to a quantum system undergoing a phase transition at a critical temperature $T_c > 0$,
because the fluctuations at the longest (diverging) length scale are always thermal unless $T=0$. Quantum fluctuations can drive continuous
transitions with $T_c=0$, as we discuss next.

\subsubsection{Quantum phase transitions}

A quantum system in $d$ spatial dimensions can be mapped through the path integral approach to an effective statistical-mechanics system with an
additional imaginary-time dimension \cite{Cardy1996,Sachdev2011}.
In some cases, this effective system is space-time isotropic beyond a simple scale factor (a velocity) that relates
space and time, but in general the time dimension is very different. The difference is quantified by the dynamic exponent $z$, which relates
the spatial correlation length $\xi$ to a relaxation time $\xi_\tau$ (a correlation length in imaginary time) according to $\xi_\tau \propto \xi^z$.
A critical real-space correlation function $C(r) \propto r^{-a}$ translates by $r \to \tau^{1/z}$ into a corresponding critical time correlation
function $C(\tau) \propto \tau^{-a/z}$. In scaling dimensions and relationships where the dimensionality appears, the classical spatial dimensionality
$d$ should be replaced by $d+z$ (where we still use $d$ for the spatial dimensionality of the quantum system), following from the scaling dimension
of the Hamiltonian density at $T=0$ (which replaces the classical free-energy density at $T>0$). With these generalizations,
the formalism discussed in the preceding section remains valid, with important extensions described below.

A classical phase transition takes place at $T_c > 0$ as a consequence of competition between the internal energy and the entropy. As
discussed in the preceding section, the energy and the entropy can both be regarded as conjugate quantities to the temperature, which
we call the thermal field when given relative to the critical temperature; $\delta=T-T_c$. In a quantum system at $T=0$ it is instead
the competition between different terms in the Hamiltonian that cause the ground state transition. The fluctuations that allow for
a continuous transition can in this case be traced to non-commuting terms of the Hamiltonian, leading to quantum fluctuations when the
ground state is expressed in some local basis, e.g., with the spin-$z$ eigenvalues $S^z_i$ ($i=1,\ldots,N$) in a spin system., 

The quantum fluctuations are regulated by changing some model parameter. A non-trivial Hamiltonian can always be formally decomposed into two non-commuting
terms, $H=aH_A+bH_B$, so that the transition of the ground state is driven by relative changes in the prefactors $a$ and $b$. Since an overall
factor does not matter for the ground state wave function, we can fix $b=1$. Then we can define the critical
Hamiltonian $H_c$ and consider deviations from it by tuning the parameter $a$ only (for simplicity):
\begin{equation}
H(\delta_a)=H_c + \delta_a H_A.
\label{htuning}
\end{equation}
Assuming that this tuning does not change the symmetry of $H$ (i.e., $H_c$ and $H_A$ have exactly the same symmetries, which normally is apparent),
$\delta_a$ is the symmetric scaling field whose inverse scaling dimension defines the conventional correlation length exponent $\nu$.
Thus, if $H_A$ consists of a sum of local terms $h_A({\bf r})$ in the same way as the generic perturbing operator
$P$ in Eq.~(\ref{msum}), we can define the correlation function, evaluated at $\delta_a=0$, and its asymptotic power-law form will be dictated
by the scaling dimension of the operator as in the classical case:
\begin{equation}
C_A({\bf r}) = \langle h_A({\bf r})h_A(0) \rangle - \langle h_A \rangle^2 \propto \frac{1}{r^{2\Delta_a}}.
\label{cadef}
\end{equation}
Then, in a generalization of the classical relationship Eq.~(\ref{ydrel2}) between scaling dimensions, the correlation-length exponent
$\nu=1/y_a$ is given by
\begin{equation}
\frac{1}{\nu}=d+z-\Delta_a.
\label{nudz}
\end{equation}
Here it should be noted that, we can also go away from the critical point by letting
$H = H_c + \delta_bH_B$, and, therefore, the scaling dimensions of the operators $H_A$ and $H_B$ must be the same; $\Delta_a = \Delta_b$. Since
it does not matter what term is tuned, we will also just use $\delta$ without subscript for the deviation from the critical point. Similarly,
we do not attach any index on $\nu$ when there is a single relevant symmetric field, i.e., at a regular critical point that can be reached
by tuning a single parameter in $H$.

The dynamic exponent
$z$ governs the low-energy excitations of the system, with the dispersion (energy--momentum) relation $\epsilon_k \propto k^z$ and a finite-size gap
\begin{equation}
\Delta_L \propto L^{-z}.
\label{gapscale}
\end{equation}
This gap scaling corresponds directly to the scaling dimension of the total Hamiltonian $H$ being $z$, i.e.,
the ground-state energy density has scaling dimension $d+z$. The corresponding quantity in the classical case is the free-energy density at
$T>0$, which has scaling dimension $d$, as stated explicitly in the finite-size form (\ref{afinitel4}). Thus, the replacement $d \to d+z$ when
converting classical scaling forms (obtained as derivatives of the free energy) to the quantum case (corresponding derivatives of the ground
state energy).

It is important to recognize that the scaling dimension of the total Hamiltonian $H=aH_a+bH_b$ is different from that of $H_a$ and $H_b$. The latter
two terms can drive a transition, as discussed above, while just changing the overall quantum-mechanical Hamiltonian by $H \to H + \delta_h H$ does
not change the ground state. It is interesting to note that the relationship Eq.~(\ref{nudz}) with $\Delta_a$ replaced by the scaling dimension of
the Hamiltonian density, $\Delta_h=d+z$, gives $1/\nu_h=0$. Thus, a scaling form $L^{-\Delta_n}g(\delta_hL^{1/\nu_h})$ of some operator $n$ under the change
of $\delta_h$ reduces to just $n=L^{-\Delta_n}g(\delta_h)$ without size dependence in $g$. In many cases the scaling function $g(\delta_h)$ must be a constant,
since the ground state does not depend on $\delta_h$. An exception is the finite-size gap, which depends on the overall scale of $H$, whence
Eq.~(\ref{gapscale}) can be written as $\Delta_L \propto (1+\delta_h)L^{-z}$, i.e., $g(\delta_h)\propto 1+\delta_H$.

A quantum phase transition takes place in the ground state, but the path integral representation is valid for any temperature, and the length
$L_\tau$ of the $d+1$ dimensional system in the imaginary-time dimension is proportional to the inverse temperature $1/T$. Finite-size scaling can in
general be formulated with both the spatial size $L$ and the temporal size $L_\tau$. In order for expectation values to converge to their ground state
limits, the temperature has to be much smaller than the gap; $T \ll L^{-z}$. This convergence and cross-over from finite-temperature behavior can be
expressed by an extended scaling function with an argument $L_\tau/L^z$:
\begin{equation}
n = L^{-\sigma/\nu}g(\delta L^{1/\nu},L_\tau/L^z).
\label{afinitelq}
\end{equation}
If $z$ of a system is not known, it can be obtained by analyzing the dependence of some suitable quantity $n$ on $L$ as well as the
inverse temperature scaled with some exponent $z'$, i.e., $1/T \propto L_\tau \propto L^{z'}$. If $z' < z$, the argument $L_\tau/L^z \to 0$
when $L$ increases, while if $z'>z$ we have  $L_\tau/L^z \to \infty$. Depending on the quantity $n$, the scaling function behaves very different
in these two limits, and the correct value $z'=z$ can be deduced as a separatrix \cite{Rieger1994}. An example of this type of analysis
for the 3D spin glass is presented in Sec.~\ref{sec:qmc}.
 
If $z$ is known, the inverse temperature can just be scaled as $1/T \propto L^z$ and the scaling function (\ref{afinitelq}) then effectively has a single
argument $\delta L^{1/\nu}$. In numerical simulations, an alternative approach if $z$ is not {\it a priori} known is to study low enough temperatures so that all 
finite-size observables have converged to their ground state values. Then, effectively the limit $L_\tau/L^z \to \infty$ is taken for each individual $L$,
and again the scaling function in Eq.~(\ref{afinitelq}) becomes one with a single argument: $g(x,y) \to \tilde g(x)$. As we will see below, the dynamic
exponent can still be accessed because it appears in scaling dimensions when the classical scaling forms are modified by setting $d \to d+z$.

An important aspect of quantum-critical scaling is that some physical observables are defined at equal time, while others are integrals over
imaginary time. Consider first the equal time expectation value $\langle m^2\rangle$ of the squared order parameter, which, following the
same procedures as in the classical case, we can express as a correlation function integrated over ${\bf r}$, resulting in
$\langle m^2\rangle \propto L^{-2\Delta_m}$. Often, this quantity is multiplied by $N=L^d$ for what is called the static structure factor,
and it scales as
\begin{equation}
S_m = \langle M^2\rangle/N=L^{d}\langle m^2\rangle \propto  L^{d-2\Delta_m} = L^{2-z-\eta},
\label{smdef}
\end{equation}
where in the last step we have used the quantum mechanical analogue of Eq.~(\ref{ydrel1}) for the scaling dimension of $m^2$;
\begin{equation}
2\Delta_m =d+z-2+\eta.
\end{equation}
The scaling form of $S_m$ also formally applies to a classical system by setting $z=0$.

The perhaps most well known example of a quantity involving integration over imaginary time $\tau$ (a Kubo integral) is the susceptibility
corresponding to the order parameter field: $\chi=d\langle m\rangle/d\delta_m$
In the common case where $m=M/N$ and $H$ do not commute
(e.g., in the case of the transverse-field Ising model), we have
\begin{equation}
\chi_m \propto \frac{1}{N}\int_0^{1/T}\langle M(\tau)M(0)\rangle d\tau,
\label{suscinteg}
\end{equation}
which can be readily demonstrated by writing ${\rm e}^{-\beta H}$ (with $\delta_m M$ included in $H$) in the thermal expectation value
$\langle m\rangle$ as $({\rm e}^{-\Delta_\tau H})^k$, with small $\Delta_\tau=\beta/k$, so that the $\delta_m$ derivative can be taken properly
in each of the factors (``time slices'') and summed up in the form of an integral when $k\to \infty$. In order for the susceptibility
to produce scaling reflecting the properties of the ground state (which is the only case we consider here, though finite-temperature
scaling can also be considered and is an important aspect of quantum criticality \cite{Sachdev2011}), $T$ has to be below the lowest
excitation gap of the system, which has the scaling form (\ref{gapscale}). Thus, the upper integration bound $1/T$ in Eq.~(\ref{suscinteg})
has to be of order $L^z$ or larger. Formally the cut-off should be set at $\tau \propto L^z$ even when the limit $T \to 0$ is taken,
because the time correlations decay exponentially for $\tau \gg L^z$.

Again writing $M$ as a sum of local operators, now $M(\tau)=\sum_{\bf r} m({\bf r},\tau)$, and integrating
$C({\bf r},\tau) =\langle m({\bf r},\tau)m(0,0)\rangle$ over both space and (imaginary) time, we have $d$ spatial coordinates
${\bf r}=(r_1,\ldots,r_d)$ and a time coordinate $\tau$. The critical correlation function has the asymptotic form
\begin{equation}
C(r,\tau)=\frac{1}{(r^2+\tau^{2/z})^{\Delta_m}}.
\label{corrtau}
\end{equation}
To integrate this over the anisotropic space-time volume, we can define $r_{d+1}\equiv\tau^{1/z}$ and  use $R$ for the length of the vector
${\bf R} = (r_1,\ldots,r_{d+1})$. Then all components are integrated up to $L$, and the scaling form of $\chi$ becomes (neglecting angular integrals that
only produce unimportant factors)
\begin{equation}
\chi_m \propto \int^L dR \frac{R^dR^{z-1}}{R^{2\Delta_m}} \propto L^{d+z-2\Delta_m} = L^{2-\eta}.
\label{chiscale}
\end{equation}
Thus, compared to the structure factor (\ref{smdef}), there is an additional factor $L^z$ arising from the time integral, and the dynamic exponent can be
singled out, e.g., $\chi_m/S_m \propto L^z$.

There is no obvious classical counterpart to the above scaling form of $\chi_m$, which is reflected in the fact that there is no $z$ left that can be set
to zero, unlike Eq.~(\ref{smdef}). The definition (\ref{suscinteg}) of the susceptibility can of course still be used classically, in which case
$M(\tau)=M(0)=M$ (because $M$ and $H$ commute) and the integral only produces a factor $1/T$. Thus, classically $\chi_{m} = S_m/T$, which holds at
any temperature.

From Eqs.~(\ref{smdef}) and (\ref{chiscale}) we see that both the exponents $z$ and $\eta$ can, in principle, be determined by investigating the size
dependence of the structure factor and the susceptibility. In current quantum annealing devices, only equal-time quantities are accessible, in the form of
the probability distribution of the spin-$z$ configurations
at the end of the annealing process. This state can contain ``memories'' of the wave function close to the quantum-critical
point, through the critical slowing-down mechanism first discussed by Zurek \cite{Zurek1985,Zurek1996} in classical statistical mechanics (following
related work on cosmological topological defects by Kibble \cite{Kibble1976}) and later generalized to quantum phase transitions as well
\cite{Polkovnikov2005,Dziarmaga2005,Zurek2005}. The dynamics of quantum annealing depends on $z$, thus offering a path to extracting this exponent
from experiments. We next discuss the generalization of FSS to a system undergoing an annealing process, first in general terms and then focusing
specifically on the spin-glass annealing experiments and simulations. 

\subsection{Dynamic finite-size scaling}\label{scaling:kz}
           
In a classical or quantum annealing process in which a parameter like $\delta$ varies linearly as a function of time $t$, the finite-size scaling
form (\ref{afinitel2}) attains a second scaling argument;
\begin{equation}
n = L^{-\Delta_n}g(\delta L^{1/\nu},vL^{\mu}),
\label{akzform}
\end{equation}
where $v$ is the velocity by which $\delta$ changes, $v=d\delta(t)/dt$ taken at $\delta=0$ (and if $v=0$ according to this definition, the approach
discussed here can be generalized to the second derivative or any higher derivative \cite{Degrandi2010,Degrandi2011}), and we consider $v>0$ for
simplicity. Since constants of proportionality do not matter in scaling forms, we can also simply define the scaling velocity as the
inverse of the total annealing time; $v=1/t_a$. The exponent $\mu$ controlling the velocity scaling is defined in the same way as in the main paper;
\begin{equation}
\mu = z+1/\nu.
\label{mudef}
\end{equation}
The FSS form (\ref{akzform}) has its origin in the KZ mechanism \cite{Zurek1996,Polkovnikov2005,Dziarmaga2005,Zurek2005} and its use for
analyzing numerical simulation data and experiments under various dynamical protocols was further elaborated and tested in numerous 
later works  \cite{Deng2008,Degrandi2011,Chandran2012,Kolodrubetz2012,Liu2014}.

We here assume that the process starts with the system in a disordered phase and approaches an ordered phase, though the opposite process can also
be treated in a similar way. The basic starting point leading to (\ref{akzform}) is that a finite system approaching a continuous classical or quantum
phase transition at rate $v$ ``freezes out'' when the remaining time to reach the critical point is less than the required time for the system
to equilibrate. In the thermodynamic limit, the correlation length reached before the system falls out of equilibrium is similarly
\begin{equation}
\xi_v \propto v^{-1/\mu}.
\end{equation}
The argument $vL^{\mu}$ in Eq.~(\ref{akzform}) is simply a power of the dimensionless ratio $L/\xi_v$.

The KZ mechanism is often discussed in terms
of defects generated by the out-of-equilibrium dynamics, but in general, except for some often studied 1D systems, it is not easy to directly relate
a defect density to observed physical quantities. However, the scaling approach relies only on the existence of some dynamic process associated with a
relaxation time $\xi_t$, which should be related to a spatial correlation length $\xi$ by $\xi_t \propto \xi^z$. No other information on the dynamic
process or the nature of the defects is required to derive universal scaling forms \cite{Chandran2012,Liu2014}.

In a quantum system undergoing an annealing process of the type considered here, $z$ is exactly the dynamic exponent discussed in the previous section.
Though there $z$ was introduced in the context of imaginary (Euclidean) time, the relevant time scale is the same in real and imaginary time:
$\xi_t \propto \xi_\tau \propto \xi^z$. At a classical equilibrium phase transition, this quantum dynamic exponent, which appears in the scaling
dimensions, e.g., in Eq.~(\ref{nudz}), is formally zero. However, when a classical system is subjected to some stochastic dynamics, e.g., in a Monte
Carlo simulation, there is a dynamic exponent corresponding to the particular process used (Metropolis or cluster updates). Then, in the KZ exponent
Eq.~(\ref{mudef}), $z$ is the stochastic dynamic exponent. A third possibility is a quantum system with intrinsic dynamic exponent $z$ that is simulated
using a Monte Carlo process, such as the SQA simulations used as benchmarks here. Then the intrinsic (quantum) dynamic exponent should still be used in
Eq.~(\ref{nudz}), but $z$ in the KZ exponent should be the relevant exponent corresponding to the imposed Monte Carlo dynamics (which also depends on
the system studied) \cite{Liu2015b}. Here, for simplicity of the notation, we will first assume an actual quantum system undergoing
Hamiltonian dynamics, where there is only a single dynamic exponent, but we will later return to the other cases as well.

The easiest way to analyze data within the dynamic finite-size scaling ansatz is to consider the quantity $n$ exactly when the
critical point has been reached: $\delta=0$ \cite{Degrandi2011,Liu2014}. Then we can simply write Eq.~(\ref{akzform}) as
\begin{equation}
n = L^{-\Delta_n}g(vL^{\mu}),
\label{akzform2}
\end{equation}
which can be tested with data for different values of $v$ and system sizes $L$. For $v \to 0$, we should of course recover the
standard FSS form $n \propto L^{-\Delta_n}$, which means $g(x) \to {\rm constant}$ when $x\to 0$. For $v > 0$, the scaling
function can be extracted by graphing $y=nL^{\Delta_n}$ versus $x=vL^{\mu}$ for sufficiently large $L$ and a range of $v$ values,
whence all the data points should collapse onto the common function $g(x)$. If the exponents are not known, they can be found by
adjusting their values for optimal collapse.

Quantities $n$ with known scaling dimension $\Delta_n=0$ are particularly useful, since any uncertainties in (or lack of knowledge of) the leading
finite-size dependence can then be avoided. A well known example is the Binder ratio defined with two different powers of the order parameter, Eq.~(\ref{rkdef}),
for which the scaling dimensions of the numerator and denominator cancel. Normally the $k=2$ case, $R_2=\langle m^4\rangle/\langle m^2\rangle^2$, is
considered, and a corresponding cumulant is defined as $U\equiv U_2 = a(1-bR_2)$, where the coefficients $a$ and $b$ are determined using the symmetry
of the order parameter such that $U \to 0$ in the disordered phase and $U \to 1$ in the ordered phase. The definition of this quantity
in a system with random couplings is further discussed below in Sec.~\ref{sec:bindercumulant}.

The asymptotic form of $g(x)$ in Eq.~(\ref{akzform2}) can be argued as follows: For $v > 0$ the correlation length $\xi_v$ is finite. In limit
$L \gg \xi_v$ the system is disordered and the size dependence becomes trivial, though its form depends on the quantity $n$ considered. In the case of a
squared order parameter, we have $\langle m^2\rangle = \langle M^2\rangle/N^2 \propto 1/N=L^{-d}$, which can be easily seen by expressing $\langle M^2\rangle$
using the exponentially decaying (asymptotically) correlation function $C_m({\bf r})$ of corresponding local operators [defined with a generic operator $P$
in Eq.~(\ref{msum})]:
\begin{equation}
\langle M^2\rangle = N\sum_{\bf r}C_m({\bf r}) \propto N~~~~({\text{for~short-ranged~}}C_m). 
\label{m2corrsum}
\end{equation}
In order to obtain the $L^{-d}$ dependence of $\langle m^2\rangle$ from Eq.~(\ref{akzform2}), the scaling function must take an asymptotic power-law form,
$g(x) \to x^{-c} \tilde g(x)$ with the function $\tilde g(x)$ approaching a constant when $x \to \infty$ and the exponent $c$
satisfying $c\mu+\Delta_{m^2}=d$, with $\Delta_{m^2}=2\Delta_m$, i.e.,
\begin{equation}
\langle m^2\rangle \propto L^{-2\Delta_m}(vL^{\mu})^{-c}\tilde g(vL^{\mu}) \to L^{-d}v^{-c},~~~~c=\frac{d-2\Delta_m}{\mu}=\frac{d-2\beta/\nu}{\mu}.
\label{mcexp}
\end{equation}
This is the KZ FSS form far from equilibrium, which must eventually break down when $v$ is very large, i.e., when the correlation length $\xi_v$ is of order
unity (the lattice spacing in a lattice model). The ultimate high-velocity form is just $\langle m^2\rangle = L^{-d}g(v)$, where the function $g$ is
analytic in $1/v$, and this form crosses smoothly over into the algebraic KZ form (\ref{mcexp}) as $v$ is reduced. When $v$ is further reduced, for
given $L$ this form turns smoothly into the equilibrium form of Eq.~(\ref{akzform2}), $\langle m^2\rangle \propto L^{-2\beta/\nu}$, when $v \to 0$.

For the excess energy density defined in Eq.~(\ref{rhoedef}), the above arguments can be repeated with the modification that now $\rho_E$ must be
size independent at high velocities since it is a simple density with only exponentially small finite-size corrections in the disordered initial
state. Thus, the algebraic KZ form
for large $vL^{\mu}$ (but still $v \ll 1$) is 
\begin{equation}
\rho_E \propto L^{-\Delta_{\rho_E}}(vL^{\mu})^{\kappa}\tilde g(vL^{\mu}) \to v^{\kappa},~~~~\kappa=\frac{\Delta_{\rho_E}}{\mu}=\frac{d+z-1/\nu}{\mu}.
\label{ecexp}
\end{equation}
This form evolves gradually into a constant when $v \to \infty$, and in the other extreme it crosses over to the equilibrium form
$\rho_E \propto L^{-\Delta_e}= L^{-(d+z-1/\nu)}$
when $v$ is of order $L^{-\mu}$ (i.e., when $\xi_v$ grows to order $L$). Note that the exponents $c$ and $\kappa$ in Eqs.~(\ref{ecexp}) and (\ref{mcexp}) are
both positive but they appear with different signs when the powers of $\delta$ are taken, reflecting the decrease in the residual energy and increase in
the order parameter as the annealing velocity is reduced.

For the quantities $n$ considered above, and many others, the algebraic KZ scaling form can be investigated by graphing $y=nL^{\Delta_n}$ versus
$x=vL^{\mu}$ after a process stopping at the critical point.
The power-law forms discussed above will then cross over into constants for sufficiently low $v$, which  is possible because of the function
$\tilde g(vL^{\mu})$ in Eqs.~(\ref{mcexp}) and (\ref{ecexp}). Upon increasing $v$, data for small systems will successively peel off from the common scaling
function when the correlation length $\xi_v$ is of the order of the lattice spacing (loosely speaking, when $v$ is of order one). In an alternative method,
$y=n$ (in the case of the residual energy density) or $y=nL^d$ (in the case of the squared order parameter) can be graphed versus $v$ or $v^{-1}$,
in which case the power-law behaviors in Eqs.~(\ref{mcexp}) and (\ref{ecexp}) are observed in the algebraic KZ regime but data collapse does not
apply when equilibrium is approached (where curves for different $L$ gradually peel off from the common power-law form). Instead, data collapse now
applies also in the high-$v$ limit, where constant behaviors pertain. The two ways of analyzing SA data have been demonstrated for uniform 2D and
3d classical Ising models (using both local and cluster Monte Carlo updates) \cite{Liu2014} as well as 2D \cite{Rubin2017,Xu2017} and 3D \cite{Liu2015}
classical spin glasses. Similar methods were also applied to imaginary-time QA (implemented in quantum Monte Carlo simulations with imaginary-time
dependent interactions) of the transverse-field Ising model with uniform \cite{Degrandi2011} and random (3-regular graphs) \cite{Liu2015b} interactions.

In the case of the Binder cumulant (or ratio), the scaling dimension is zero but the asymptotic form of the equilibrium scaling function
$U=u(\delta L^{1/\nu})$ in the disordered phase is non-trivial and still not well understood \cite{Privman1990,Shao2020}. The size dependence of $U$
in a disordered state is not just a constant [as in Eq.~(\ref{ecexp})] or $\propto L^{-d}$ [as in Eq.~(\ref{mcexp})] but a different power-law
$L^k$ that is not known generically. The exponent $b$ governing the KZ power law $U \propto v^{-b}$ when $L \gg \xi_v$ can therefore also not be
obtained in a simple manner using just the arguments leading to Eqs.~(\ref{mcexp}) and (\ref{ecexp}). The scaling form $U=u(v L^\mu)$ still of course
applies (including an asymptotic power-law form with some exponent $b$) and can be extracted empirically as in other cases by graphing $y=U$
versus $x=vL^{\mu}$ with $\mu$ optimized.

When annealing slightly into an ordered phase (taken as $\delta > 0$) the extended KZ form (\ref{akzform}) can be Taylor expanded as long as $\delta L^{1/\nu}$
remains small, i.e., asymptotically for large $L$ the system tends to the critical point. When annealing more significantly into the ordered phase,
but still within the critical window $\delta \propto L^{-1/\nu}$, power laws appear similar to those discussed above on the disordered side of the
transition. When annealing far into an ordered state the situation is much more complicated and non-generic, requiring additional knowledge of the
dynamical processes inside the phase. While some progress on this front has been reported recently \cite{Schmitt2022}, in general the issue remains
largely open. We next discuss annealing into a spin-glass phase and present insights and hypotheses of direct relevance to the quantum annealing
experiments and simulations reported in this work.

\subsection{Application to annealing experiments and simulations}
\label{scaling:qa}

For a system annealed through a phase transition and slightly into an ordered phase, we can expect the extended KZ finite-size scaling form (\ref{akzform})
to apply. In principle, once the system is deep enough in the ordered phase, some other dynamical process beyond the critical fluctuations underlying
the KZ mechanism will take over and eventually dominate. In some cases, a dynamical process in the ordered phase is faster than the KZ dynamics, e.g.,
in the case of coarsening dynamics of a system with no randomness---see Ref.~\cite{Blanchard2017} for discussions of how the KZ scaling is violated in
the ordered phase in a simple Ising model. Then KZ dynamics should in general not be expected, though recent work suggests that an extension of the
scheme, with modified power laws, is valid slightly inside in ordered phase in some cases like the standard uniform 2D Ising model in a transverse
field \cite{Schmitt2022}.

Inside a classical or quantum glass phase, all dynamical processes should be extremely slow. Therefore, KZ scaling can be anticipated
even if the annealing process stops (and the outcome is observed only) far beyond the transition point. In the QA device used in our experiments, there are
also technical reasons related to the shape of the annealing protocol (see Fig.~\ref{fig:1}b) why the system should ``freeze out'' and not evolve
significantly at the latter stages of the process ending at the classical point $s=1$ (because the transverse field becomes very small already
at $s \approx 0.7$).  In MC simulations, KZ scaling can be measured by observing the system exactly at the critical point (as was done in an SA study
of the simple-cubic 3D Ising spin-glass \cite{Liu2015}). However, in this work we anneal deep into the glassy phase even in SA and SQA for two
reasons. First, we wish to qualitatively reproduce the QA protocol. Second, annealing toward the $T=0$, $\Gamma=0$ point guarantees a final equilibrium
near the classical ground state.

Fig.~\ref{fig:3}f shows quite clearly the expected crossover when the density of AFM bonds is increased from $p=1/2$, between glassy KZ dynamics with an
exponent $\mu$ that is close to its known KZ value ($\mu=z+1/\nu$) and a smaller value $\mu \approx 2$. The smaller value may indicate coarsening
dynamics inside the AFM phase for $p \agt 0.8$. However, coarsening in quantum systems has not been extensively investigated using reliable unbiased
simulations, and no direct evidence of such dynamics was found at the early stages of entering the ferromagnetic phase of the transverse-field Ising
model in Ref.~\cite{Schmitt2022}.

Focusing here on the glass phase, from the analysis of the Edwards-Anderson (EA) order parameter $\langle q^2\rangle$ (where again $\langle .\rangle$
also involves an average over disorder realizations), while the exponent $\mu$ is close to its KZ value, the scaling dimension of the quantity itself
(the exponent $r$ in Figs.~\ref{fig:ed_collapse_full} and \ref{fig:ed_collapse_half}) is far from the anticipated critical scaling dimension. We will
show here that such behavior over a range of relatively fast annealing rates and small to moderate system sizes can be explained by an extended KZ
scaling form supplemented by mild, physically reasonable assumptions pertaining to the glassy state.

\subsubsection{Spin glass order parameter}

Before considering scaling behavior, we briefly discuss the EA order parameter in light of the definitions of exponents and scaling
dimensions explained in the previous sections. We use the conventional overlap parameter of the spins $S^{(a)}$ and $S^{(b)}$ in two
replicas, $a$ and $b$,
\begin{equation}
q_{ab} = \frac{1}{N} \sum_i S_i^{(a)}S_i^{(b)},~~~S_i^{(a,b)} \in \{-1,+1\},
\label{qdef}
\end{equation}
obtained from the same realization of the Ising couplings but in different runs with the QA device or in independent MC simulations. The expectation
value (quantum or thermal, depending on the case considered) of the squared order parameter for a given coupling realization is
\begin{equation}
\langle q^2_{ab}\rangle =  \frac{1}{N^2} \sum_{i,j} \langle S_i^{(a)}S_j^{(a)}\rangle\langle S_i^{(b)}S_j^{(b)}\rangle,
\label{q2def1}
\end{equation}
where the expectation value factors because the two replicas are statistically independent. We can now take averages also over the replicas. Because
the coupling realization is the same in all replicas, the replica-averaged expectation values  in Eq.~(\ref{q2def1}) will be the same. In fact, over a
sufficiently long time a finite system is ergodic, and no average over $a$ and $b$ replicas need to be formally taken. However, in reality simulations of
glasses are extremely slow, and replicas have to be considered in order to sufficiently sample the configuration space. In annealing experiments and
simulations out of equilibrium, the systems are by definition not ergodic, and averaging over replicas is an integral aspect of such studies. Once
replica and disorder averages have been taken, we drop the indices $a$ and $b$ and have
\begin{equation}
\langle q^2\rangle =  \frac{1}{N^2} \sum_{i,j} \langle S_iS_j\rangle^2.
\label{q2def2}
\end{equation}
When averaging over coupling realizations, translational symmetry is also restored, and we can introduce a squared correlation function
$C^2({\bf r})$ as the averaged $\langle S_iS_j\rangle^2$, with ${\bf r}={\bf r}_i-{\bf r}_j$, to write
\begin{equation}
\langle q^2\rangle =  \frac{1}{N} \sum_{{\bf r}} C^2({\bf r}).
\label{q2def3}
\end{equation}
Such a sum without squaring $C({\bf r})$ would vanish because of coupling averaging when the fraction of AFM couplings is $1/2$, and the use
of the overlap $q$ defined with replicas is a convenient way to solve this problem. The fact that the squared EA order parameter formally is a
sum over squared correlation functions in Eq.~(\ref{q2def3}) then has to be taken into account when using exponent relationships such as (\ref{ydrel1}),
where $d$ and $z$ cannot depend on what correlation function is considered and, therefore, the factor in front of the scaling dimension must be $4$
instead of $2$.

Note that $\langle q^2\rangle$ is a perfectly valid order parameter also for a system without disorder, e.g., an Ising ferromagnet. Then
it is clear that the decay of the squared correlation function should be associated with four times the scaling dimension of the order parameter,
$C^2_m(r) \propto 1/r^{4\Delta_m}$, because of the exponent $2\Delta_m$ in the conventional (not squared) correlation function (\ref{cmdef}). Thus,
if we wish to use conventional definitions of the critical exponents in the case of the EA order parameter, obeying forms analogous to
Eq.~(\ref{nudz}), we would have to write the FSS form of $\langle q^2\rangle$ implied by Eq.~(\ref{q2def3}) as
\begin{equation}
\langle q^2\rangle \propto  L^{-2(d+z-2+\eta)}. 
\label{q2def4}
\end{equation}
However, normally the exponents are defined differently in spin glasses, by treating $q$ as a regular order parameter and defining $\eta$ based on
the scaling form for susceptibilities such as (\ref{chiscale}), where $\chi_q$ would be defined with $M=Nq$ in Eq.~(\ref{suscinteg}) \cite{Rieger1994}.
Such a definition of $\eta$---let us call it $\eta'$---does not satisfy the common relationship to a scaling dimension, i.e., $2\Delta_q \not= d+z-2+\eta'$.

To conform with the standard notation in the spin glass literature, we will here use the
common definition of the order parameter exponent $\beta$ for spin glasses, i.e., we treat the EA order parameter as a conventional order
parameter with the form $\langle q^2\rangle \propto \delta^{2\beta}$ in the thermodynamic limit, so that the critical finite-size form is
$\langle q^2\rangle \propto L^{-2\beta/\nu}$. Note that the above issue with exponent relationships does not affect the exponent $\nu$ when
extracted using data-collapse methods, or the intrinsic (quantum) dynamic exponent $z$ when it is extracted from the dependence on the aspect
ratio $L_\tau/L^z$ in Eq.~(\ref{afinitelq}). The KZ exponent $\mu=z+1/\nu$ is also not affected by the specific definitions of $\beta$ and $\eta$.
We will use a conventional scaling dimension further below when discussing the residual energy, whose definition is independent of the use
of replicas.

\subsubsection{Scaling of the spin-glass order parameter}

Applying the extended KZ form (\ref{akzform}) to the EA order parameter, we have
\begin{equation}
\langle q^2\rangle = L^{-2\beta/\nu}g(vL^{\mu},\delta L^{1/\nu}),
\label{qqscale1}
\end{equation}
where $\delta=s-s_c$ and $s(t)\in [0,1]$. In theoretical work and simulations, the simple linear protocol $J \propto s$, $\Gamma \propto 1-s$
is often used for transverse-field Ising models. In the QA experiments, the protocol is nonlinear (Fig.~1d) but linearity still holds for $s$
in the close neighborhood of the critical point and Eq.~(\ref{qqscale1}) is valid with $v$ defined as the actual coupling derivative, or, for scaling
purposes, the inverse of the total annealing time.

Though there are some claims in the literature that $z \to \infty$ at the glass transition in the 2D and 3D Ising spin glasses considered here
\cite{Miyazaki2013,Fernandez2016} (similar to random-field models \cite{Pich1998,Nishimura2020}), in our opinion the available numerical evidence leans in the favor of
finite, relatively small values of $z$ \cite{Rieger1994,Singh2017}. In our own simulations and QA experiments (SI Sec.~II and Figs.~E6-E8), we also do not see any
evidence of large or significantly flowing (with increasing system size) values of $z$. Thought the possibility of an infinite-disorder fixed point (where
formally $z=\infty$) cannot be completely ruled out, we here consider only scaling with finite $z$.

As discussed in the preceding sections, the thermodynamic-limit form $\langle q^2 \rangle \propto \delta^{2\beta}$ demands that $g(0,x)$ with
$x=\delta L^{1/\nu}$ reduces to $x^{2\beta}$ when $x \to \infty$ inside the spin-glass phase close to the transition point. Moving further inside the glass
phase, this power-law form can no longer hold. To enable the cross-over to a different (unspecified) form in equilibrium, we can consider a more
flexible version of Eq.~(\ref{qqscale1}),
\begin{equation}
\langle q^2\rangle = L^{-2\beta/\nu}f(vL^{\mu},\delta, L),
\label{qqscale2}
\end{equation}
where $\delta$ and $L$ appear as two different arguments in the function. For small $\delta$ we know that the function must simplify to the
standard scaling form (\ref{qqscale1}), where the two arguments $\delta$ and $L$ are replaced by just the single argument $\delta L^{1/\nu}$.

We can express the cross-over from the two arguments $(L,\delta)$ to effectively a single argument by introducing another function as an argument to
the scaling function,
\begin{equation}
\langle q^2\rangle = L^{-2\beta/\nu}g(vL^{\mu},h(\delta,L)\delta L^{1/\nu}),
\label{qqscale3}
\end{equation}
where $h \to 1$ in the regime of $\delta$ and $L$ where the simpler form (\ref{qqscale1}) applies. This functional form is still as general as
Eq.~(\ref{qqscale2}) and can in principle describe the behavior for any $\delta$ and $L$. To recover the form $\langle q^2\rangle \propto \delta^{2\beta}$
from (\ref{qqscale3}) in the thermodynamic limit when $v$ is small, a power-law of the second argument must again form
\begin{equation}
\langle q^2\rangle \to  L^{-2\beta/\nu}h^{2\beta}(\delta,L)(\delta L^{1/\nu})^{2\beta}g(vL^{\mu}) = \delta^{2\beta}h^{2\beta}(\delta,L)g(vL^{\mu}),
\label{qqscale4}
\end{equation}
where we have also assumed that, to a good approximation, we do not need to keep a second argument of the function $g$ when the power law form has
set in, which is equivalent to saying that the function $g$ now is of the form $g(vL^{\mu},k(\delta,L))$, where the function $k$ is almost constant
in the relevant regime of $L$ and $\delta$.

To simplify further, we can re-define the function $h$ in Eq.~(\ref{qqscale4}) without the exponent $2\beta$, and we can also absorb the factor
$\delta$. Then
\begin{equation}
\langle q^2\rangle = h(\delta,L)g(vL^{\mu}).
\label{qqscale5}
\end{equation}
This form is clearly valid for $v\to 0$, since it can
capture any dependence on $\delta$ and $L$, and we know that $h(\delta,L) \to h(\delta L^{1/\nu})$ close to the critical point. We have of course
made assumptions leading to $g(vL^{\mu})$ appearing as a separate factor, for which we have no proof but which can be tested in experiments
and simulations.

Inside the glass phase, the squared order parameter converges with $L$ to a non-zero value when $v \to 0$. Thus, for sufficiently
large $L$ inside the ordered phase, in Eq.~(\ref{qqscale5}) $h(\delta,L) \to k(\delta)$ and
\begin{equation}
\langle q^2\rangle = k(\delta)g(vL^{\mu}),
\label{qqscale6}
\end{equation}
where the function $k$ represents whatever the $\delta$ dependence is in equilibrium,
\begin{equation}
k(\delta)\equiv \langle q^2(\delta,v=0)\rangle,
\label{kdel}
\end{equation}
and $g(x\to 0) \to 1$.

The final forms (\ref{qqscale5}) and (\ref{qqscale6}) cannot be strictly correct, as KZ dynamics should at least not be valid at the final stage 
(the most extreme case) when the order saturates. However, this just means that there is some other (assumed to be slower) dynamics that takes over,
but we have assumed only one dynamical mechanism. The assumptions and resulting forms may be valid as long as the annealing time is not too long,
so that the dynamics associated with the glass phase can be considered as a weak perturbation to the critical dynamics with exponent $z$. Then
also $\langle q^2\rangle$ will remain small though its scaling may reflect (as we will see) a cross-over from critical to ordered behavior. 

For $\delta$ not small and $L$ not large enough to eliminate the $L$ dependence of $h$ in (\ref{qqscale5}), we must still keep the 
function with two arguments. It is not clear  what the functional form of this $h(\delta,L)$ is, but it is at least plausible that a
power law applies for some range of system sizes, i.e.,
\begin{equation}
h(\delta,L) \to \tilde k(\delta)L^{-r}.
\label{hform}
\end{equation}
Here, for $\delta\to 0$ (i.e., in the limit of conventional FSS and KZ scaling) we must have $r=2\beta/\nu$ and $k(\delta)$ approaching a non-zero constant.
For $L \to \infty$ and $\delta \not=0$, the exponent $r$ must tend to zero and $\tilde k(\delta)$ the takes the equilibrium form as in Eq.~(\ref{kdel}).
For fixed $\delta$ inside the glass phase, the exponent $r$ must then be flowing but may still evolve only slowly with $L$, thus essentially
constant when observed only over a small range of system sizes. Indeed, in our experiments and simulations, we observe an effective exponent
$r$ between $0$ and $2\beta/\nu$.

\subsubsection{The Binder ratio}
\label{sec:bindercumulant}

According to our arguments above, an effective scaling dimension different from that at the critical point can be manifested when
annealing a system into a spin glass phase;
\begin{equation}
\langle q^2\rangle = \tilde k(\delta)L^{-r_L}g(vL^{\mu}),
\label{qqscale7}
\end{equation}
with an exponent $r_L$ that changes only slowly with $L$ from $r_L=0$ to $r_L=2\beta/\nu$. As in the case of conventional critical scaling, the effective
scaling dimension of $\langle m^4\rangle$ should be twice that of $\langle m^2\rangle$, and the Binder ratio (and cumulant) then remains dimensionless.
Thus, it is the preferred quantity to analyze when the goal is to extract the exponent $\mu$ in annealing experiments where the state exactly at the transition
point $s_c$ is inaccessible.

The Binder ratio for a system with intrinsic randomness (e.g., a spin glass) can be defined  with the average over realizations taken either
before or after evaluating the ratio in Eq.~(\ref{rkdef}). Now using $[.]$ to denote a disorder average and $\langle .\rangle$ for only thermal and
quantum averages (including replicas), we can define the following two ratios:
\begin{subequations}
\begin{eqnarray}
R_a & = & \left [\frac{\langle q^4\rangle}{\langle q^2\rangle^2}\right], \label{rata} \\
R_b & = & \frac{[\langle q^4\rangle]}{[\langle q^2\rangle]^2}. \label{ratb}
\end{eqnarray}
\end{subequations}
While Eq.~(\ref{rata}) has been frequently used (e.g., Ref.~\cite{Rieger1994}), Eq.~(\ref{ratb}) has become more common (see also Ref.~\cite{Marinari1997}).
They are both dimensionless quantities and in principle equally valid for use in scaling analysis.

Equation (\ref{rata}) has a drawback of being biased when evaluated with a finite number of measurements to estimate the expectation values.
Consider statistical deviations $\epsilon_2$ and $\epsilon_4$ of $\langle q^2\rangle$ and $\langle q^4\rangle$ from their exact values
$\langle .\rangle_{\rm ex}$ for a given coupling realization, based on $\Lambda$ measurements. Then the ratio Eq.~(\ref{rata}) is
\begin{equation}
R=\frac{\langle q^4\rangle}{\langle q^2\rangle^2}=\frac{\langle q^4\rangle_{\rm ex}+\epsilon_4}{(\langle q^2\rangle_{\rm ex}+\epsilon_2)^2}
= \frac{\langle q^4\rangle_{\rm ex}}{\langle q^2\rangle_{\rm ex}^2} + O(\epsilon_2^2,\epsilon_2\epsilon_4),
\label{rerror}
\end{equation}
where odd powers of the error vanish on average. Since the magnitudes of $\epsilon_4$ and $\epsilon_2$ scale as $\Lambda^{-1/2}$, the expected bias in the realization-averaged ratio $R_a$, Eq.~(\ref{rata}), scales as $\Lambda^{-1}$. While this bias decays faster than the expected overall $\propto \Lambda^{-1/2}$ statistical error in $R$, if the number of measurements $\Lambda$ is small there can still be a non-negligible effect of bias left. This bias is smaller by a factor $N_{\rm r}^{-1/2}$ in $R_b$, Eq.~(\ref{ratb}), when both the numerator and denominator are averaged over $N_{\rm a}$ independent disorder realizations.

The leading bias can in principle be removed by combining results for different $\Lambda$. In the simulations and experiments for the cumulant $U=(3-R)/2$ presented here, $\Lambda$ is sufficiently large for the bias to be insignificant. We nevertheless use Eq.~(\ref{ratb}), though (\ref{rata}) gives compatible results.

\subsubsection{Residual Ising energy}\label{sm:kappa}

For the residual Ising energy $E$, we proceed in a slightly different way. Since the Ising part of the Hamiltonian drives the transition, its scaling
dimension $\Delta_e=d+z-1/\nu$ is given by Eq.~(\ref{nudz}) with $\nu$ being the conventional correlation-length exponent. The KZ-FSS scaling form is
\begin{equation}
\rho_E = L^{-\Delta_{\rho_E}}f(vL^{\mu},\delta L^{1/\nu}),
\end{equation}
where it is assumed that the regular part of the energy, i.e., the ``background'' that is not part of the singular behavior, has been subtracted off.
The need for a subtraction here is an important aspect of the residual energy that is absent in the case of the order parameter. The subtracted part
should normally be the equilibrium critical Ising energy in the thermodynamic limit, as in Eq.~(\ref{rhoedef}). However, nothing prohibits us 
from instead subtracting the equilibrium finite-size Ising energy $E(\delta,L)$, which has the same scaling dimension as $\rho_E$. Then we have the
following expected FSS form [keeping the same symbol $\rho_E$ as before even though its meaning has changed slightly]:
\begin{equation}
\rho_E(L,\delta,v) = L^{-\Delta_{\rho_E}}[g(vL^{\mu},\delta L^{1/\nu})-g(0,\delta L^{1/\nu})].
\end{equation}
Here the difference between the two scaling functions is just another scaling function, and we can write
\begin{equation}
\rho_E(L,\delta,v) = L^{-\Delta_{\rho_E}}k(vL^{\mu},\delta L^{1/\nu}).
\label{rhoe2}
\end{equation}
For any $\delta$, by definition (because of what we have decided to subtract), this $\rho_E$ must vanish when $v \to 0$. That
means that a positive power of $v$ must appear as a factor, and $\delta L^{1/\nu}$ can only appear in a function multiplying this
overall $v$ dependence. The most natural way to express this behavior is for $k$ to develop a power law of its first argument, i.e.,
\begin{equation}
k(vL^{\mu},\delta L^{1/\nu}) \to (vL^{\mu})^\kappa h(vL^{\mu},\delta L^{1/\nu}).
\label{kvwithh}
\end{equation}
Note that the new scaling function $h$ can still also depend on $vL^{\mu}$, and mathematically there is no approximation here
because the right-hand side still depends only on the two arguments $x=vL^{\mu}$ and $y=\delta L^{1/\nu}$ (i.e., it is some function of
these two arguments, as on the left-hand side).

The exponent $\kappa$ can again be extracted as in Eq.~(\ref{ecexp}) if we demand no $L$ dependence apart from that in the function $h$,
and we now call this critical exponent $\kappa_c$. We know that this scaling must hold when $\delta=0$ and the KZ correlation length is smaller
than the system size. At least for small $\delta$, as long as the function $h(x,y)$ is well behaved, the exponent on $v$ cannot change.
Then, the final residual Ising energy in a QA process stopping inside the glassy phase takes the form
\begin{equation}
\rho_E(L,\delta,v) = v^{\kappa_c} h(vL^{\mu},\delta L^{1/\nu}),
\label{rhoeinglass}
\end{equation}
with the same exponent $\kappa_c$ as when stopping at the critical point,
\begin{equation}
\kappa_c = \frac{\Delta_{\rho_E}}{z+1/\nu} = \frac{d+z-1/\nu}{\mu}.
\label{kappaform}
\end{equation}
It should be emphasized that Eqs.~(\ref{rhoeinglass}) and Eq.~(\ref{rhoe2}) are two completely equivalent forms of $\rho_E$, and whichever
form to use just depends on the type of scaling analys one wishes to perform. They are both scaling functions of the single variable $x=vL^\mu$
at the critical point $\delta=0$.

For a process not stopping exactly at the critical point, a scaling function of a single variable can still be obtained when
considering the thermodynamic limit of Eq.~(\ref{rhoeinglass}). For $\delta \not=0$, when $L \gg \xi$ the system size can no longer appear.
$L$ can only be eliminated from the scaling function $h(x,y)$ with $x=vL^\mu$ and $y=\delta L^{1/\nu}$ if the function of two argument
evolves into one of a single argument, $h[z(x,y)]$, where $z$ does not depend on $L$. To cancel $L$, we can take $z=x^{1/\mu\nu}y^{-1}$ and
obtain a scaling function $e(z)$ of a single variable, whence Eq.~(\ref{rhoeinglass}) becomes
\begin{equation}
\rho_E(\delta,v) = v^{\kappa_c} e(v^{1/\nu\mu}\delta^{-1}).
\label{rhoeinglass2}
\end{equation}
Recall that here we must have $\delta \not=0$ so that the limit where $L$ exceeds $\xi$ exists. A physical argument for the new scaling variable 
is that $\delta$ in this limit should not be compared with $L^{-1/\nu}$ as in equilibrium finite size scaling (with the single scaling argument
$\delta L^{1/\nu}$). Instead, $L$ must be replaced by the finite velocity-dependent KZ length scale $\xi_v$, which is smaller than $L$. Thus,
$L^{1/\nu}$ is replaced by $\xi_v^{1/\nu}=v^{1/\mu\nu}$ and the proper scaling argument is $\delta v^{-1/\mu\nu}=z^{-1}$, or, alternatively, 
the inverse thereof as in (\ref{rhoeinglass2}), where the low-velocity limit corresponds to $e(z \to 0)$.

The way in which we have defined the excess Ising energy $\rho_E$, it vanishes when $v \to 0$ for any $\delta$. However, pure critical KZ behavior,
$\rho_E \propto v^{\kappa_c}$, should only apply very close to $\delta=0$ for some range of velocities before a cross-over into a form governed by
the dynamics of the glassy state when $v \to 0$. For a process stopping significantly inside the glassy state, there may be no range of velocities
for which the exponent $\kappa_c$ strictly applies. However, if $v$ is not too low, i.e., the argument $x=v^{1/\nu\mu}\delta^{-1}$ is relatively large
in $e(x)$ in Eq.~(\ref{rhoeinglass2}), it is plausible that the scaling function takes a power-law form, $e(x) \sim x^k$ with some unknown exponent
$k$ (and also note that the scaling function cannot be analytic in the glass phase, because $\kappa_c$ does not asymptotically describe the dynamics).
Since $\delta$ is just a fixed number, this behavior would imply a change in the exponent on $v$ in Eq.~(\ref{rhoeinglass2}), $\rho_E \sim v^{\kappa_f}$
with $\kappa_f = \kappa_c+k/\mu\nu$. Given that the dynamics of the quantum spin glass should be slower than the KZ dynamics, it is natural to expect
$\kappa_f < \kappa_c$, as we indeed also found in our QA experiments (Fig.~4b).

The above analysis presumes that $\delta$ drives a quantum phase transition at which the Hamiltonian is the sole provider of dynamics.
At a classical phase transition of a model with no intrinsic dynamics, $d+z$ in the expression for $\kappa_c$, Eq.~(\ref{kappaform}), should be
just $d$, and the dynamic exponent in the KZ exponent $\mu=z+1/\nu$ should be that of the stochastic process used to evolve the system with SA,
\begin{equation}
\kappa_{\text{SA}} = \frac{d-1/\nu}{z_{\text{SA}}+1/\nu},
\label{kappaform2}
\end{equation}
where $z_{\text{SA}}$ also depends on the system studied (i.e., it is a universal exponent given a critical point and a type of stochastic process).

The case of SQA is a mix of QA and SA in the sense that it implements the imaginary-time dimension through a path integral, hence $d+z$, with
the same $z$ as in QA, appears in the numerator of $\kappa_{\text{SQA}}$. However, the $d+1$ dimensional effective statistical-mechanics system
is simulated by an MC procedure as in SA, and it is the exponent corresponding to that process that allows the simulation to relax toward
its equilibrium. Thus, in this case, the exponent contains two dynamic exponents
\begin{equation}
\kappa_{SQA} = \frac{d+z-1/\nu}{z_{\text{SQA}}+1/\nu},
\label{kappaform3}
\end{equation}
where $z_{\text{SQA}}$ is not {\it a priori} known, and, like the classical exponent $z_{\text{SA}}$, depends on both the system studied and the MC procedures
applied. To our knowledge, this exponent had not been computed previously for the 3D Ising spin glass in a transeverse field. In our SQA work, we extracted
$\mu_{\rm SQA}=z_{\text{SQA}}+1/\nu$ from the data collapse of the Binder cumulant (Fig.~\ref{fig:ed_collapse_full}).

\section{3D spin glasses}\label{sec:3d}

\subsection{Construction}

The instances we study for size $L$ are on $L\times L\times \min(L,12) \times 2$ spins, indexed by $(x,y,z,w)$ with $x,y\in\{0,\ldots,L-1\}$, $z\in\{0,\ldots,\min(L-1,11)\}$, and $w\in{0,1}$.  For a given 3D coordinate $x,y,z$, sites $(x,y,z,0)$ and $(x,y,z,1)$ are coupled with a strong ferromagnetic coupling, $J=-J_{\text{FM}}$; in this work $J_{\text{FM}}=2$.  For $L>9$, instances contain site vacancies due to inoperable qubits.

In the $x$ dimension, couplings use only the $w=0$ sites; the coupling between $(x,y,z,0)$ and $(x+1,y,z,0)$ for $x<L$ is $\pm J_{G}$: $J_G$ with probability $p$, and $-J_G$ otherwise.  Random choices of spin-glass couplings are independent.  Except where stated, $p=0.5$. The $x$ dimension has open boundaries.

In the $y$ dimension, couplings use only the $w=1$ sites; the coupling between $(x,y,z,1)$ and $(x,y+1,z,1)$ for $y<L$ is $\pm  J_{G}$.  The $y$ dimension has open boundaries.

The $z$ dimension is periodic.  For $z' \equiv z+1$ (mod $L$), the coupling between $(x,y,z,1)$ and $(x,y,z',0)$ is equal to the coupling between $(x,y,z,0)$ and $(x,y,z',1)$; each is $\pm J_{G}/2$, meaning that their total is $\pm J_{G}$.  Said differently, the four spins $(x,y,z,0)$, $(x,y,z,1)$, $(x,y,z',0)$, and $(x,y,z',1)$ induce an unfrustrated loop.

In the limit $J_{\text{FM}}/J_{G}\rightarrow \infty$, FM-coupled dimers can be identified into single ``logical'' spins, so the model is equivalent to a bimodal spin glass (random bond model) on a standard 3D cubic lattice with open $x$- and $y$-boundaries and periodic $z$-boundaries.

The $x$- and $y$- dimensions are isotropic with respect to one another; the $z$-dimension is anisotropic in the following three ways.  First, the pair-to-pair couplings are split between two couplers.  Second, the dimension is periodic, while the $x$- and $y$-dimensions are open.  Third, for $L>12$, the $z$-dimension has length bounded by $12$.  This is because the ``Pegasus'' \cite{Boothby2020} layout of qubits is roughly described as a $16\times 16$ grid of $24$-qubit unit cells, with some loss around the boundaries; the $24$ qubits are used for 12 2-qubit dimers, forming the $z$ and $w$ dimensions.  For this reason when studying finite-size scaling, we restrict our attention to $L\leq 12$.  When studying energy scaling, we use the largest size: $15\times 15\times 12\times 2$, with inoperable qubits in the particular processor used leaving only $5374$ of $5400$ qubits in the inputs.

\subsection{Broken FM dimers in the 3D ground state}

There are differences between the 3D spin glasses studied in this work and the corresponding glasses on the simple cubic lattice that one obtains by contracting each two-qubit dimer into a single spin.  In Section \ref{sec:qmc} we will discuss the effect on critical exponents, which explains some of the deviation seen between QA values of the KZ exponent $\mu=z+1/\nu$ from previously reported values for 3D quantum Ising spin glasses \cite{Guo1994}.  Here we consider the separate question of how the classical ground states relate to one another between the ``embedded'' lattice that we study, versus the ``logical'' counterpart in a simple cubic lattice.  

Let us consider a single spin-glass realization, which has an embedded version and a logical version.  When $J_{\text{FM}}> 3J_{G}$, it is easy to show that no ground state in the embedded problem can have a broken dimer, since flipping a spin to repair a broken dimer (one spin up, one spin down) in such a state would reduce the energy.  Likewise, when $J_{\text{FM}}\geq  3J_{G}$ it is easy to see that the problem must have at least one ground state with no broken dimers.  However, when $J_{\text{FM}}< 3J_{G}$ it is possible that all ground states could have at least one broken dimer.

As an example, we begin with a completely ferromagnetic instance, where all couplings have negative sign.  Now, for some fixed $x_0$, $y_0$, and $w_0$, the spins with coordinates $(x_0,y_0,z,w_0)$ form an independent set.  We flip the sign of all glass couplings incident to this set.  It is a simple exercise to confirm that this instance is valid in the sense that it has a well-defined logical counterpart.  Furthermore the ground state is twofold degenerate, with broken dimers (with $x=x_0$ and $y=y_0$) in each ground state.  We confirmed this with a computer search for the small case $L=4$.

Therefore when $J_G > J_{\text{FM}}/3$ we cannot guarantee the existence of a ground state of the embedded lattice that maps to a ground state of the logical lattice.  However, this construction is pathological, and appears to rely on a specific configuration spanning the periodic $z$-dimension.  Thus it seems to be highly unlikely in random inputs, and we have not observed such an input in this study.

\subsection{Estimating ground state energies}\label{sec:gse}

\begin{figure}
  \includegraphics[scale=0.8]{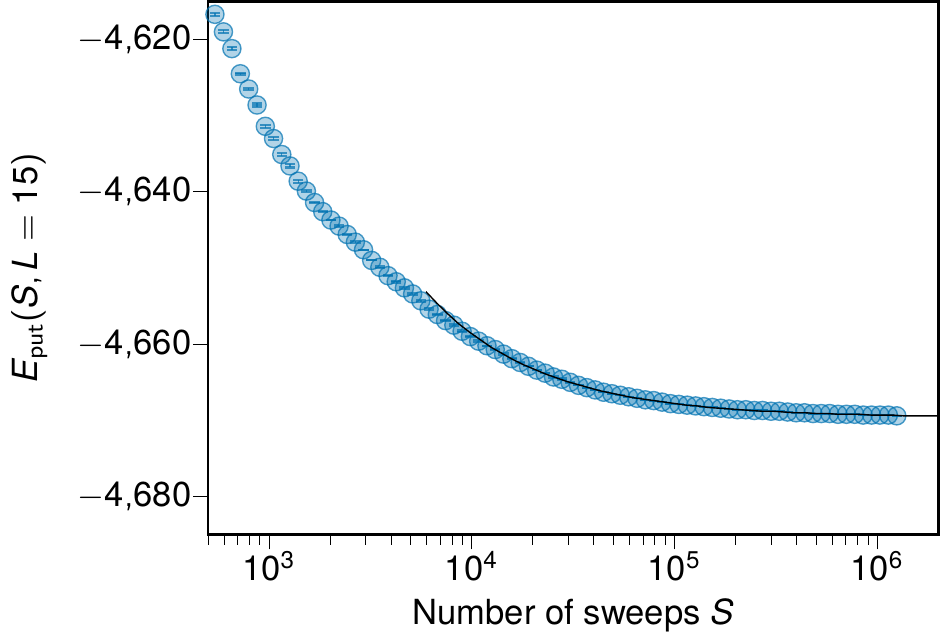}
  \caption{{\bf Convergence of mean ground-state energy under the PT-ICM heuristic.}  Shown are average energies over 300 spin-glass realizations with $L=15$ for a parallel tempering method used to estimate ground-state energies, running for up to $S \geq 10^6$ sweeps.  The asymptote (power-law fit to last 50 data points, shown in black, with limit $-4669.6$) appears to be well established with bias and uncertainty negligible in comparison to the experimentally relevant residual energies.}\label{fig:pticmgs}
\end{figure}

To understand mean residual energy in the context of approximate optimization, we must estimate ground state energy---this task is NP-hard for cubic lattices and practically challenging in the typical case for the largest system sizes we consider.  Bearing in mind the preceding discussion, we simplify the task by searching for ground states in the logical model, i.e., spin glasses on a simple cubic lattice.

The spin glasses studied here are too large to solve with exhaustive approaches such as branch-and-cut \cite{Charfreitag2022}.  We instead use a parallel tempering algorithm with isoenergetic cluster moves (PT-ICM)~\cite{PhysRevLett.115.077201} to find putative ground states for each instance, and show by a self-consistent method that any bias introduced by failure to uncover the exact ground state in some subset of instances is negligible with respect to the average residual energies studied in this work.

We employ an adaptive form of the algorithm. The temperature range is initialized so that the lowest and highest chains sample at $T=\infty$ (spin flip proposals accepted with $50\%$ probability) and $T=0$ (no upward energy proposals accepted). Intermediate temperatures are then inserted iteratively, so as to approach replica exchange rates limited to the range $[0.2, 0.5]$.  The temperature set stabilizes after several thousand sweeps of all replicas. One sweep entails updating all spins in all replicas in some fixed order. Isoenergetic cluster moves are performed for all temperatures every 10 sweeps.  After $S$ sweeps of the algorithm we can estimate the ground state as the best state observed across the entire algorithm run.

This random estimator produces an $S$-dependent upper bound on the ground state $E_{put}(S)$ in the lowest temperature chain that tightens with $S$. For our analysis we require a model average per system size, $E_{put}(L,S) = [ E_{put}(S) ]_L$ where square brackets denote our expectation with respect to the set of instances used experimentally. This estimator is subject to a variance and bias, which increase with system size.  We can determine confidence intervals by considering many independent instantiations of the algorithm at each $S$. The bias is expected to decrease with $S$ as shown in Fig.~\ref{fig:pticmgs}; $E_{put}(L,S)$ appears to converge and we can assume the limit $E_{put}(L)= \lim_{S\rightarrow \infty} E_{put}(L,S)$ with reasonable confidence.  The black line in Fig.~\ref{fig:pticmgs} indicates a power-law fit to the last 50 data points, which has a limit $E_{put}(15) = -4669.6$, which would imply that our estimate of the mean ground state energy is correct to within $0.00005$ relative error, or $0.00004$ energy per site---negligible in comparison to experimental values.

We needn't (and cannot) claim to have found the ground state for every instance, at every size by this method for any finite $S_{max}$.  We only need convincing evidence that the error in the estimate of ground state energy is negligible compared to the residual energies that we use to compare SA, QA, and SQA.

We can assume $E_{put}(L,S)$ converges smoothly in $S$ to obtain a sufficient estimator for our mean residual energy analysis.  Assuming a power-law form of residual energy gives one estimate of $E_{put}(L,S)$, which describes the data closely (see Fig.~\ref{fig:pticmgs}).  In practice, we find it sufficient to approximate
\begin{equation}
  E_{put}(L) \approx \min_{\textrm{independent runs}} E_{put}(L,S_{max}),
\end{equation}the best energy observed amongst 5 independent replicas at some large (but experimentally reasonable) number of sweeps. The convergence $E_{put}(L)$ is shown in Figure \ref{fig:pticmgs} for the most challenging problem scale ($L=15$), and it is clear the error in the ground state energy across $300$ instances is a negligible factor in our analysis.

We emphasize that an error in ground state energy affects the computed residual energy for all dynamics equally.

\subsection{Quantum and classical phase transitions}

Here we briefly describe the phase diagram for the 3D spin glass shown in Fig.~\ref{fig:1}e.  We first focus our attention on the $\pm J$ spin glass on a simple cubic lattice.  The classical case has been studied in detail using the Janus special-purpose computer \cite{Baity-Jesi2013}.  Although it was not initially clear \cite{Binder1986,Kawashima1996}, it has been established that there is a finite-temperature transition at $T_c^{\rm{3D}}\approx 1.1019$ \cite{Baity-Jesi2013}.  The critical correlation exponent and anomalous dimension have been estimated, respectively, as $\nu=2.562$, $\eta=-0.39$ \cite{Baity-Jesi2013}.  In the main text we referenced an estimate of the dynamic exponent $z=5.93$ for the $\pm J$ case \cite{Liu2015}, but in that paper\cite{Liu2015} (Table I) it is pointed out that a range of estimates has been reported over the last 20 years.

Our picture of the quantum phase transition is mostly informed by a study of Guo, Bhatt and Huse from 1994 \cite{Guo1994}, which gives estimates of $z \approx 1.3$, $\nu \approx 1/1.3$.  Making use of relatively recent algorithmic advances---primarily, continuous-time PIMC---we perform similar MC simulations on both the simple cubic lattice and the embedded 3D lattices studied in this work.  We discuss these simulations in the next section.

\section{Equilibrium analysis of critical phenomena by Monte Carlo methods}\label{sec:qmc}

As highlighted in the main text, the universal and non-universal critical behavior of 2D and 3D transverse field Ising spin glass models has been studied in previous works~\cite{Baity-Jesi2013,Guo1994}.  In this section we use path-integral Monte Carlo---in the limit of continuous imaginary time---to establish the equilibrium critical behavior of 3D spin glasses both on simple cubic lattices and on the embedded models studied in the main text, as well as 2D square ferromagnets.  We generate data for ferromagnetic and spin-glass models at comparable scales to those employed in the QA studies.  Our aims in this section are: to verify previously reported critical exponents with an independent numerical analysis; to test the viability of finite-size scaling analysis at the experimented scales; and to verify universality of the embedded 3D spin glasses via the universal correlation exponent $\nu$, regardless of microscopic model details.  We use a methodology inspired by previous studies of universal behavior, but adapted to allow for a continuous imaginary-time integral limit of the Trotterized model~\cite{Guo1994}.

\subsection{Finite-size scaling in classical and path-integral models}

As discussed in Section \ref{sec:fss}, the properties of a classical or quantum spin-glass phase transition may be determined by a finite-size scaling analysis of $\langle q^2\rangle$~\cite{newman1999monte,Rieger1994,Guo1994}.

In the classical case we have a model over $N$-spin states $\vec s$:
\begin{equation}
  P(\vec s) = \exp\left(-\beta \sum_{i<j} J_{ij}s_{i}s_{j} \right),\label{eq:Prob_s}
\end{equation}
with
\begin{equation}
\langle q^2\rangle =\sum_{\vec s^{(1)}} \sum_{\vec s^{(2)}} P(\vec s^{(1)})P(\vec s^{(2)}) \left( \frac{1}{N}\sum_i s^{(1)}_{i}s^{(2)}_{i}\right)^2.
\end{equation}
The anticipated scaling form is
\begin{equation}
\langle q^2 \rangle = L^{b} B(\delta L^{1/\nu}) \label{classical_scaling}
\end{equation}
with $\delta = (T-T_c)/T_c$.

Critical behavior in the quantum model may be established by analysis of the density matrix $\exp(-\beta \mathcal H(s))$, where $\mathcal H(s)$ is the Hamiltonian of our experimental Ising system (\ref{eq:ham}) and $\beta$ is the inverse temperature. By a process of Trotterization this density matrix can be transformed into a classical model, sufficient to establish the distribution of projected states and other statistics. The Trotterized model is defined by a modified Hamiltonian over a space of worldlines, in which each qubit $i$ is replaced by $L_z$ time-indexed classical spin-states $s_{i,t}$. For each time slice, variables interact in accordance with the problem Hamiltonian, scaled to $L_z$. Spins are coupled ferromagnetically in imaginary time, subject to periodic boundary conditions $s_{i,L_z}=s_{i,0}$. The probability distribution is given by
\begin{equation}
  P(\vec s,L_z) \propto \exp\left( \left(-\frac{1}{2}\log\tanh\left(\frac{\beta \Gamma(s)}{L_z}\right)\right)\sum_{t=0}^{L_z-1} \sum_{i} s_{i,t}s_{i,t+1} - \frac{\beta J(s)}{L_z}\sum_{i<j} J_{ij}s_{i,t}s_{j,t} \right). \label{eq:Prob_w}
\end{equation}
An order parameter $q$ can be defined as the overlap of two states projected into the computational basis, the projected state distribution is determined by the statistics of a single time slice (say t=0) in our model~\cite{Rieger1994,Guo1994}. The spin-glass susceptibility is defined as the expected square value of this quantity, averaged over spin-glass realizations, i.e., $\langle q^2 \rangle$:
\begin{equation}
  \langle q^2\rangle = \sum_{\vec s^{(1)}} \sum_{\vec s^{(2)}} P(\vec s^{(1)},L_z)P(\vec s^{(2)},L_z) \left(\frac{1}{N}\sum_i s^{(1)}_{i,0}s^{(2)}_{j,0}\right)^2.
\end{equation}
Statistics of the quantum model are correctly established by taking the continuous limit in imaginary time\footnote{Numerically we work at $L_z=2^{16}$ with varying $\beta$.  Because $\beta \ll L_z$ in all experiments conducted, this captures the continuous-time limit.}, $L_z\rightarrow \infty$.  
In this limit the scaling of susceptibility is expected to take the form
\begin{equation}
\langle q^2 \rangle = L^{b_q} B(\delta L^{1/\nu},\beta/L^z)
\label{eq:chiSG},
\end{equation}
for a two-parameter collapse function $B$.  Our analysis of the phase transition involves variation of $\Gamma = \Gamma(s)/J(s)$ at fixed $J(s)$, so that the reduced transverse field is defined as $\delta = (\Gamma - \Gamma_c)/\Gamma_c$ for a critical field $\Gamma_c$. 

\subsection{Monte Carlo sampling}

Monte Carlo methods can be used to sample both the classical model (\ref{eq:Prob_s}) and the Trotterized model (\ref{eq:Prob_w}). For the latter, we take the numerically convenient value of $L_z=2^{16}$, which is indistinguishable from the continuous-time limit ($L_z\rightarrow \infty$) for the relevant parameter ranges. Since in this limit the coupling strength in the imaginary time direction is strong, it is necessary to employ Swendsen-Wang cluster updates~\cite{Rieger1999}. Our algorithm thus proceeds by iterating over sites $i$, and applying a cluster update (single-site, many-times) to $s_{i,\cdot}$ for each $T$ or $\Gamma$. For our study of embedded problems where disjoint pairs of qubits (dimers) are strongly ferromagnetically coupled (with relative strength $J_{\text{FM}}/J_G$---see Fig.~\ref{fig:1}) we can apply a multi-site multi-time update per set; clustering occurs both in space and time with respect to each dimer. The cluster updates are described, and code published, in previous studies~\cite{King2018,King2021a}. Since we are interested only in equilibrium behavior it is sufficient to consider a fixed sequence of updates covering all sites; one such sequence is called a sweep.

For the classical systems, we employ a standard parallel tempering (exchange Monte Carlo) method~\cite{Baity-Jesi2013}. 
In the quantum model we must sample across a larger parameter space, and with a slower worldline sampling update. For efficiency of data collection we replace the parallel tempering routine with an annealing procedure. For a fixed model (disorder realization, at some size $L$) and fixed $\Gamma$ we vary inverse temperature at each sweep according to a geometric schedule from a low value (where worldlines are fast-mixing) through the critical region to a low-enough temperature for purposes of establishing critical scaling. In this way we collect single-sample (equilibrated, but correlated) data at a wide range of temperatures. Statistics are averaged based upon $400$ independent anneals drawing one realization of the bond disorder per anneal. Uncertainty is estimated with bootstraps with respect to the disorder realizations. In both the classical and quantum cases, we can maintain two independent Monte Carlo processes (replica) in parallel, from which the overlaps can be calculated.

In the parallel tempering routine temperatures are spaced inclusive of high temperature fast mixing models and critical region models to ensure sufficient replica exchange rates. Statistics are collected sequentially with exponentially increasing burn-in and sampling times until convergence through the critical region is apparent. In the annealing procedure fair sampling through the critical region requires a sweep rate (sweeps per unit change in $\log(\beta)$) decreasing with system size. In our methodology $\langle q^2\rangle$ converges from below, and this process becomes slow for large $L$, small $\Gamma$ and low temperature. We truncated our data collection at small size in part for this reason, and ensured that statistics were for practical purposes insensitive in the data presented to variation by a factor of $4$ in the sweep rate (estimators agreeing within confidence intervals at each system size and parameter set presented). The rate of progress, $d\log(\beta)$ per sweep, was chosen between between $2 \times 10^{-6}$ and $7\times 10^{-4}$, for the data presented. 

\subsection{Classical critical scaling}

\begin{figure}
\includegraphics[scale=1]{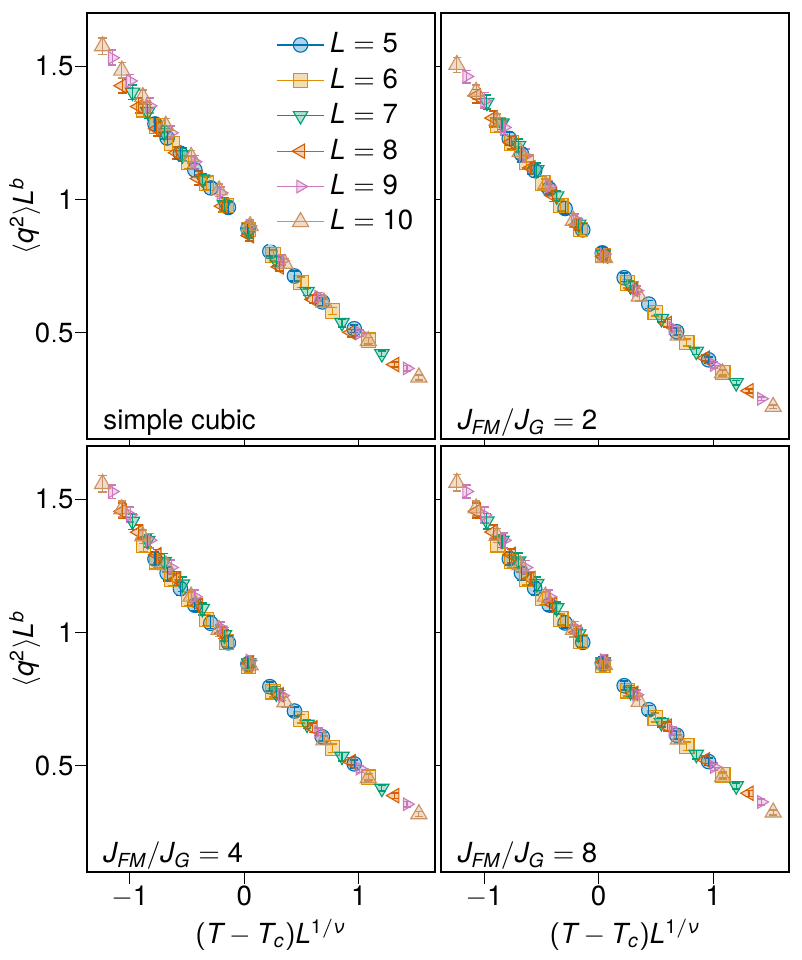}
\caption{{\bf Collapse of $\langle q^2\rangle$ in the critical region for classical spin glasses.}  Finite-size collapses are shown for classical 3D spin glasses on the simple cubic lattice, along with the dimer-embedded lattices with varying $J_{\text{FM}}/J_G$, using identical parameters $T_c=1.1$, $\nu=1.5$, $b=0.61$.  The value of $\nu$ is far from recent estimates but provides a good collapse for these small systems, regardless of embedding, suggesting that dimer-embedding does not cause a large change to $T_c$.  The largest deviations from the simple cubic model are seen for small $J_{\text{FM}}/J_G$, as one would expect.}\label{fig:sm_classical_collapse}
\end{figure}

Before showing collapse results in the quantum model, we digress to consider the classical model. In this case we can collect data by a similar methodology but set $\Gamma=0$ and employ in independent code a Metropolis method supplemented with cluster flip moves for the dimers.  Estimates of $T_c$, $\nu$, and $\eta$ satisfying the classical scaling form (\ref{classical_scaling}) have evolved over the years \cite{Kawashima1996, Palassini1999, Hasenbusch2008, Baity-Jesi2013}, with the most recent of these papers \cite{Baity-Jesi2013} estimating $T_c\approx 1.1$, $\nu\approx 2.56$, and $\eta\approx -0.39$.  Earlier estimates of $\nu$ \cite{Kawashima1996,Palassini1999} were affected by unaccounted corrections to finite-size scaling, leading to significantly lower estimates of $\nu$.  In Fig.~\ref{fig:sm_classical_collapse} we show collapses of $\langle q^2 \rangle L^{b}$ as functions of $tL^{1/\nu}$ for $T_c=1.1$, $b=0.61$, and $\nu=1.5$, for the logical (simple cubic) model and the embedded model with $J_{\text{FM}}=2$, $4$, and $8$.  The value of $\nu$ is far from recent estimates, but it provides a good collapse for the modest system sizes shown.  For larger systems, the appropriate collapse is achieved for $\nu \approx 2.5$.  Here, the collapses simply illustrate the consistency of critical scaling with respect to the dimer embedding and $J_{\text{FM}}$.  The evidence is consistent with little or no change in $T_c$, $\eta$, and $\nu$ between the logical and embedded classical 3D spin glass models.

\subsection{Quantum critical scaling}

We can now move on to collapses for the quantum model. An additional difficulty in performing a collapse of data for the quantum model arises from the need to collapse a two-dimensional form (\ref{eq:chiSG}). In order to reduce this to a one-dimensional problem we adapt the method of Guo, Bhatt and Huse~\cite{Guo1994} to the continuous-time setting. The effect of increasing inverse temperature in our model (\ref{eq:Prob_w}) is a nonlinear decrease in coupling strength in imaginary time, with a linear increase in the coupling strength between sites. Therefore, at zero temperature the correlation length $\xi_z$ is much smaller than $L_z$ in imaginary time, whereas spatial correlation length $\xi$ approaches zero at high temperature.
Between these two extremes there exists---near the critical point---a value for which $\xi_z\approx L_z$ and $\xi\approx L$ can be determined.  This defines the point of maximum susceptibility which we presume (and show) to be approached smoothly from high temperature; consequently $\beta \sim L^z$ for this point.

\begin{figure}
\includegraphics[scale=1]{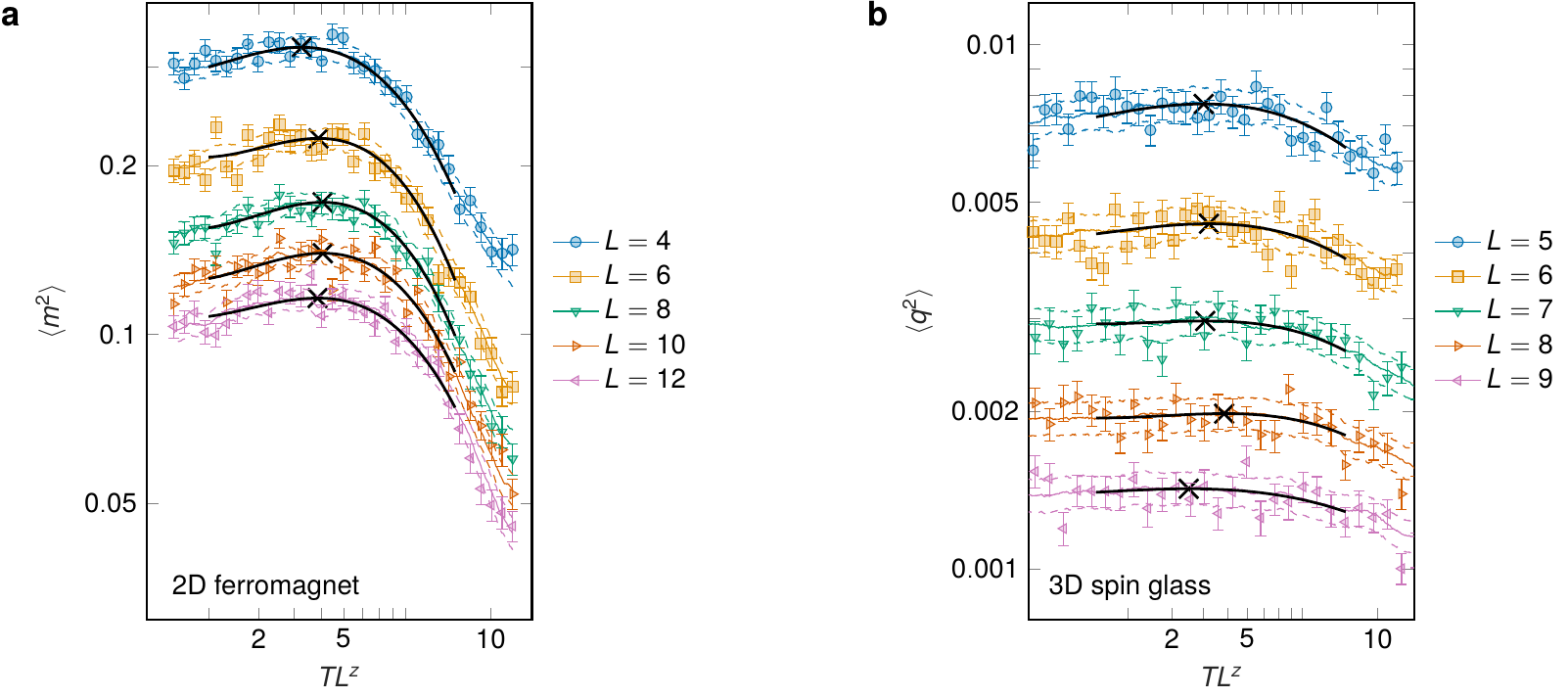}    
\caption{{\bf Susceptibility as a function of temperature for quantum models.} {\bf a}, 2D ferromagnet ($\Gamma=3.06$) and {\bf b}, embedded 3D spin glass with $J_{\text{FM}}/J_G = 2$ ($\Gamma=2.9$). In the case of the ferromagnet, we use the standard linear susceptibility $\langle m^2\rangle$ in place of $\langle q^2\rangle$. Near the critical $\Gamma$, susceptiblity is small at high temperature and reaches a peak value as temperature is decreased.  Each model and size has data for 1024 temperatures.  Solid colored lines indicate a 64-temperature moving average, dashed lines indicate moving averages of bootstrap 95\% confidence intervals, and data with error bars indicate a subset of individual temperatures and bootstrap confidence intervals, selected for visual clarity.  Black lines indicate polynomial fits to log-log data, with maxima marked with $\times$.  Bootstrapping is done over 400 independently annealed realizations (all realizations being identical for the ferromagnet).  Rescaling the temperature by $L^z$ with $z=1$ (left) and $z=1.3$ (right) leads to reasonably well horizontally-aligned maxima, consistent with previous studies.}\label{fig:sm_susc_peak}
\end{figure}

\begin{figure}\includegraphics[scale=0.8]{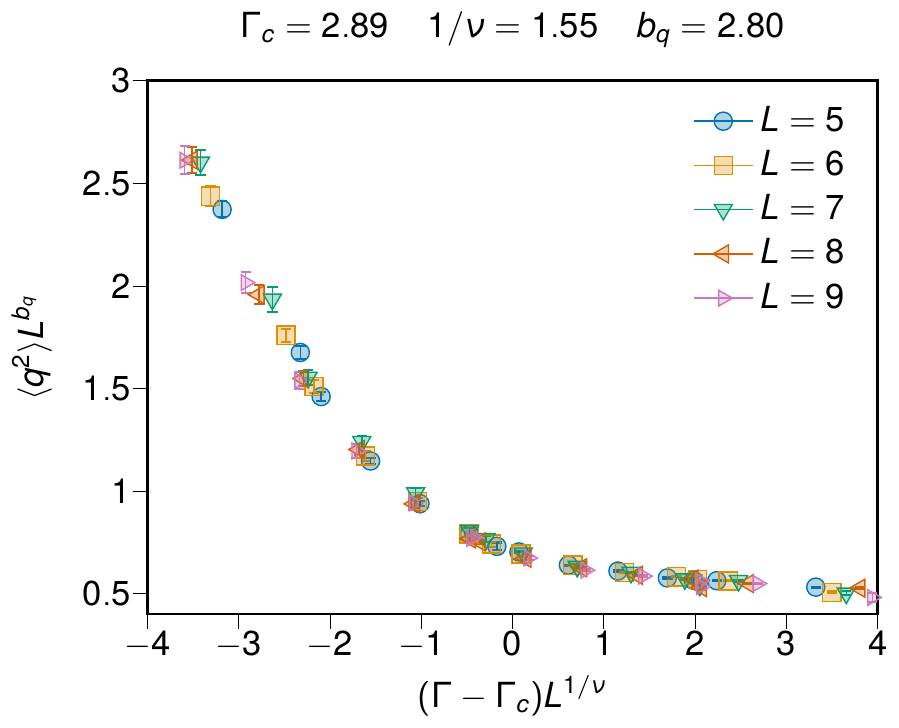}
  \caption{{\bf Collapse of peak susceptibility for embedded 3D quantum spin glass.} Scaled susceptibility $\langle q^2\rangle L^{b_q}$ collapses as a function of reduced transverse field $(\Gamma-\Gamma_c)L^{1/\nu}$ in the vicinity of the QPT, for the embedded 3D system with $J_{\text{FM}}/J_G=2$.  From fitting the collapse we can determine $\nu$, $\Gamma_c$, and $b_q$.  Error bars indicate $S+1$ goodness-of-fit intervals (see text).}\label{fig:collapse_nu}
\end{figure}

\begin{table}
  \begin{tabular}{|l|l|l|l|l|l|}
    \hline
   Model & $\Gamma_c$ & $1/\nu$ & $b_q$ & Goodness-of-fit $S$\\
    \hline
  2D Spin Glass & 2.11(1) & 1.02(16) & 1.76(3)&2.05\\
  3D Spin Glass & 3.02(2) & 1.52(15) & 2.79(6)&1.01\\
  3D Spin Glass ($J_{\text{FM}}/J_G=8$) & 5.07(2) & 1.56(12) & 2.77(3)&0.84\\
  3D Spin Glass ($J_{\text{FM}}/J_G=4$) & 3.75(1) & 1.55(6) & 2.79(3)&0.87\\
  3D Spin Glass ($J_{\text{FM}}/J_G=2$) & 2.89(1) & 1.55(11) & 2.80(3)&0.99\\
\hline
    \end{tabular}
  \caption{Optimized collapse parameters for quantum spin glasses, with restriction of data to a rescaled interval $|\Gamma-\Gamma_c|/\Gamma_c < 1$. Collapse values are determined by a $\chi^2$ method~\cite{Melchert2009}.  Parameters $b$ and $\nu$ deviate slightly downward from expectations based on the literature \cite{Guo1994,Rieger1994}, both for embedded and unembedded models.  Aside from a large shift in the non-universal critical point, foreseeable based on modification of local tunneling rates, fit quality is similar across all 3D models.}\label{tab:fits}
\end{table}

We verify this intuition in a ferromagnetic model, then use it in a spin-glass model, as shown in Fig.~\ref{fig:sm_susc_peak}.  Since we can determine by this method a $\beta$ sufficient for the correlation length in imaginary time to match $L_z$, we are able to reduce (\ref{eq:chiSG}) to a one-dimensional fit $\langle q^2 \rangle = L^{b_q} \max_{\beta'} B(\gamma L^{1/\nu},\beta')$.  As in the classical case we can extract parameters $\Gamma_c$, $\nu$, and $b_q$ from the fit.  We used an open-source library, autoScale.py~\cite{Melchert2009}, to fit this data, with the $J_{\text{FM}}=2J_G$ example shown in Figure \ref{fig:collapse_nu}.  For a goodness of fit $S$ (with small $S$ indicating a good fit; see \cite{Melchert2009}), error bars indicate values that lead to goodness of fit at most $S+1$.  Outcomes for other models are shown in Table~\ref{tab:fits}.  We note that all models show $1/\nu$ near $1.55$, significantly different from the previously reported $1/\nu \approx 1.3$ \cite{Guo1994}, and we use this new estimate in the main text.

As shown in Figure \ref{fig:sm_susc_peak} the peak susceptibility in $T$ for a given $L$ and $\Gamma$ can be established with reasonable confidence.  We expect $b_q$ to describe the scaling of the peak height in the critical region, i.e. $\langle q^2\rangle_{\text{max}} \sim L^{-b_q}$; indeed we find very good agreement with this power-law form and find very consistent estimates $b_q \approx 2.8$ for both logical and embedded 3D quantum spin glasses (Table~\ref{tab:fits}).  In these simulations we were restricted in the largest value of $L$ by convergence requirements at low $T$ and small $\Gamma$. 

Similarly to the peak height, the peak location (in $T$) is expected to scale as $L^{-z}$, so in principle we could use this information to extract an estimate of the dynamic exponent $z$.  Although the data in \ref{fig:sm_susc_peak}b are consistent with the reported value of $z\approx 1.3$ \cite{Guo1994}, the data are not reliable enough to confidently estimate $z$, owing to the flat and noisy nature of the peak for large systems. We note that in a random ferromagnetic transverse field model study based on a similar methodology, an alternative collapse form in temperature was successful \cite{Pich1998}.  If a similar {\em infinite randomness fixed point} applied in the spin glass model, an exponential collapse form might prove valuable, but we were unable based upon our data to establish $z$ with confidence based on such a hypothesis.  (Our collapses use a peak susceptibility value, and hence are robust with regards the scaling form in temperature, a strength of the method.)  We maintain the hypothesis that $z\approx 1.3$, which despite the challenges of classical simulation, agrees well with our quantum simulation data from the QA processor.  The observed value of $b_q\approx 2.8$ differs from Guo, Bhatt, and Huse's estimate $b_q\approx 3.4$ \cite{Guo1994}.

In general, the FSS analysis appears to provide a good qualitative description of the quantum models at these small scales. As in the classical case, collapses indicate that the scaling of the embedded models is very close to that of the unembedded model. For these unembedded models we find values that are---with the exception of $\eta$---not far from those reported in the literature, but with important deviations in $\nu$, where we use our estimate $1/\nu\approx 1.55$ rather than the previous estimate $1/\nu\approx 1.3$.  Differences in methodology may account for these deviations.

\section{Nonequilibrium dynamics of SA and SQA}\label{sec:mcdynamics}

\subsection{Freezing of dynamics in SA}\label{sec:freezing}

\begin{figure}
\includegraphics{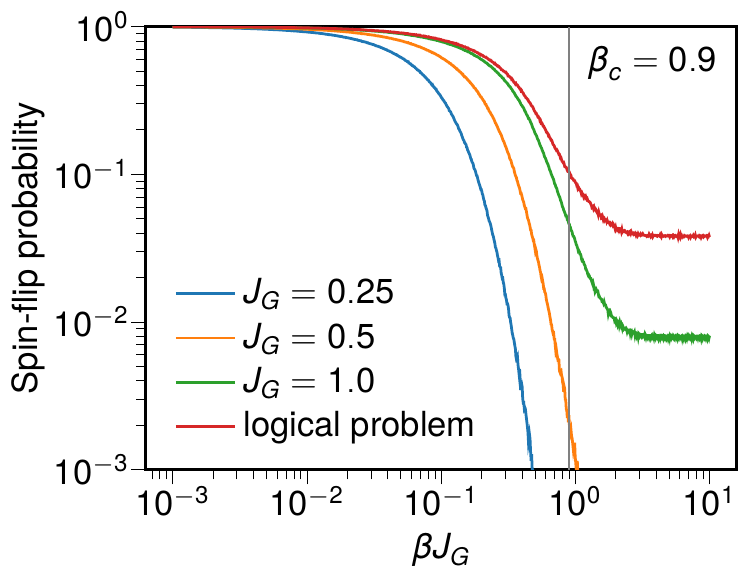}
  \caption{{\bf Freezing in SA.} Spin-flip acceptance probability varies as a function of $J_G$, resulting in a freezing of dynamics at high temperature for small $J_G$.  Three values of $J_G$ are shown, along with data for the logical model on a simple cubic lattice.}\label{fig:sm_freezing}
\end{figure}

The SA dynamics studied here proposes and accepts or rejects one spin update at a time, and proceeds along a geometric schedule in $\beta/J_G$.  When $J_{\text{FM}}\gg J_G$, dynamics ``freeze''---spin flips become very unlikely to be accepted---at a high temperature.  To probe this phenomenon, we annealed $L=9$ embedded 3D spin glasses with $J_G\in\{1/4,1/2,1\}$, as well as the corresponding logical (simple cubic) spin glasses, and tracked the probability of a spin flip being accepted at each sweep.  Data are shown in Fig.~\ref{fig:sm_freezing}.

For sufficiently long anneals and large systems, this effect should not change extracted KZ exponents.  However, there is a significant effect for the system sizes and anneal lengths probed in the main text, as evidenced by Extended Data Figs.~\ref{fig:ed_collapse_full}---\ref{fig:ed_collapse_logical}.

\begin{figure}
\includegraphics{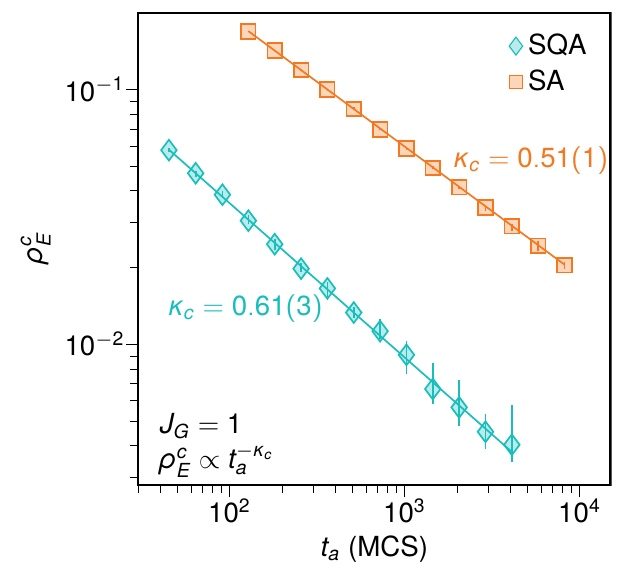}
\caption{{\bf Scaling of residual Ising energy density at the critical point.} Results shown are analogous to Fig.~\ref{fig:4}a, with SQA and SA anneals stopped at their respective critical points.  SQA has $\kappa_c = 0.61(3)$ with equilibrium Ising energy per site $\approx -0.63$, whereas SA has $\kappa_c = 0.51(1)$ with equilibrium Ising energy per site $\approx -1.79$.  Error bars indicate $95\%$ bootstrap confidence intervals.
}\label{fig:sm_kappac}
\end{figure}

\subsection{Energy decay at the critical point and inside the glass phase}\label{sm:kappac}

As discussed in Section \ref{sec:fss}, the form of energy decay at the critical point is subject to corrections within the glass phase, and it is not obvious that Eq.~(\ref{eq:kappa}) should hold even approximately.  Here we investigate this question.

In Fig.~\ref{fig:4}, mean Ising energy at the end of the anneal for QA, SA, and SQA is compared to the putative ground-state energy, giving a final residual Ising energy density $\rho_E^f$.  To study energy decay at the (quantum or classical) critical point, we stop the anneal at the critical point---here this is $T_c = 1.1$ and $\Gamma_c/J = 2.89$ for SA and SQA respectively, with $J_G=1$ (see Section \ref{sec:qmc}).  We take the mean Ising energy for a given anneal time $t_a$, and subtract off the equilibrium Ising energy to obtain a critical Ising residual energy density $\rho_E^c$.

Here, the equilibrium Ising energy is estimated as a fitting parameter to a best-fit power-law form $\rho_E^c \propto t_a^{-\kappa_c}$.  This fit is shown in Fig.~\ref{fig:sm_kappac}.  Based on the correction to Eq.~(\ref{eq:kappa}), we may expect both SA and SQA to show $\kappa_f < \kappa_c$.  Indeed, for SA we see $\kappa_c \approx 0.51$, $\kappa_f \approx 0.42$.  For SQA we see $\kappa_c \approx 0.61$, $\kappa_f \approx 0.51$.  We note that the equilibrium Ising energy per site is estimated at $-0.63$ at the quantum critical point for SQA, much higher than the value $-1.79$ at the thermal critical point for SA.  Eq.~(\ref{eq:kappa}) predicts $\kappa_c$ to be very close to the measured $\kappa_f$ rather than the measured $\kappa_c$; we attribute this mainly to finite size effects. 

Although we are not able to projectively read out QA states at the critical point, this evidence suggests that one might expect QA to show $\kappa_c > \kappa_f$, more in line with Eq.~(\ref{eq:kappa}).

\section{Data collapse domains}\label{sec:windows}

All dynamics in this work are studied on the same geometries with the same boundary conditions (one periodic dimension, $2$ open dimensions).  For this reason collapses are ideally performed over a range of $t_a$ such that the observable, e.g., $U$, follows a consistent power-law scaling.  Thus we discard fast annealing times that deviate from this scaling.  We also discard long annealing times where correlation length approaches system size, since this can lead to anomalous boundary effects.  This window of $t_a$ varies from one dynamics to another.  In particular, we restrict our collapse for QA to the region $t_a\leq \SI{30}{ns}$ to minimize complications arising from decoherence and noise, which causes a smooth increase in the observed KZ exponent $\mu$.

\begin{figure}
\includegraphics{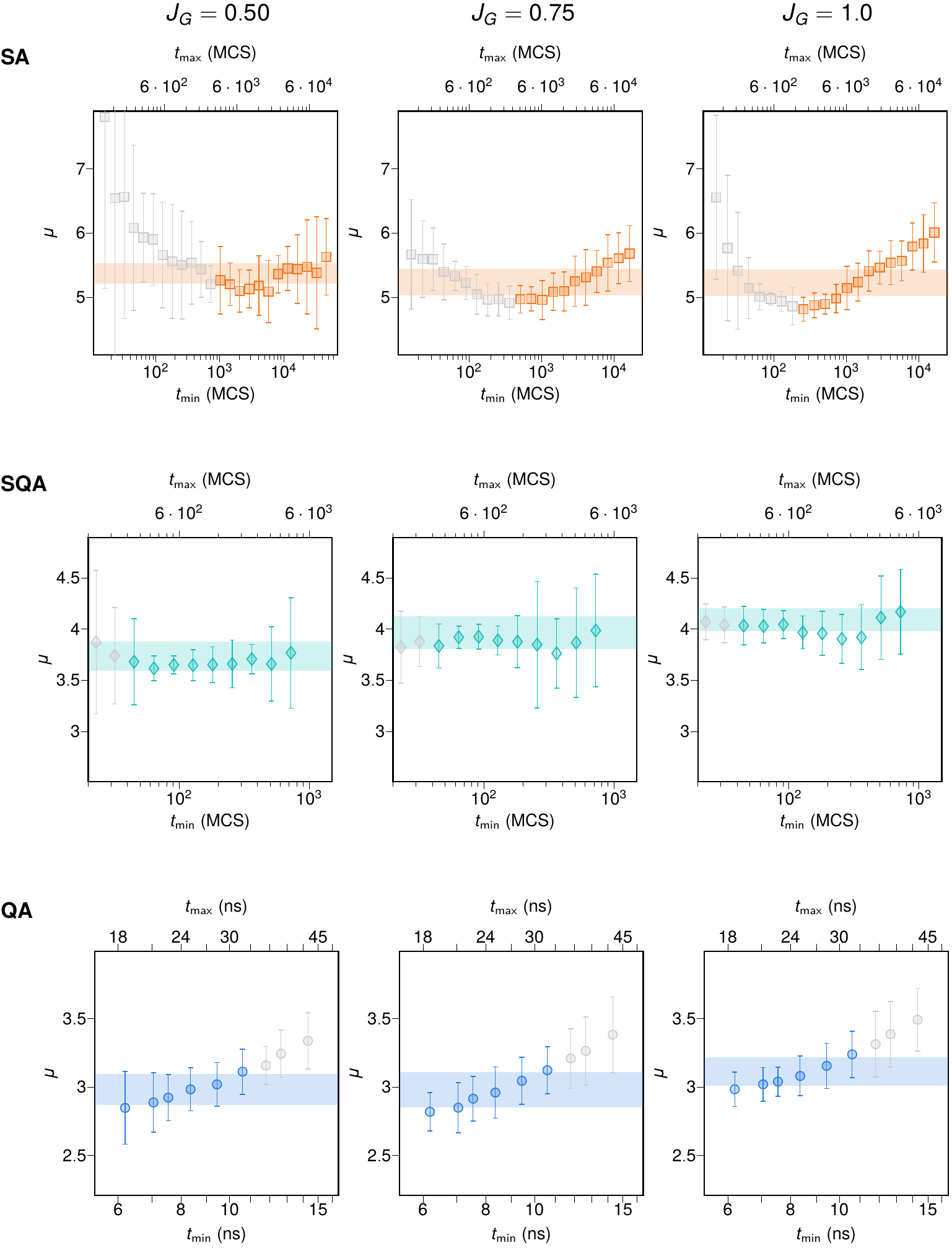}
  \caption{{\bf Data collapse windows.} Each dynamics (SA, SQA, and QA) is run for varying annealing times; cumulants and order parameters are collapsed over a subset of these times.  For each dynamics and each $J_G\in \{0.50,0.75,1.0\}$, the 95\% confidence interval for the extracted exponent $\mu$ (Fig.~\ref{fig:3}e) is shown as a shaded region.  To test self-consistency, we extract $\mu$ for annealing time windows $[t_{\rm{min}},kt_{\rm{min}}]$ where $k=3$ for QA, $k=6$ for SA and SQA (see text).  Windows within the collapse range used to determine $\mu$ are shown in color; windows intersecting unused annealing times are shown in gray.}\label{fig:sm_windows}
\end{figure}

In Fig.~\ref{fig:sm_windows} we plot $\mu$ for QA, SA, and SQA across sliding windows of $t_a$.  These windows are chosen to span a given dynamic range in $t_a$: $t_{\text{min}} \leq t_a\leq  3t_{\text{min}}$ for QA and $t_{\text{min}} \leq t_a\leq  6t_{\text{min}}$ for SA and SQA; using different dynamic ranges is justified due to the polynomial speedup in QA compared to the software solvers.  Only windows containing at least 6 measured annealing times are considered.  As explained in the Methods section, confidence intervals are generated from combined jackknife standard errors across system sizes and annealing times.  The confidence intervals for the individual windows overlap the confidence intervals for the overall region of data collapse; annealing times that are deemed too fast or too slow are shown in gray.

There are several potential causes for the observed increase in $\mu$ for increasing $t_a$ in QA.  Decoherence and noise, as mentioned above, are leading causes.  Deviation between the rf-SQUID flux qubit model and the transverse-field Ising model may also play a role.

\section{Effect of temperature on QA}\label{sec:temperature}

\begin{figure}
\includegraphics[scale=0.8]{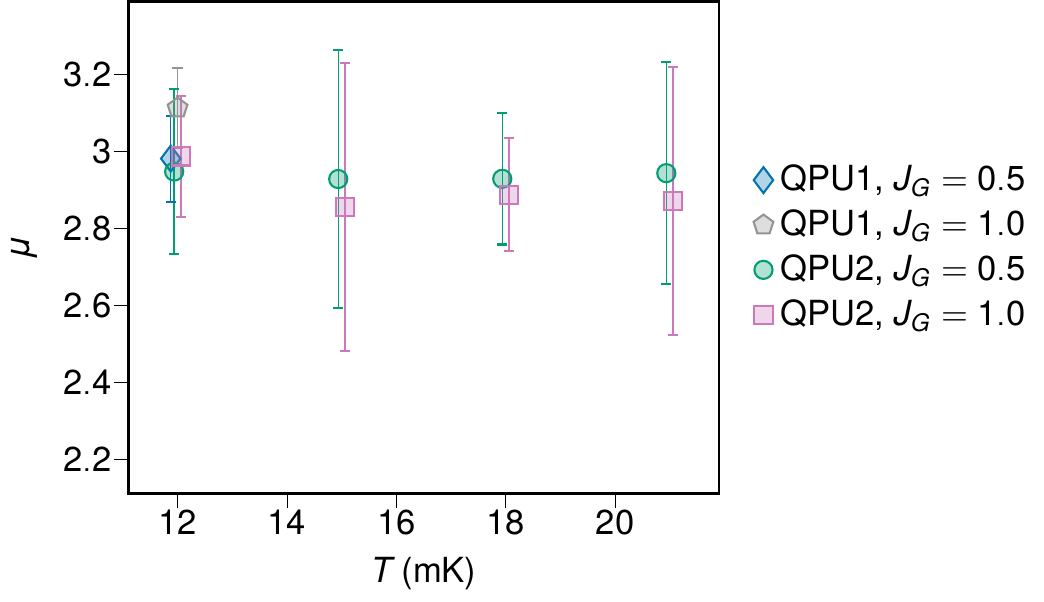}
\caption{{\bf Kibble-Zurek exponents for varying temperature.}  Main-text measurements on QPU1 at $\SI{12}{mK}$ are compared against a second QA processor, QPU2, operating at a range of temperatures.  Error bars indicate jackknife 95\% confidence intervals over system sizes and anneal times.}\label{fig:sm_temperatures}
\end{figure}

Measurements in the main text were collected on a first quantum processing unit (QPU1) at a cryostat set-point of $\SI{12}{mK}$.  To probe the effect of temperature, we performed measurements on a second QPU (QPU2) at temperatures varying from $\SI{12}{mK}$ to $\SI{21}{mK}$.  KZ exponents $\mu$ were extracted using annealing times ranging from $\SI{8.1}{ns}$ to $\SI{30}{ns}$, with results shown in Fig.~\ref{fig:sm_temperatures}.  Measurements are insensitive to temperature over this parametric range, supporting a hypothesis of a coherent regime with negligible interaction with the environment.

\section{Calibration refinement}\label{sec:shim}

Symmetries in the Ising Hamiltonian provide an opportunity to suppress calibration imperfections.  This has been shown to be very effective for geometrically-frustrated low-dimensional systems \cite{King2018,Nishimura2020,Kairys2020,King2021,King2021a}.  In this work we study thousands of spin-glass realizations, and it is impractical to extensively refine the calibration for each one.  Instead, we tune only two aspects of the calibration: First, we balance qubits at degeneracy---with average magnitude zero---using flux offsets.  Second, we synchronize the eight annealing lines that control the annealing schedule of eight sets of qubits, using anneal offsets.  The latter is most relevant for the fastest anneals, since desynchronization between annealing lines is on the order of $\SI{1}{ns}$ or less.  For both of these refinements we use the same approach as was taken in Ref.~\cite{King2022}, without tuning individual couplings; we refer the interested reader to the supplementary material of Ref.~\cite{King2022} for more detail.  For each selection of parameters ($L$, $J_G$, $t_a$, and $p_{\rm{AFM}}$) we perform an independent iterative shim for both flux offsets and anneal offsets; each of these offsets is programmable on a per-qubit basis.

For $i\in\{1,\ldots, 8\}$ let $V_i$ denote the set of qubits on annealing line $i$, and for $i,j\in\{1,\ldots,8\}$ let $E_{ij}$ denote the set of couplings coupling a qubit in $V_i$ to a qubit in $V_j$.  

\begin{figure}
\includegraphics[scale=1]{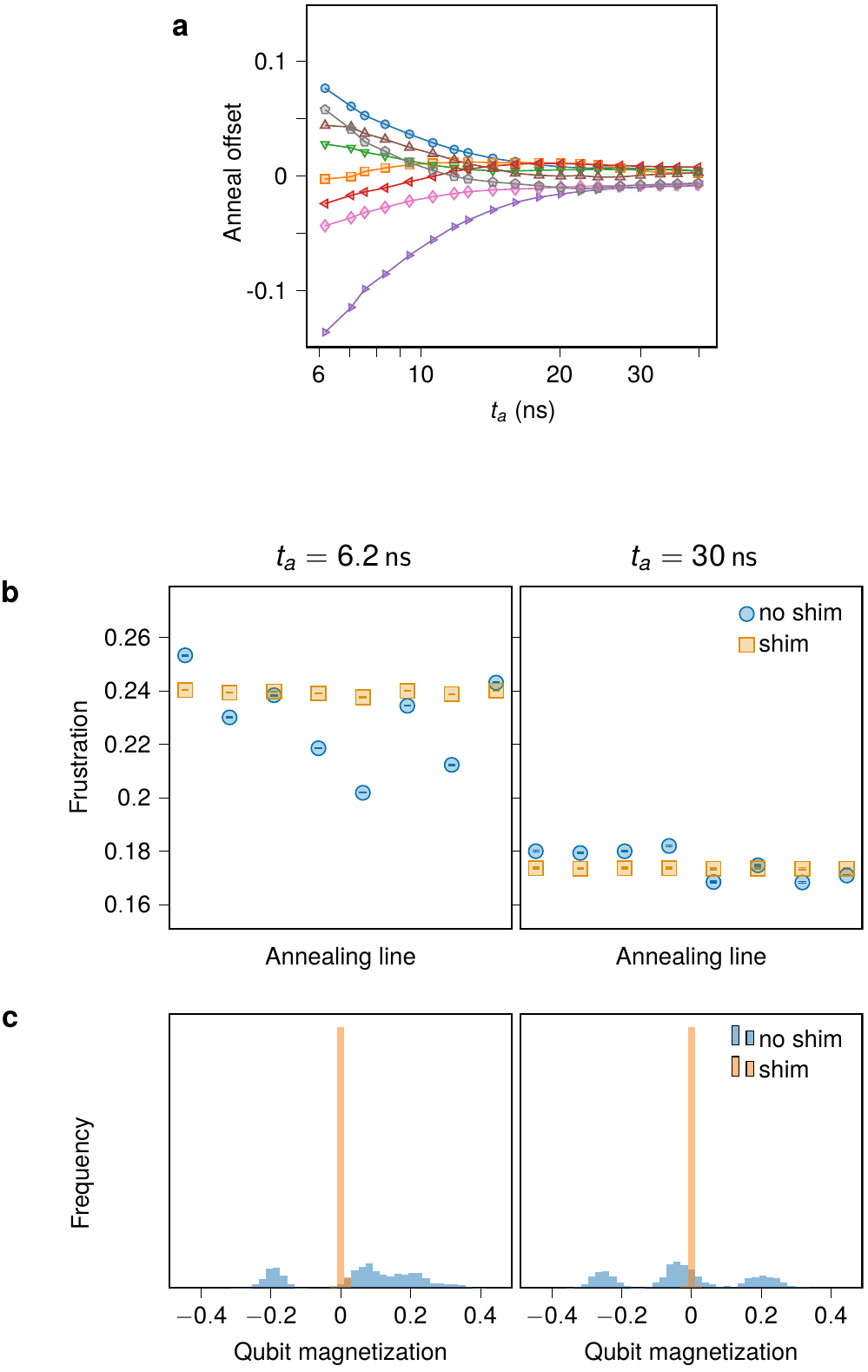}
\caption{{\bf Calibration refinement.}  {\bf a}, Anneal offsets (unitless, in $s$) are tuned to synchronize eight annealing lines for fast anneals.  {\bf b}, Anneal offsets are learned through a loss function related to inhomogeneity of frustration of couplings incident to the qubits on each annealing line.  Flux offsets are learned through a loss function that minimizes nonzero average magnetization of individual qubits.  Tuning these improves the homogeneity of frustration with respect to annealing line.  {\bf c}, Systematic nonzero magnetization for the qubits is also reduced.}\label{fig:sm_shim_meta}
\end{figure}

\begin{figure}
\includegraphics[scale=1]{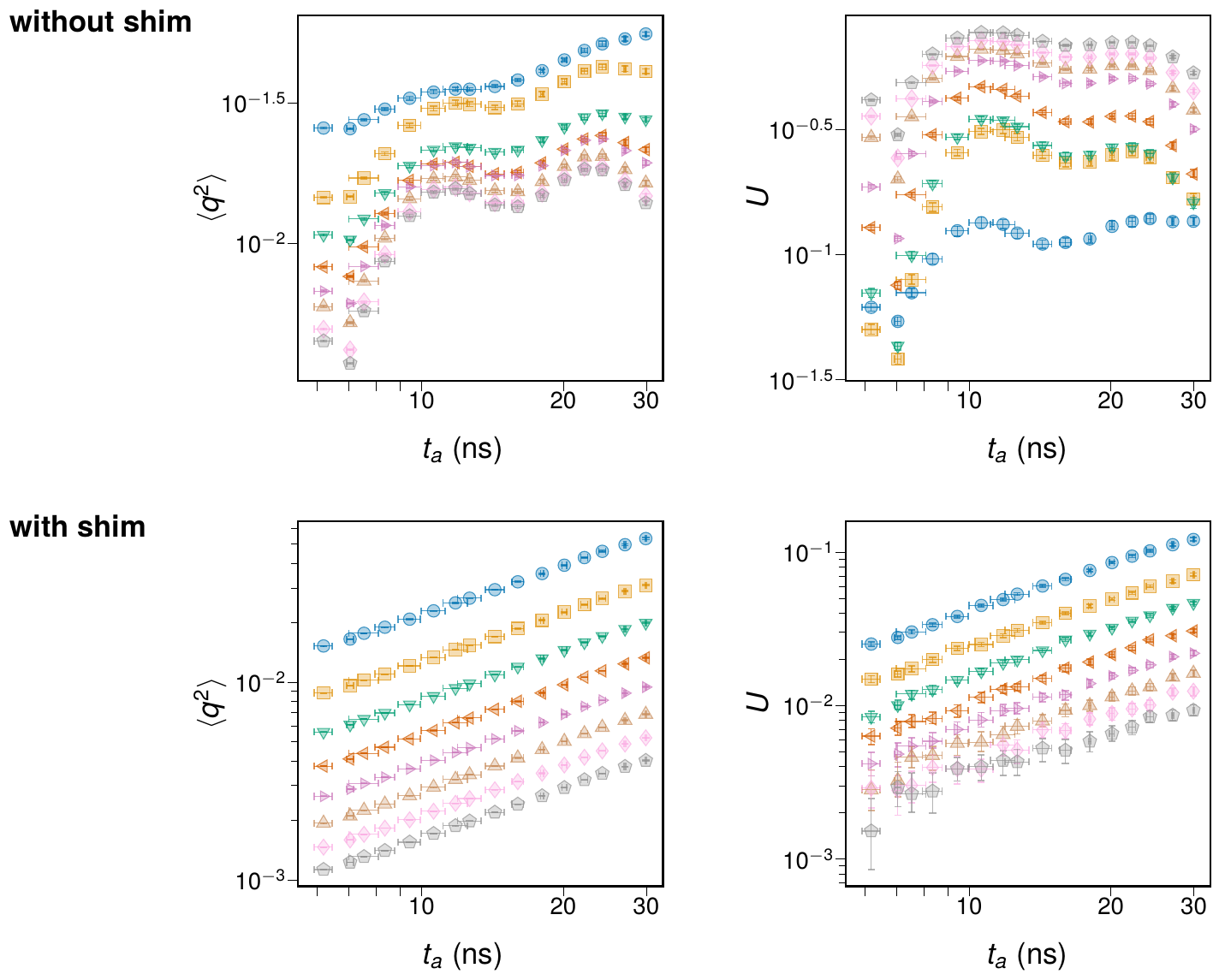}
\caption{{\bf Effect of shim on dynamic scaling.}  Shown in the top and bottom rows, respectively, are Binder cumulant $U$ and order parameter $\langle q^2\rangle$ with and without the anneal offset (synchronization) and flux offset (balancing) shim.  Data are for 3D lattices, $J_G=0.5$.}\label{fig:sm_shim_scaling}
\end{figure}

We perform the shim based on two assumptions:
\begin{itemize}
\item All qubits should have average magnetization zero.
\item The average frustration probability of a coupler in $E_{ij}$---that is, the average observed probability of a nonzero coupler between $V_i$ and $V_j$ having a positive contribution to the energy, where the average is taken over both samples and realizations---should be effectively independent of the choice of $i$ and $j$.
\end{itemize}
The first assumption is trivially justified because there are no longitudinal fields used in the Ising Hamiltonian in this work.  For the second, we assume that the sets $E_{ij}$ are large and sufficiently spatially uncorrelated from the position in the 3D lattice position.  This assumption is reasonable because the annealing line assignments follow a regular geometric pattern and the 3D lattice embeddings are determined with a heuristic random approach.

In Fig.~\ref{fig:sm_shim_meta}a we show the final anneal offsets for the eight annealing lines after 1200 iterations on $L=12$ 3D spin glasses.  In Fig.~\ref{fig:sm_shim_meta}b we show the distribution of average frustration in a glass coupling incident to each annealing line.  Data are shown with and without the anneal offset and flux offset shim, for $t_a=\SI{6.2}{ns}$ and $\SI{30}{ns}$.  The calibration refinement shim has a clear homogenizing effect on the per-line frustration.  In Fig.~\ref{fig:sm_shim_meta}c we show the average magnetization of each qubit over the final 300 of 1200 iterations with and without the shim, which has a clearly beneficial effect in balancing the qubits at zero magnetization.

In Fig.~\ref{fig:sm_shim_scaling} we show the effect of calibration refinement on the dynamic scaling of the order parameter and Binder cumulant.  These data make it clear that the calibration refinement is necessary to obtain reasonable estimates of critical exponents.  However, we are focusing on a regime far outside the specifications of the calibration being refined, which is intended for $t_a\geq \SI{500}{ns}$.

\section{Residual energy decay}\label{sec:energy}

\begin{figure}
\includegraphics[scale=1]{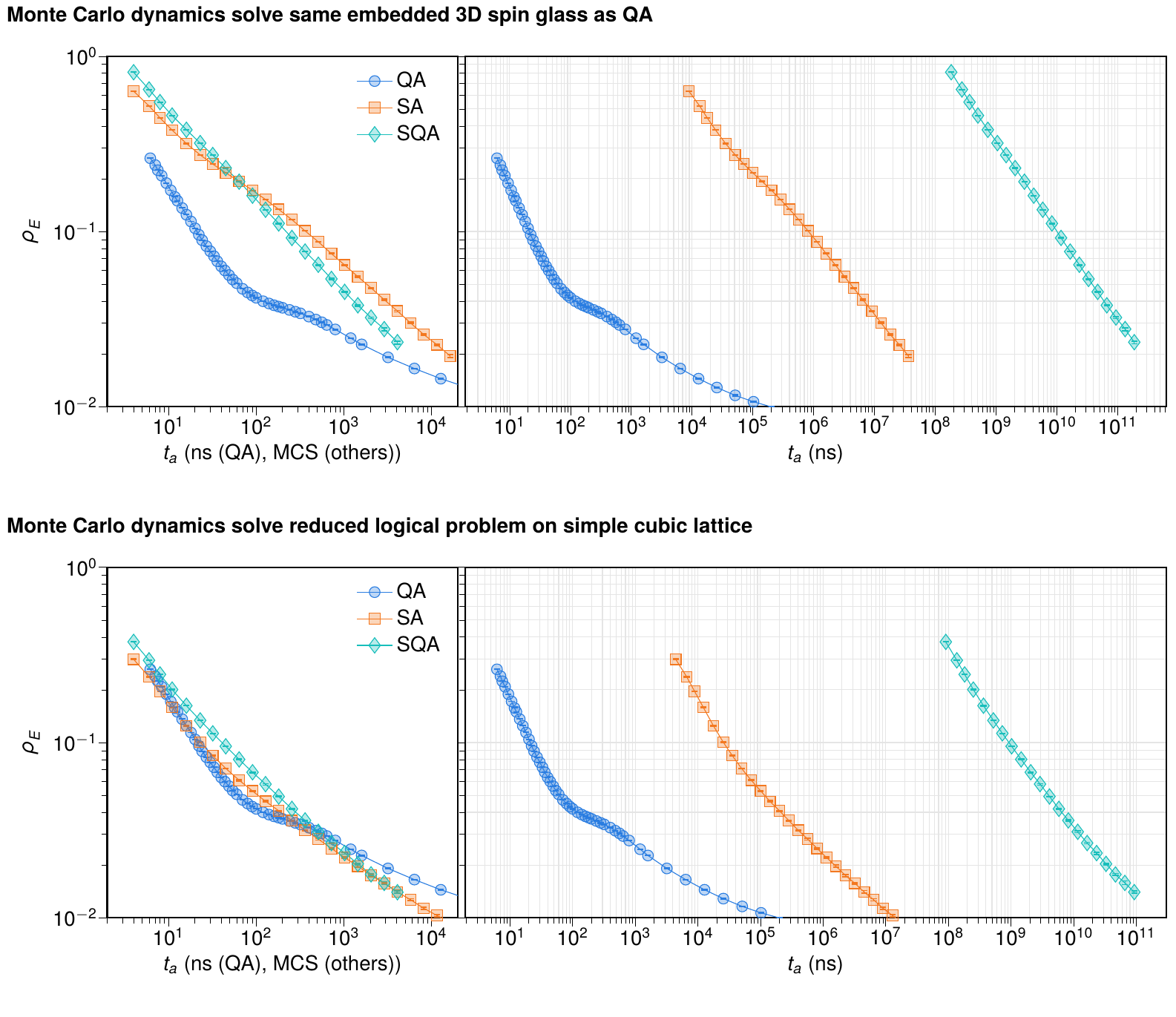}
\caption{{\bf Residual energy decay.}  We compare $\rho_E$ as a function of $t_a$ for QA, SA, and SQA.  The top row shows data for MC dynamics running on the same Ising model as QA, with two-qubit FM dimers.  The bottom row shows the same data for QA, but with MC dynamics running on a reduced Ising model on a simple cubic lattice, in which two-qubit FM dimers have been contracted into single logical variables, reducing both time per sweep and the number of sweeps required to reach a given $\rho_E$.  The right-hand plots show annealing time in nanoseconds.  For SQA we use measured sweep times; for SA we use a time of $\SI{0.4}{ns}$ per spin update as reported in Ref.~\cite{Isakov2014} for a highly optimized SA code.  Our SQA timescales rely on worldline updates, for which open-source code is provided \cite{King2021a}.}\label{fig:sm_burndown}
\end{figure}

In Fig.~\ref{fig:sm_burndown} we show decay of residual energy for three dynamics: QA, SA, and SQA.  For the MC solvers (i.e., SA and SQA), we show this both in terms of MCS and in terms of computation time.  We measured times per MCS on a CPU (Intel® Core™ i7-7700HQ CPU @ 2.80GHz): $\SI{0.4}{ms}$ for SA and $\SI{8.5}{ms}$ for SQA. 
We call out two caveats:
\begin{itemize}
\item SQA time per sweep is approximately linearly dependent on $\beta$, and we have used a high value $\beta=64$ throughout (relative to crossing point $\Gamma(s)=J(s)$; see Methods), to minimize thermal effects.  However, even with a lower $\beta$, SQA is not competitive with SA on the systems studied here.
\item The codes used are reasonably fast but are written to be general, without optimizations such as lattice-specific memory structure, function lookup tables, static spin ordering for sweeps (which deviates from standard interpretations of quasi-physical dynamics), and random number reuse.  SA in particular can be sped up significantly, so we show SA annealing time using the quoted time of $2.42$ spin flips per nanosecond ($\SI{0.4}{ns}$ per spin flip) for a highly optimized version of SA measured on an Intel® Xeon™ E5-2670 CPU @ 2.60GHz (\cite{Isakov2014}, Table~4, \texttt{an\_ss\_ge\_fi}).
\end{itemize}

As previously noted in Section \ref{sec:gse}, the ground state energies of these ``embedded'' spin glasses, which use two-qubit ferromagnetic dimers, can generally be found by instead solving a reduced ``logical'' problem on a simple cubic lattice.  Therefore we also compare QA performance on the embedded spin glasses against MC dynamics solving the logical spin glasses.  Although the primary aim of this work is to study quantum critical spin-glass dynamics, it is notable that QA outperforms SA---both in scaling (in the coherent QA regime) and in absolute terms---even when QA has an embedding overhead that does not affect SA.

\end{document}